\documentclass[useAMS,usenatbib]{mn2e}
\usepackage{amsmath}
\usepackage{amssymb}

\usepackage{graphics}
\usepackage{graphicx}
\usepackage{epstopdf}
\usepackage{footnote}
\usepackage{tablefootnote}
\usepackage{txfonts}
\usepackage{lipsum}
\usepackage{float}
\usepackage[draft]{hyperref}
\usepackage{macros}
\usepackage{fixltx2e}

\makesavenoteenv{tabular}

\DeclareSymbolFont{cmletters}{OML}{cmm}{m}{it}
\DeclareMathSymbol{v}{\mathalpha}{cmletters}{"76}

\usepackage[usenames,dvipsnames,svgnames,table]{xcolor}
\usepackage{hyperref}
\definecolor{darkblue}{rgb}{0.0,0.0,0.3}
\hypersetup{colorlinks,breaklinks,
            linkcolor=darkblue,urlcolor=darkblue,
            anchorcolor=darkblue,citecolor=darkblue}

\title[Feeding Sagittarius A*]{Hydrodynamic Simulations of the Inner Accretion Flow of Sagittarius A* Fueled By Stellar Winds
 }
\author[S. M. Ressler, E. Quataert, J. M. Stone ]{S. M. Ressler$^{1},$  E. Quataert$^{1},$ J. M. Stone$^{2}$\\
$^{1}$Departments of Astronomy \& Physics, Theoretical Astrophysics Center, University of California, Berkeley, CA 94720 \\
$^{2}$Department of Astrophysical Sciences, Princeton University, Princeton, NJ 08544}

\begin{document}

\maketitle

\begin{abstract}
We present {\tt Athena++} grid-based,  hydrodynamic simulations of accretion onto Sagittarius A* via the stellar winds of the $\sim 30$ Wolf-Rayet stars within the central parsec of the galactic center.  These simulations span $\sim$ 4 orders of magnitude in radius, reaching all the way down to 300 gravitational radii of the black hole, $\sim 32$ times further in than in previous work.  We reproduce reasonably well the diffuse thermal X-ray emission observed by \emph{Chandra} in the central parsec. The resulting accretion flow at small radii is a superposition of two components: 1) a moderately unbound, sub-Keplerian, thick, pressure-supported disc that is at most (but not all) times aligned with the clockwise stellar disc, and 2) a bound, low-angular momentum inflow that proceeds primarily along the southern pole of the disc. We interpret this structure as a natural consequence of a few of the innermost stellar winds dominating accretion, which produces a flow with a broad distribution of angular momentum.  Including the star S2 in the simulation has a negligible effect on the flow structure. Extrapolating  our results from simulations with different inner radii, we find an accretion rate of $\sim$ a few $\times 10^{-8} M_\odot$/yr at the horizon scale, consistent with  constraints based on modeling the observed emission of Sgr A*.  The flow structure found here can be used as more realistic initial conditions for horizon scale simulations of Sgr A*.
\end{abstract}

\section{Introduction}
Both the Event Horizon Telescope (EHT, \citealt{Doeleman2008}) and GRAVITY \citep{Gillessen2010} will soon reach resolution comparable to the event horizon scale of the supermassive black hole at the center of our own galaxy, Sagittarius A* (Sgr A*), at 230 GHz and in the infrared, respectively.  The primary source of emission in Sgr A* is believed to be a combination of thermal and nonthermal particles in either an accretion disc or the strongly magnetized outflow fed by a disc.   The properties of the plasma immediately surrounding the black hole are then coupled with the properties of the black hole itself in determining what we will actually observe.  In order to properly interpret current and forthcoming observations and to be able to infer physical parameters from the data, it is of paramount importance to have theoretical and computational models of the inner accretion flow.

Because the luminosity of Sgr A* is well below the Eddington limit, it is classified as a Radiatively Inefficient Accretion Flow (RIAF).  RIAFs are well suited for numerical simulation because they are geometrically thick, meaning that they can more easily be resolved than their thin-disc counterparts. To date, a number of groups have simulated RIAFs around rotating black holes in the Kerr metric using general relativistic magneto-hydrodynamic simulations (GRMHD,\citealt{Komiss1999,DeVilliers2003,Gammie2003,White2016}).  However, due to the large temperatures and low densities inherent in RIAFs, a variety of collisionless effects not captured in the standard ideal MHD framework may be dynamically important.  Recent work has made great strides in this respect by incorporating increasingly sophisticated physics into simulations.  This includes considering the plasma as a two-temperature fluid  \citep{Ressler2015,Sadowski2017,Ressler2017,Chael2018}, fully coupling radiation to the MHD equations \citep{Koral,BHLIGHT,Sadowski2017,Ryan2017,Chael2018}, injecting nonthermal particles into the fluid \citep{Ball2016,Chael2017}, as well as adding the effects of anisotropic electron conduction \citep{Ressler2015,Ressler2017}, anisotropic ion conduction, and anisotropic viscosity \citep{Foucart2016,Foucart2017}.

These simulations, however, predominantly use a fairly standard set of initial conditions. An equilibrium, constant angular momentum torus (e.g., \citealt{Fishbone1976}, though see also \citealt{Penna2013}, \citealt{Witzany2017}), surrounded by empty space is seeded with a magnetic field, a configuration which is unstable to the magneto-rotational instability (MRI).  As the instability grows, enough angular momentum is transported outward so that the torus is able to accrete and eventually reach an approximate steady state in which the magnetic energy is comparable to the thermal energy of the disc.  The flow structure can, however, depend strongly on the initial conditions.  For instance, if there is a net vertical magnetic flux in the equilibrium torus, an entirely different evolution is seen in which the flux threading the black hole eventually becomes large enough to halt accretion, leading to a violently time-variable, magnetically arrested disc (MAD, \citealt{Narayan2003,Sasha2011}). In contrast to the growing body of work on plasma microphysics, there has been much less work done studying the effect of varying the initial conditions on GRMHD simulation results; much of what has been done has focused in the possibility that the angular momentum vector of the disc is misaligned with the spin of the black hole \citep{Fragile2005,Liska2017}.

In general, not much is known about the feeding of black holes in galactic nuclei.  For the case of the galactic center in particular, however, we have a unique opportunity to actually determine a proper set of initial conditions, as the source of accretion is believed to be known. This source is the stellar winds of the $\approx 30$ Wolf-Rayet (WR) stars orbiting within $\sim 1$ pc of the black hole.  Though there are over a million other stars in the central nuclear star cluster \citep{FK2017}, including the well known ``S-stars'' whose orbits have been used to significantly improve estimates of the mass of Sgr A* \citep{Ghez2008,Gillessen2009}, these stars are generally fainter and less massive, with mass loss rates orders of magnitude smaller than the WR stars (see e.g. \citealt{Vink2001,Habibi2017}).  Since the mass loss rates and wind velocities \citep{Martins2007,YZ2015}, as well as the positions and orbital velocities \citep{Paumard2006,Lu2009} of the WR stars have been well constrained by both infrared and radio observations, this problem is well posed.   Moreover, both simple 1D calculations \citep{Quataert2004,Shcherbakov2010} and 3D smoothed particle hydrodynamic (SPH) simulations \citep{Rockefeller2004,Cuadra2008} have shown that the observed stars provide more than enough mass to explain the observed accretion rate onto Sgr A* and the diffuse X-ray emission in the galactic center observed by 
\emph{Chandra} \citep{Baganoff2003}.  

In this work we seek to better inform initial conditions of GRMHD simulations of Sgr A* by directly simulating the accretion flow produced by the winds of the WR stars in the galactic center. Though this calculation is similar to the work of \citet{Cuadra2008} (see also, \citealt{Rockefeller2004,Cuadra2015,Russell2017}), we use completely different numerical methods, probe even smaller radii, and focus especially on the properties of the innermost accretion flow, which has not been a primary focus of previous work. 

To do this, we employ three dimensional hydrodynamic simulations with $\sim$30 independent orbiting stars as sources of mass, momentum, and energy.  While it is almost certainly true that on scales comparable to the event horizon of Sgr A* magnetic fields play an important role in the transport of angular momentum, and thus, in determining the structure of the accreting plasma, here we focus on a purely hydrodynamic calculation. This is primarily because, even if magnetic fields are important for the gas near the horizon, the properties of the flow at larger radii may be set by strictly hydrodynamic considerations.  Furthermore, in order to properly evaluate the effects of magnetic fields in the future, we must first understand the detailed properties of the hydrodynamic simulation, meaning that this work will serve as a basis for comparison to subsequent calculations.  In addition, both the direction and the magnitude of the magnetic fields in the WR stellar winds are unconstrained observationally, so that a full treatment will require a larger exploration of parameter space than is needed in the purely hydrodynamic case. 

The paper is organized as follows.  \S \ref{sec:Model} describes the physical model and numerical methods, \S \ref{sec:tests} describes two tests of our implementation of the subgrid stellar wind model, \S \ref{sec:application} details the properties of the full 3D simulation of stellar wind accretion onto Sgr A*, \S \ref{sec:Xray} compares the X-ray luminosity of our simulation to \emph{Chandra} observations, \S \ref{sec:imp} discusses the implication of these results for GRMHD simulations, \S \ref{sec:comp} compares our results to previous work, and \S \ref{sec:conc} concludes.


\section{Model and Computational Methods}
\label{sec:Model}
\subsection{Equations Solved}
We perform our simulations with {\tt Athena++}, a 3 dimensional grid-based scheme that solves the equations of conservative hydrodynamics. {\tt Athena++} is a complete rewrite of the widely used {\tt Athena} code \citep{Stone2008} optimized for the c++ coding language. We use a point source gravitational potential for the central black hole. The code is 2nd order in space and time and adopts piece-wise linear reconstruction with the Harten-Lax-van Leer-Contact (HLLC) Riemann solver.  

In addition to the basic equations of hydrodynamics, we include the effect of the stellar winds emitted by stars orbiting the black hole by adding source terms in mass, energy, and momentum.  Each star is assumed to orbit in a Keplerian orbit as described in \S \ref{sec:orbits}.  The wind of each star is given an effective radius of $r_{wind} \approx 2$ cells centered on the position of the star's orbit (more precisely, twice the length of the diagonal of a cell determined by the local level of mesh refinement).  Inside this radius the wind is assumed to supply a constant source of mass that is determined by the observed mass loss rate, $\dot M_{wind}$: $\dot \rho_{wind}  = \dot M_{wind} / V_{wind}$, where $V_{wind} = 4\pi /3$ $ r_{wind}^3$.  Furthermore, the wind is assumed to have a constant radial velocity in the frame of the star, $v_{wind}$, and a negligible pressure. To calculate the net source terms for the finite volume, conservative equations solved by {\tt Athena++} we break each cell that intersects a stellar wind into a $5 \times 5 \times 5$ subgrid and integrate over the whole cell. For a wind which occupies a fractional volume $f$ of a cell, this amounts to source terms in mass, momentum, and energy of $f \dot \rho_{wind}$, $f \dot \rho_{wind} \langle {\mathbf v_{wind,net}}\rangle$, and $1/2 f \dot \rho_{wind} \langle { |\mathbf v_{wind,net}}|^2\rangle$, respectively, where ${\mathbf v_{wind,net}}$ is the wind velocity in the fixed frame of the grid and $\langle \rangle$ denotes an average over the volume of the star contained in the cell.  Though similar in purpose, we note that this model differs from \citet{Lemaster2007} in that the stellar winds are treated as source terms as opposed to ``masked regions,'' within which the fluid quantities are over-written by an analytic solution.  The benefit of treating the winds as source terms is that we can accommodate scenarios where multiple stellar winds overlap.   

As the stellar winds interact and shock-heat, radiative losses due to optically thin bremsstrahlung and line cooling are expected to become significant. To account for this, we use the optically thin cooling routine described in \citet{Townsend2009}, which analytically integrates the energy equation over a single time step using a piece-wise power law approximation to the cooling curve.  This avoids any limitation on the accuracy or time step when the cooling time is short compared to the dynamical time of the fluid.  The piece-wise power law approximation to the cooling curve is obtained from a tabulated version of the exact collisional ionization equilibrium cooling function (as is appropriate for the hot $\sim 10^7$ K gas in the Galactic Center; see the next section for details).  

To summarize, the equations we solve are the equations of conservation of mass, momentum, and energy, with source terms to account for the gravity of the supermassive black hole, optically thin radiative cooling, and the stellar winds of the orbiting stars:
\begin{equation}
\label{eq:hydro}
\begin{aligned}
\frac{\partial \rho}{\partial t} + \mathbf{\nabla} \cdot \left(\rho \mathbf{v}\right) &= f \dot \rho_{wind} \\
\frac{\partial \left(\rho \mathbf{v}\right)}{\partial t} + \mathbf{\nabla} \cdot \left(P \mathbf{I} + \rho \mathbf{v}\mathbf{v}\right) &= -\frac{\rho GM_{BH}}{r^2} \hat r  \\&+ f \dot \rho_{wind} \langle {\mathbf v_{wind,net}}\rangle \\
\frac{\partial \left(E \right)}{\partial t}  + \mathbf{\nabla} \cdot \left[(E+P)\mathbf{v}\right] &= - \frac{\rho G M_{BH}}{r} \mathbf{v}\cdot \hat r \\ &+ \frac{1}{2} f \dot \rho_{wind} \langle { |\mathbf v_{wind,net}}|^2\rangle - Q_{-}, 
\end{aligned}
\end{equation} 
where $\rho$ is the mass density, $P$ is the pressure, $\mathbf{v}$ is the fluid velocity, $E = 1/2 \rho v^2 + P/(\gamma-1)$, $\gamma=5/3$ is the adiabatic index of the gas, and $Q_{-}$ is the cooling rate per unit volume.  The calculation of $Q_{-}$ is described in the next section. Note that in equation \eqref{eq:hydro}, we have neglected the effect of the central nuclear star cluster on the gravitational potential.  For the galactic center, the gravitational contribution from these stars is negligible for $r \lesssim 5'' \approx 0.2$ pc but is non-negligible ($\sim 25 \%$) for $r\gtrsim 10'' \approx 0.4$ pc \citep{Genzel2003cusp}.  In the innermost regions of the domain that are the primary focus of this work, neglecting the stellar contribution to gravity is a good approximation.  
   
\subsection{Calculating The Cooling Function}
\label{sec:cool}

We define the cooling function, $\Lambda$, such that the cooling rate per unit volume is $Q_{-} = n_e \frac{\rho}{m_p} \Lambda$, where $n_e = \rho/\mu_e$, $m_p$ is the mass of a proton, and $\mu_e$ is the mean molecular weight per electron. For the conditions in the galactic center, the dominant cooling mechanisms are line emission in collisional ionization equilibrium and thermal bremsstrahlung. The cooling function is thus a function not only of temperature but also of the relative abundances of the elements.
To calculate $\Lambda$ for a given set of hydrogen, helium, and metal mass fractions ($X,Y$ and $Z$, respectively), we first calculate the cooling curve for the photospheric solar abundances presented in \citet{Lodders2003}, that is, $X_\odot = 0.7491$, $Y_\odot = 0.2377$,and $Z_\odot =0.0133 $.  We do this using the spectral analysis code {\tt SPEX} \citep{SPEX} in the manner of \citet{Schure2009}, and calculate separately the contributions from H, $\Lambda_{H,\odot}$,  He, $\Lambda_{He,\odot}$, and metals, $\Lambda_{Z,\odot}$.  Then we can write the cooling curve for arbitrary abundances as a linear combination of these solar quantities as
\begin{equation}
\Lambda = \frac{X}{X_\odot} \Lambda_{H,\odot} + \frac{Y}{Y_\odot} \Lambda_{He,\odot}+ \frac{Z}{Z_\odot} \Lambda_{Z,\odot}.
\label{eq:Lambda}
\end{equation}
The mean molecular weight per electron, $\mu_e$, and the mean molecular weight per particle, $\mu$, are directly related to $X$ and $Z$ by \citep{Townsend2009}
\begin{equation}
\label{eq:mu}
\begin{aligned}
\mu_e =& \frac{2 m_p}{1+X} \\
\mu   =& \frac{m_p}{2X + 3(1-X-Z)/4+Z/2},
\end{aligned}
\end{equation} 
where we have made the approximation that the majority of the mass in metals is provided by oxygen, and that the mean molecular weights are constant. The former is a good approximation assuming that the relative abundance of metals are roughly solar, while the latter is a good approximation for $T\gtrsim$ a few $ \times 10^4$ K for a gas composed of mostly hydrogen or $T\gtrsim 10^5 $ K for a gas composed of mostly helium. At lower temperatures, where hydrogen/helium become less ionized, the approximation breaks down.  This introduces an error in the cooling curve at lower temperatures, but this error only increases the sharpness at which $\Lambda \rightarrow 0$ and is thus limited to a small range in temperatures.  Furthermore, most of the gas in our simulation is above $10^5$ K, so this approximation does not significantly affect our results.

Once we have calculated the cooling curve, we then approximate it as a piece-wise power law composed of 12 carefully chosen segments over the range $10^4$ and $10^9$ K.  Above $10^9$ K we use a single power law, which is reasonable because at such high temperatures $\Lambda$ is dominated by thermal bremsstrahlung of electrons with either H or He. 

\begin{figure}
\includegraphics[width=0.45\textwidth]{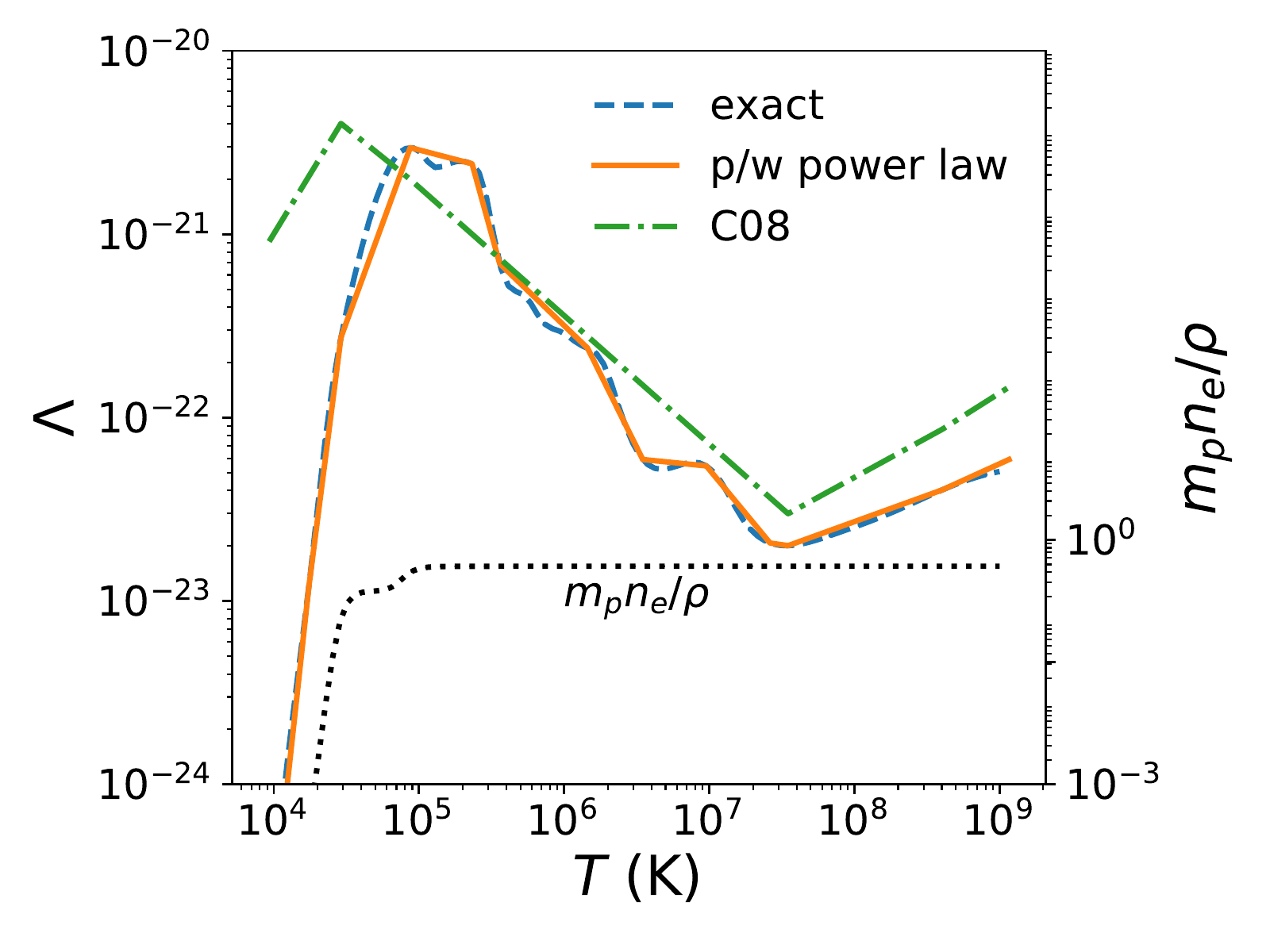}
\caption{Piece-wise power law approximation used in this work (solid) compared to the full cooling curve calculated by {\tt SPEX} (dashed, see \S \ref{sec:cool}) and the cooling curve used by \citet{Cuadra2008} (C08, dot-dashed). Our cooling curve is for hydrogen free, $Z = 3 Z_\odot$ gas appropriate for the Wolf-Rayet star winds near Sgr A*. Also plotted is the number of free electrons relative to $\rho/m_p$.  This shows that 1) the piece-wise power law does an excellent job capturing the shape of the full cooling curve, and 2) the simplification that $\mu_e \approx $ const. is well motivated for all temperatures in which the cooling curve is non-negligible.  The main difference between our cooling function and that of C08 is our choice to use $X=0$, which reduces the high temperature bremsstrahlung tail and moves the cut-off at low temperatures to slightly higher temperatures.  }
\label{fig:cc}
\end{figure}

The values of $X$ and $Z$ in the stellar winds is somewhat uncertain.  However, WR stars are typically bereft of Hydrogen, having ejected their outer hydrogen envelopes in earlier stages of stellar evolution.  We would thus expect their stellar winds to be composed of very little hydrogen and a higher fraction of metals.  Indeed, by fitting the spectra, \citet{Martins2007} find that the H/He ratio is small in most of the stars and suggest that higher values of $Z$ might be appropriate.  Therefore, for this work we adopt $X=0$ and $Z = 3 Z_\odot$. Note that this is also the metallicity assumed in several previous works (e.g., \citealt{Cuadra2008}, \citealt{Calderon2016}).  The resulting cooling curve is plotted in Figure \ref{fig:cc} along with the piece-wise power law approximation that we employ in our simulation. The agreement is excellent.  Also plotted in Figure \ref{fig:cc} is the ratio between the number of free electrons and $\rho/m_p$, which shows that the approximation of $\mu_e \approx$ const. is good for $T\gtrsim 10^5$ K.  For lower temperatures, Helium becomes mostly neutral and that approximation breaks down.  However, the cooling curve also rapidly decreases below $10^5$ K so this is not a significant source of error.

\subsection{Computational Grid and Boundary/Initial Conditions}
Our simulations are performed on a Cartesian grid to avoid the severe time step restriction inherent in 3D spherical-polar coordinates caused by the limited azimuthal extent of the cells near the pole. In addition, there is not necessarily an a priori symmetry axis in our problem, limiting the utility of spherical-polar coordinates.  To effectively resolve the smaller spatial scales of interest, we utilize nested levels of static mesh refinement (SMR) to resemble logarithmic spacing in radius.  Furthermore, to avoid an unphysical build-up of material in the cells near the origin, we remove a sphere of radius $r_{in}$ equal to twice the width of the smallest grid cell, replacing it with a region of negligible pressure, negligible density, and zero velocity. This allows material to flow into the ``black hole'' while limiting unphysical boundary effects to only a few cells outside of $r_{in}$.   Tests demonstrate that this effective inner boundary condition correctly reproduces the Bondi accretion solution.  The outer boundary condition is outflow in all directions.  

\subsection{Floors and Ceilings}
Since {\tt Athena++} evolves the conservative variables of mass density, momentum density, and total energy density, occasionally the primitive variables of $\rho$ and $P$ can reach unphysical (i.e., negative) values.  When this occurs, to prevent code failure, we utilize floors on the density and pressure such that if $\rho<\rho_{ \rm floor}$, we set $\rho = \rho_{\rm floor}$, and if $P<P_{\rm floor}$ we set $P=P_{\rm floor}$.  In particular, we adopt the values of $\rho_{\rm floor} = 10^{-7} M_{\odot}$pc$^{-3}$ and $P_{\rm floor} = 10^{-10} M_{\odot}$pc$^{-1}$kyr$^{-2}$ .  In runs with radiative cooling, we impose a minimum temperature of $10^4$ K which acts as an additional, density-dependent floor on pressure. The aforementioned floors are activated sufficiently rarely that they do not affect our results.

Additionally, unphysically large temperatures or velocities that occur in a handful of problematic cells can severely limit the time step of the simulation, which is set by the Courant$-$Friedrichs$-–$Lewy number multiplied by the maximum wave speed over all cells in the domain.  To limit the effect of these isolated cells, we impose a ceiling on both the sound speed and the velocity that is equal to $10$ times the free-fall velocity at the inner boundary.  If the sound speed of a cell exceeds this value, we reduce the pressure in that cell such that the new sound speed is equal to the ceiling.  When the magnitude of one of the components of the velocity exceeds the ceiling, we reduce the magnitude to the ceiling while keeping the sign fixed. In practice, we find that these ceilings are only necessary during the first time step of our simulations for cells located within the stellar wind source term. This is because the initial time step, which is set by the initial conditions of a cold, low density gas, is large compared to the wind crossing time in these cells. The time step is appropriately reduced after the first time step and the ceilings are no longer needed.

\section{Tests of Implementation}
\label{sec:tests}
In this section we describe two hydrodynamic simulations to both test and demonstrate the implementation of the model described above. 
\subsection{Stationary Stellar Wind} 
\label{sec:wind_test}
In order to test that our subgrid model for the stellar winds produces the desired effect, we place a single, stationary star with $v_{wind} =$ 1 pc/kyr $\approx 1000$ km/s and $\dot M_{wind} = 10^{-5}$ $M_{\odot}/yr$ at the center of a uniform, low density, low pressure medium. The grid is a cube of $128^3$ cells with three levels of mesh refinement, so that the box size is $\approx 300 r_{wind}$. We run the test for $\approx 2$ times the wind crossing time of the box. Absent gravity, as time evolves a steady state should be reached where the star drives a global wind with $\mathbf{v} = v_{wind} \hat r$ and $\rho = \dot M_{wind}/(4 \pi v_{wind} r^2)$.  

\begin{figure*}
\includegraphics[width=0.45\textwidth]{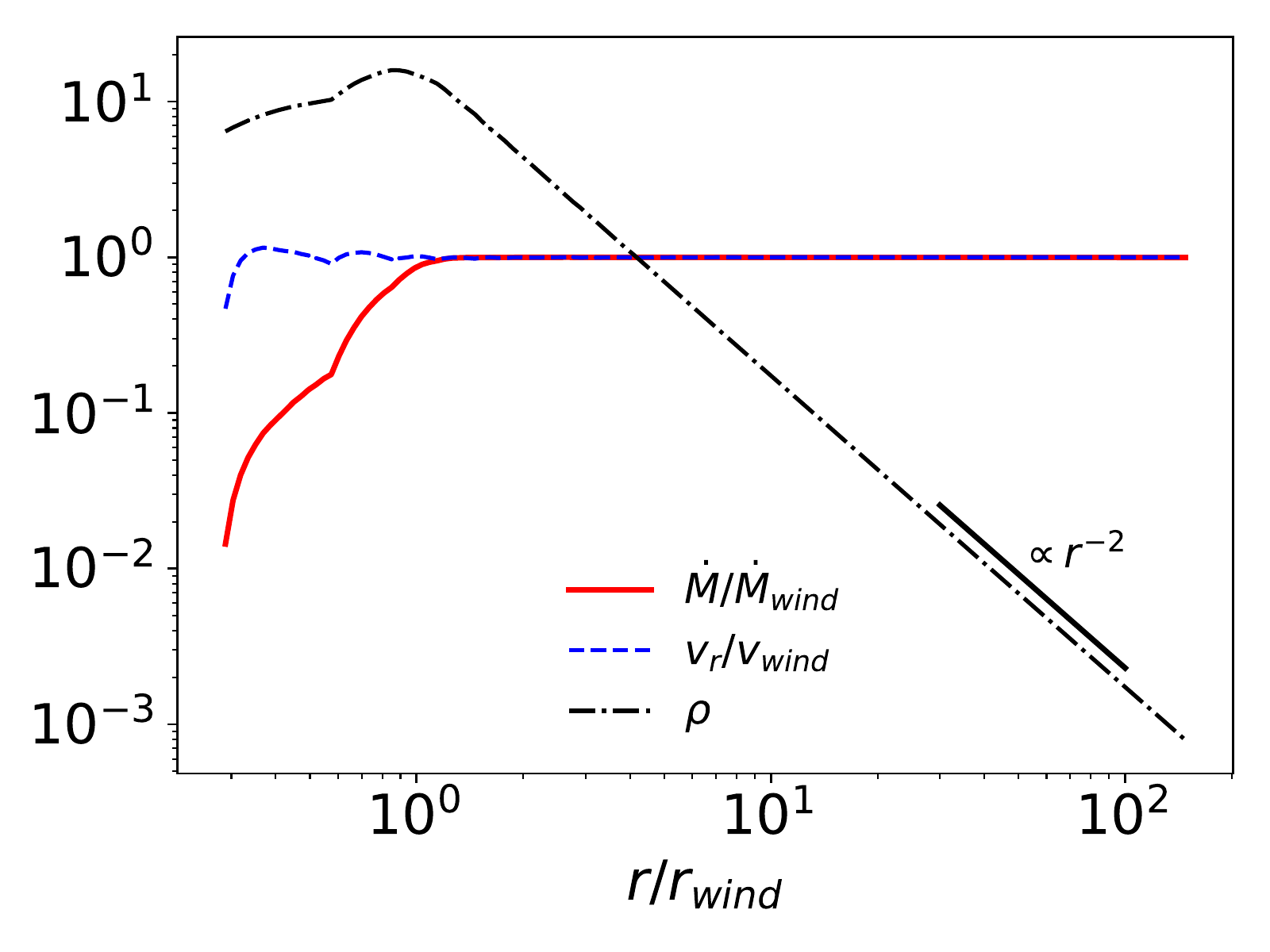}
\includegraphics[width=0.49\textwidth]{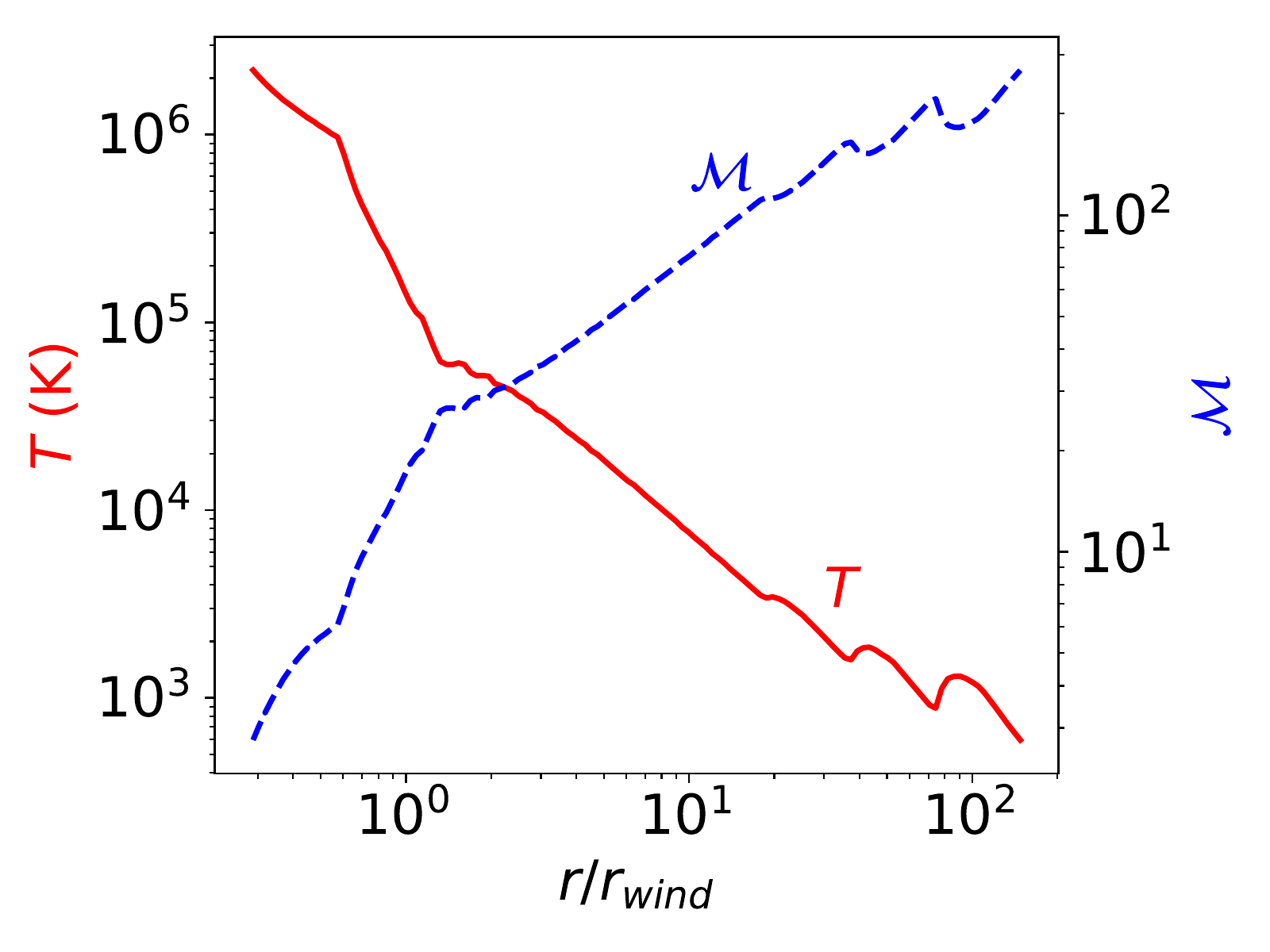}
\caption{Left: Angle averaged outflow rate, radial velocity, and mass density profiles for a single, stationary stellar wind at the center of a uniform, low pressure medium (\S \ref{sec:wind_test}). The wind is sourced in a sphere of radius $r_{wind} \approx$ 2 cells. Right: Temperature profile and mach number, $\mathcal{M}\equiv v_{r}/c_s$, in the same test.  For $r>r_{wind}$, the angle-averaged  $\dot M$, $v_r$, and $\rho$ match nearly perfectly with the desired solution. As desired, the wind is also cold, with the Mach number at the base of the wind of $\approx 30$ and rising with increasing distance from the base. Note that the bumps in the temperature profile for $r/r_{wind} \gtrsim 20$ (directly corresponding to the bumps in Mach number) are caused by truncation error as the internal energy of the gas drops to the level of the numerical precision of the total energy.   }
\label{fig:wind_test}
\end{figure*}

\begin{figure}
\includegraphics[width=0.49\textwidth]{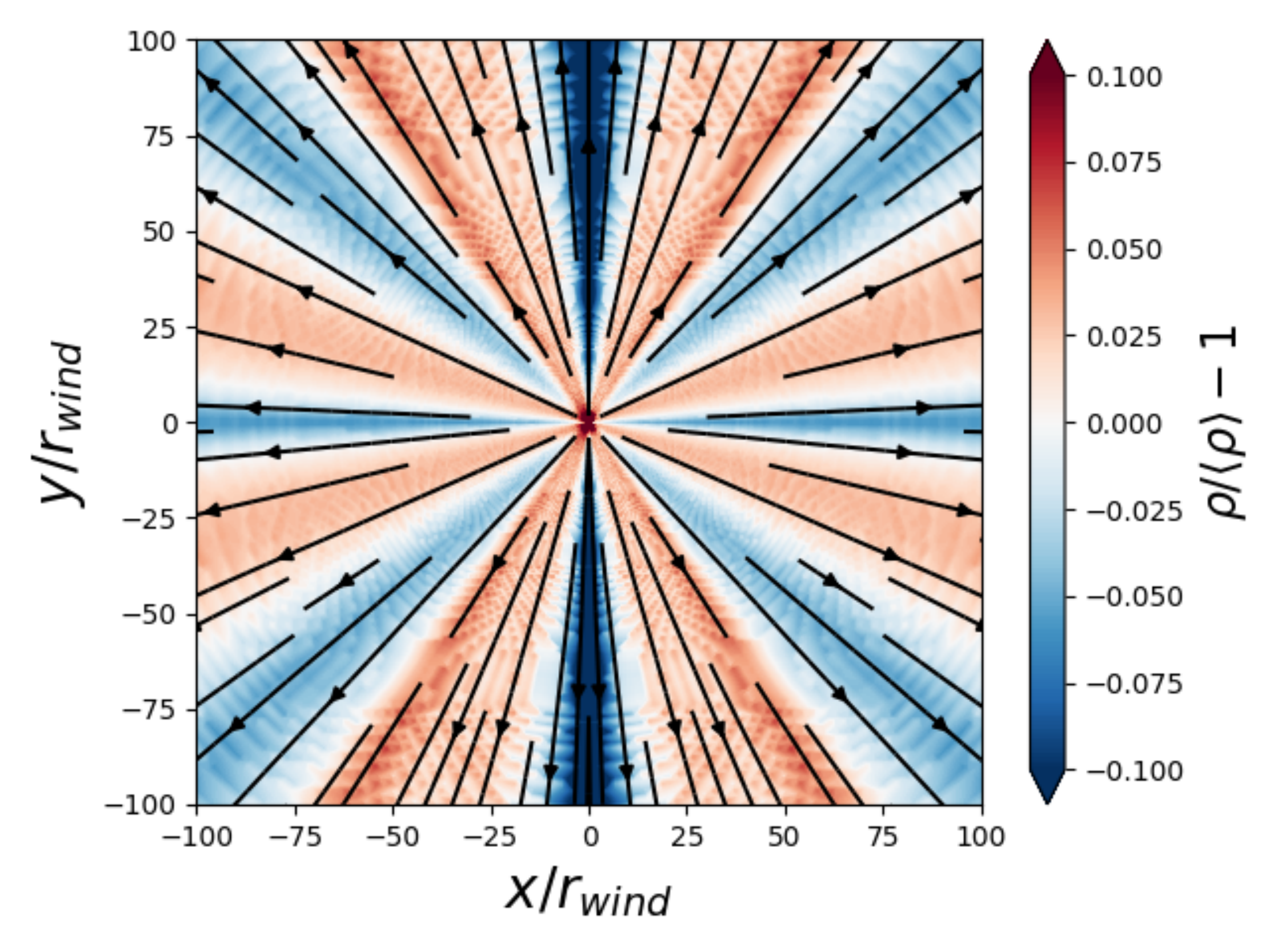}
\caption{Deviation of the $\varphi$-averaged density from spherical symmetry for the stationary stellar wind test (see \S \ref{sec:wind_test}, Figure \ref{fig:wind_test}), over-plotted with velocity streamlines.  These deviations caused by the Cartesian grid are acceptably small, $<10 \%$ everywhere.}
\label{fig:wind_test_contour}
\end{figure}

Our simulation shows excellent agreement with the analytic solution, as shown in the left panel of Figure \ref{fig:wind_test}, where the angle averaged density, outflow rate, and radial velocity are all essentially equal to the expected values for $r>r_{wind}$.  In principle, the temperature of the wind should be $\approx 0$, but in practice there is a finite amount of thermal pressure added by the model described in \S \ref{sec:Model} due to the difference between $|\langle\mathbf v_{wind,net} \rangle |^2$ and $\langle |\mathbf v_{wind,net}|^2\rangle$ caused by the averaging of a purely radial velocity over a Cartesian grid cell. This effect, however, is sufficiently small for our purposes, as shown in the right panel of Figure \ref{fig:wind_test}.  At the base of the wind, the radial Mach number of the flow is $\approx 30$ and increases due to adiabatic cooling as $\sim (r/r_{wind})^{2/3}$, showing that the thermal pressure is a negligible contribution to the wind dynamics.  For this particular choice of $v_{wind}$ and $\dot M_{wind}$, which are typical of the values of the stars contributing to accretion onto Sgr A*, this corresponds to a base wind temperature of $\approx 2 \times 10^{4}$ K that decreases as $\sim (r/r_{wind})^{-4/3}$.  Furthermore, despite the Cartesian nature of the grid, the generated wind is still approximately spherically symmetric, as seen by the relatively small deviations ($<10 \%$) from spherical symmetry shown in Figure \ref{fig:wind_test_contour}.

\subsection{Isotropic Stars on Circular Orbits}
\label{sec:rand_stars}
In order to test our  implementation of the stellar winds in a more complicated and dynamic problem, we seek to reproduce the results of \citet{Quataert2004}, in which the winds of the stars orbiting Sgr A* were modeled in spherical symmetry using a smooth source term in mass and energy.  To do this, we place $ 720$ stars in circular orbits in a point source gravitational potential, roughly uniformly distributed in solid angle and uniformly spaced in radius between 2$^{\prime\prime}$ (0.08 pc) and 10$^{\prime\prime}$ (0.4 pc).  Each star has the same stellar wind velocity, namely, $1000$ km/s, and mass loss rate determined by requiring the total mass loss rate to be 10$^{-3}$ $M_{\odot}$/yr. Furthermore, we neglect radiative cooling. In the limit of an infinite number of stars, this should be equivalent to a smooth source term between 2$^{\prime\prime}$ (0.08 pc) and 10$^{\prime\prime}$ (0.4 pc) that depends only on radius and supplies a net addition of mass and energy without adding momentum (corresponding to $\eta = 2$ in the notation of \citealt{Quataert2004}).   Since we consider orbiting and not stationary stars, in order to make a proper comparison we add an additional source term to \citet{Quataert2004}'s spherically symmetric calculation to account for the extra kinetic energy in the injected gas due to orbital motion: $1/2$ $q(r)$ $GM_{BH}/r$ where $q(r)$ is the stellar mass loss rate per unit volume and $M_{BH} \approx 4 \times 10^6 M_\odot$.  The 3D simulation is run for 7 kyr, and performed with a base resolution of $128^3$ with 6 levels of nested mesh refinement on a $5^3$ pc$^3$ Cartesian grid, resulting in an inner boundary of $r_{in} \approx 2.4 \times 10^{-3}$ pc.

\begin{figure*}
\includegraphics[width=0.49\textwidth]{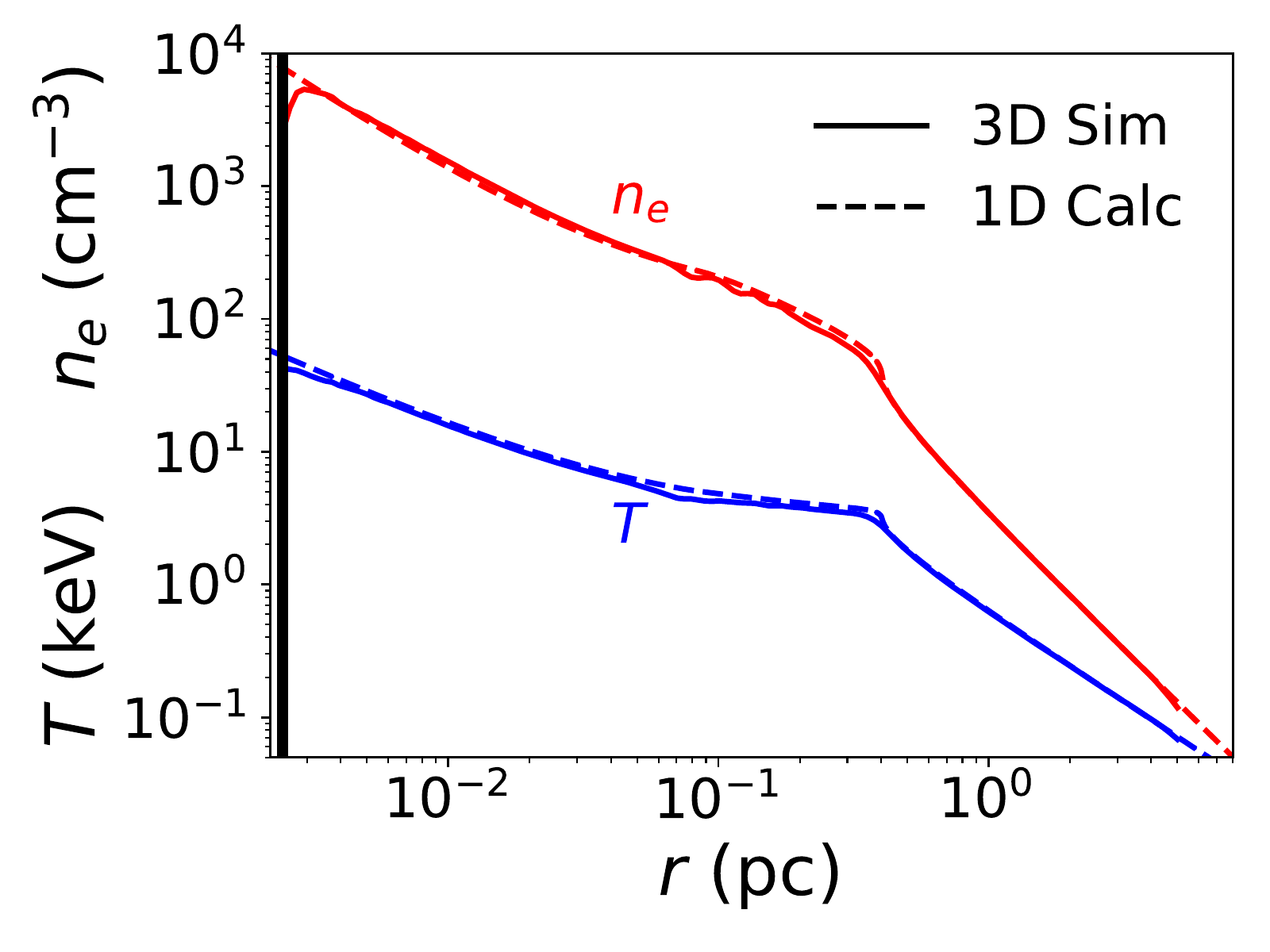}
\includegraphics[width=0.49\textwidth]{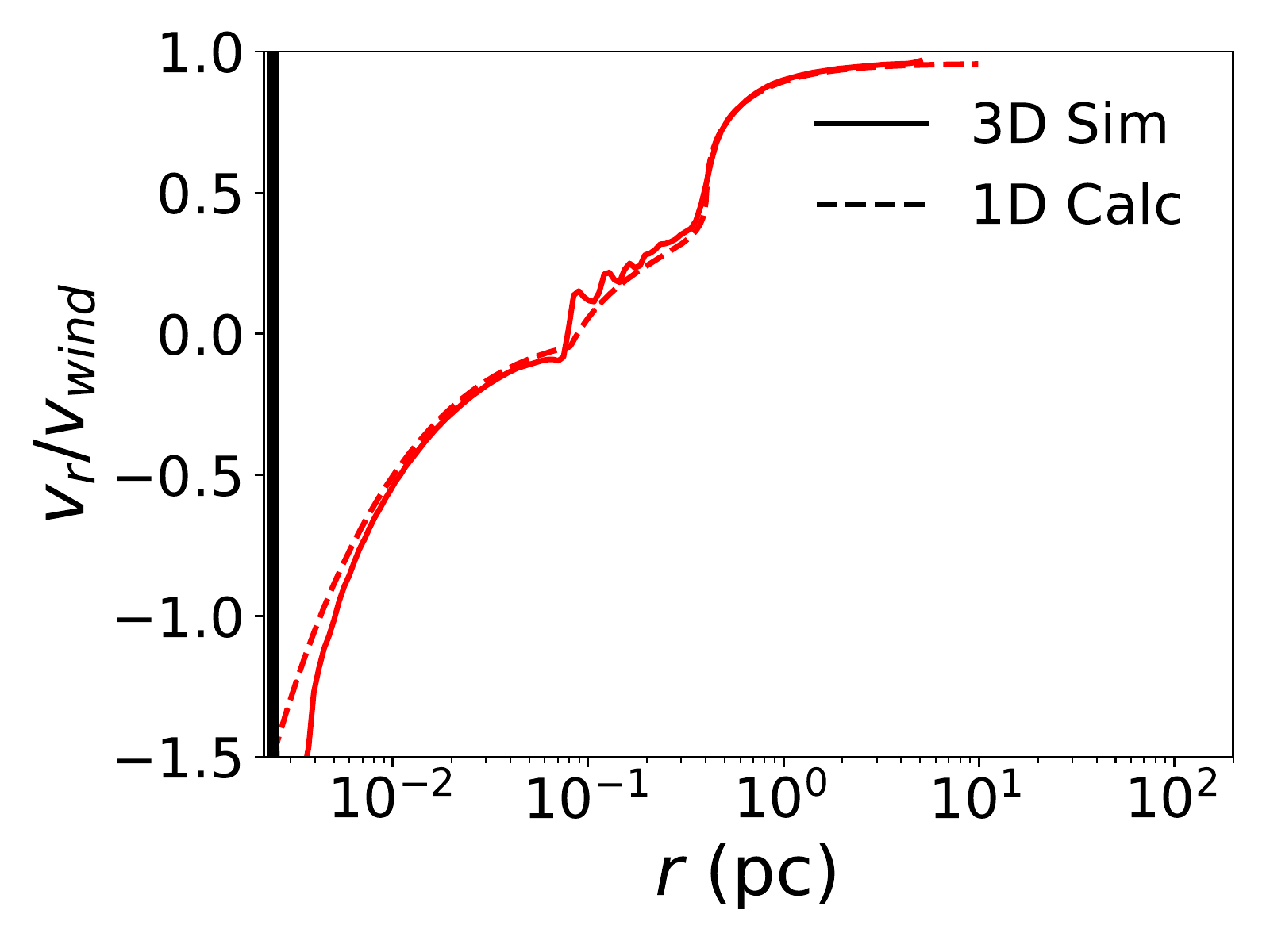}
\caption{Electron number density, $n_e$, temperature, $T$, (left) and radial velocity normalized to the stellar wind velocity, $v_r/v_{wind}$, (right), all averaged over solid angle for the isotropic circular orbits test described in \S \ref{sec:rand_stars}. Dashed lines represent the high resolution spherically symmetric solution with a smooth radial source of mass and energy, while solid lines represent a full 3D simulation of 720 stars in circular orbits uniformly spaced in solid angle and radius between 0.08-0.4 pc ($2-10''$). The black vertical line is the location of the inner boundary, $r_{in}$, for the 3D simulation.  We find excellent agreement between the two calculations, which verifies that our model for injecting stellar winds (\S \ref{sec:Model}) produces the desired results.   The small differences we find are due to both the finite number of stars and minor effects of the absorbing inner boundary condition.}
\label{fig:rand_stars}
\end{figure*}

The angle-averaged results for the density, temperature, and radial velocity in this test are shown in Figure \ref{fig:rand_stars}, over-plotted with the results of a high resolution 1D calculation using the smooth source term described in the preceding paragraph.  We find excellent agreement between the two calculations.  The small differences are (i) small variations in the  region where mass is injected due to the finite number of stars and (ii) small differences in the few cells closest to the absorbing inner boundary. We have verified that by moving the inner boundary to smaller radii, the agreement improves. These results verify that 1) our subgrid model for the stars effectively drives stellar winds with the desired accretion rate and wind speed and 2) the effects of the inner boundary condition are limited to only a few cells and do not affect the rate at which mass is captured by the black hole or the flow structure in the majority of the computational domain.

\section{3D Simulation of Accreting Stellar Winds Onto Sgr A*}
\label{sec:application}
In this section we focus on the problem of accretion onto Sgr A* as fed by the stellar winds of the 30 Wolf-Rayet stars and describe in detail the resulting flow properties.    

\subsection{Stellar Winds and Orbits}
\label{sec:orbits}
Before describing the simulation itself, we first briefly summarize the stellar wind parameters and orbits of the stars that we include as sources of mass, momentum, and energy.

Of the hundreds of stars orbiting Sgr A* at distances less than about a parsec, we include in our simulation only the $\approx 30$ Wolf-Rayet stars identified in \citet{Martins2007} as strong wind emitters. The wind speeds and mass loss rates that we set for each star are taken directly from Table 1 of \citet{Cuadra2008}, which summarizes \citet{Martins2007}. The locations of the stars are determined by solving Kepler's equation at each time step for the set of orbital elements corresponding to the present day location and velocities with respect to Sgr A*.  Unfortunately, while the proper motions, radial velocities, and positions in the plane of the sky for the stars are precisely measured \citep{Paumard2006,Lu2009}, their location in the plane of the sky (i.e. the $z$ direction) is undetermined because the acceleration measurements for nearly all of stars are consistent with 0 \citep{Lu2009}.  It was noted by \citet{Levin2003} (and later confirmed by \citealt{Belo2006,Lu2009,Bartko2009}), however, that the velocities of some of the stars lie within a thin planar structure, which allowed them to perform a likelihood analysis to precisely determine the $z$-coordinates of the disc-stars.  Some have proposed the existence of a second stellar disc (e.g. \citealt{Paumard2006}), but this disc remains uncertain \citep{Belo2006,Lu2009}. Thus, in order to determine the orbits of the remaining, non-disc-stars, we require a prescription for their $z$-coordinates.  For simplicity and ease of comparison to previous calculations, we adopt the ``1-disc'' model of \citet{Cuadra2008}\footnote{Note, however, that in this work we assume a black hole mass of 4.3 $\times 10^6$ $M_\odot$, resulting in orbits that are not quite identical to those in \citet{Cuadra2008}, who assumed a black hole mass of 3.6 $\times 10^6$ $M_\odot$.}, where the $z$-coordinate of the stars outside the stellar disc are determined by minimizing the eccentricity of the implied stellar orbit.  

In summation, for each disc-star as identified by \citet{Belo2006}, we use the velocities and three dimensional positions as listed in Table 2 of \citet{Paumard2006} to determine the stellar orbits, while for the remaining stars we use the velocities and two dimensional positions from Table 2 of \citet{Paumard2006} with $z$-coordinate determined by minimizing the eccentricity.  The single exception to this is the star S97 (aka E23), whose orbit has a short enough period to have been precisely determined (e.g., \citealt{Gillessen2009,Gillessen2017}).  For this star we use the orbital elements listed in Table 3 of \citet{Gillessen2017}.

In addition to the winds of the WR stars, we also perform one simulation that includes the stellar wind of the star S2. For this star we use the orbit given by \citet{Gillessen2017} and theoretical estimates of its mass-loss rate and stellar wind speed.  This is described in more detail in \S \ref{sec:S2}.

\begin{figure*}
\includegraphics[width=0.45\textwidth]{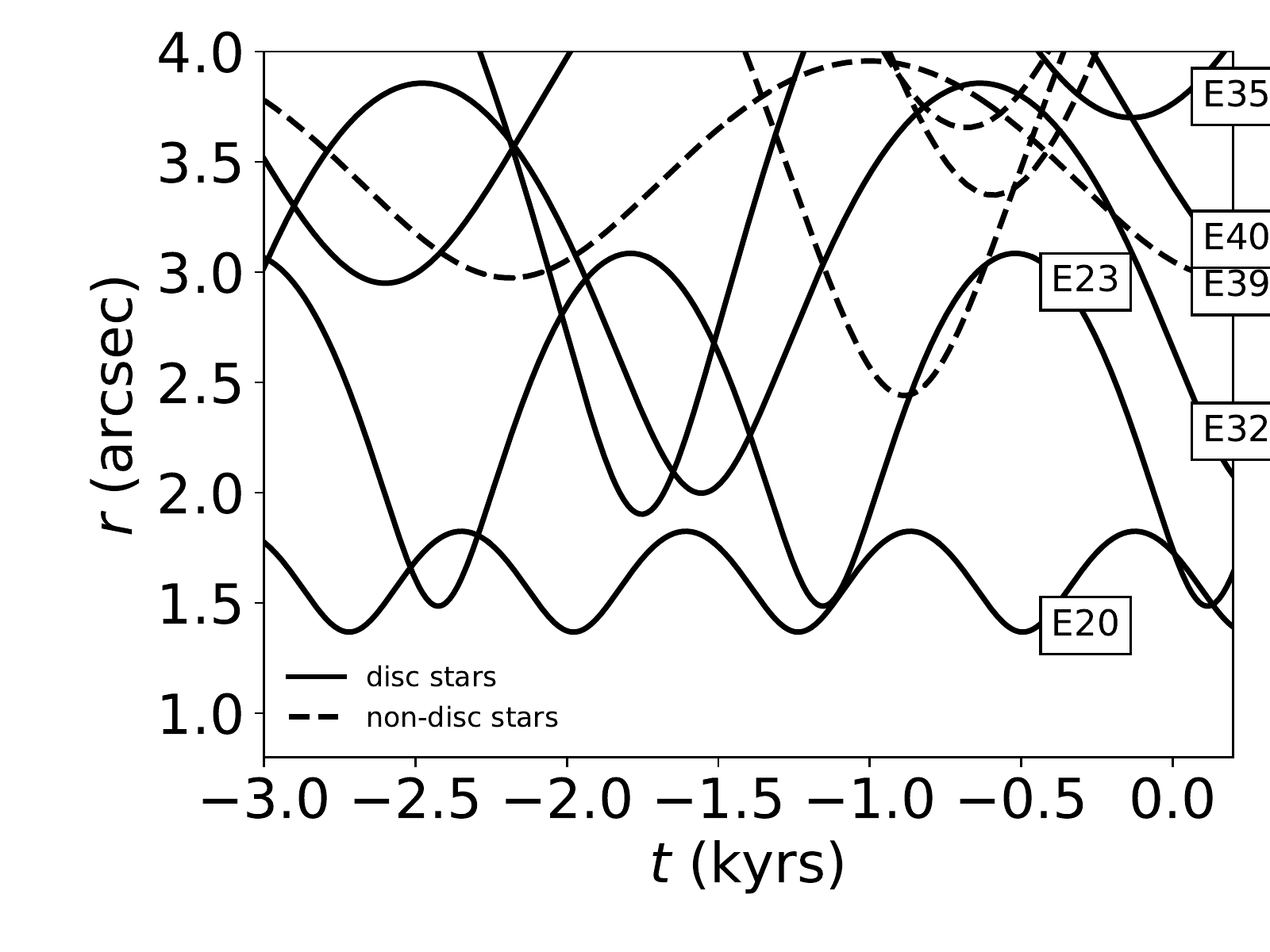}
\includegraphics[width=0.45\textwidth]{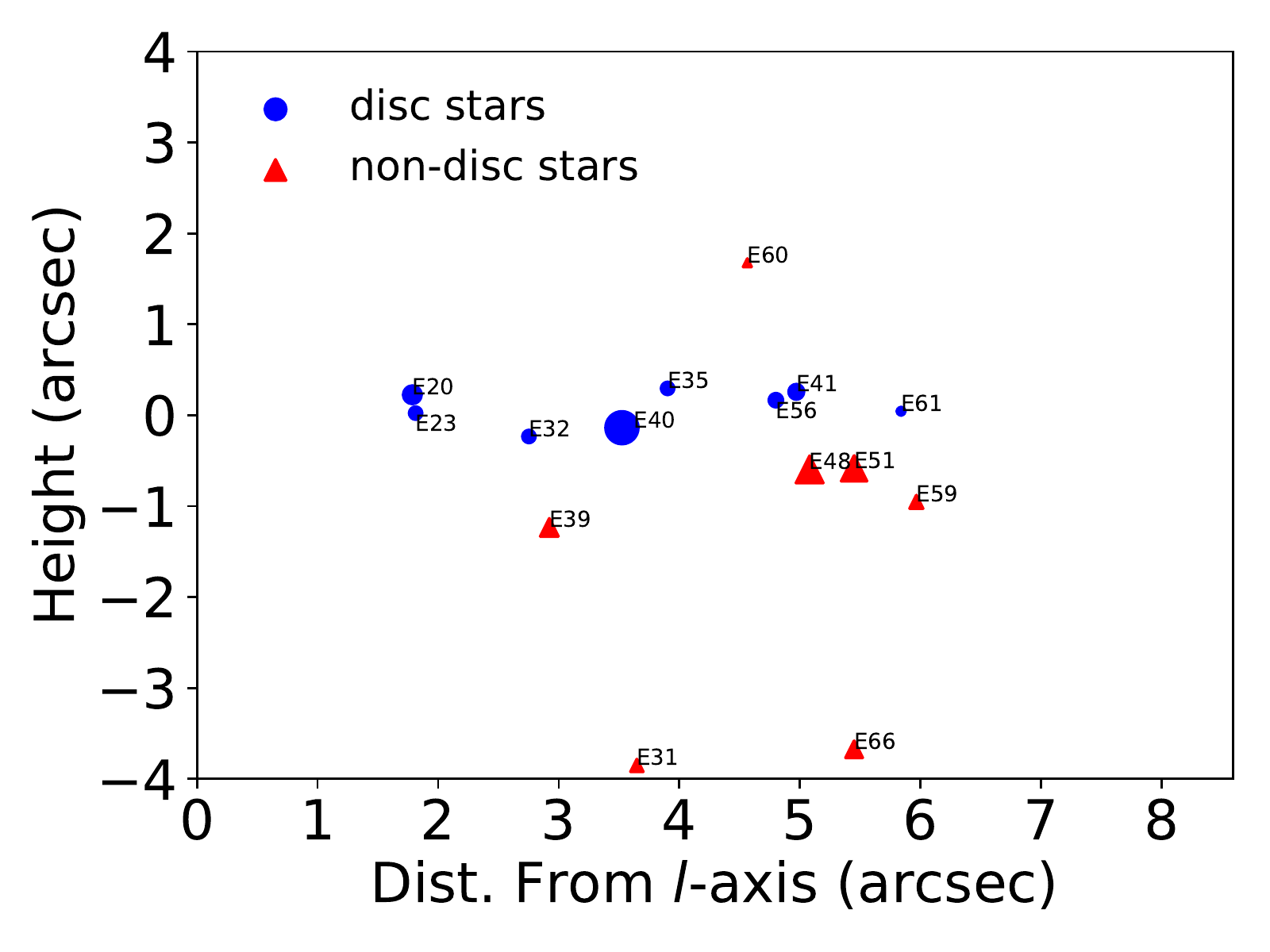}\caption{Left: Radial distance from the black hole for the innermost stars in our simulation. Right: Present day height and cylindrical radius of the innermost stars defined with respect to the anuglar momentum axis of the clockwise stellar disc described by \citet{Belo2006}, where the sizes of the circles/triangles are linearly proportional to the mass-loss rate of the stars. Each star is labeled using the `E' notation of \citet{Paumard2006}.  Solid lines (blue circles) denote stars identified with the stellar disc, while dashed lines (red triangles) denote stars outside the disc that fall into a more isotropic distribution of angular momentum.   The line-of-sight position for the latter stars are chosen by minimizing the eccentricity of the orbit, as in \citet{Cuadra2008}.   At the present day, five out of the six stars  in the inner $4''$ are disc-stars, which suggests that the angular momentum of the inner accretion flow might preferentially be aligned with the angular momentum of the stellar disc.  Moreover, within the central $6''$, at the present day, five of the six non-disc stars are located below the midplane of the stellar disc, which further suggests that the accretion flow structure might be asymmetric in polar angle (defined with respect to the angular momentum vector of the stellar disc). Indeed, we find both of these to be true in our simulations, as discussed in \S \ref{sec:results}.}
\label{fig:orbits}
\end{figure*}

The radii of the resulting stellar orbits (not including S2) as a function of time  as well as their height and cylindrical radius defined with respect to the stellar disc at the present day are shown in Figure \ref{fig:orbits} for the inner few arc seconds (inner few $\sim 0.1$ pc). Here and throughout we define $t=0$ as December 2017.  We expect the handful of stars in the inner few arc-seconds region to be the dominant source of accretion, as their stellar winds are more gravitationally bound to the black hole than the winds of the stars orbiting at larger radii. From Figure \ref{fig:orbits}, note first that, at the present day, a majority of the innermost stars are disc-stars, which is encouraging for the robustness of our calculation of the inner accretion flow as these orbits are better constrained than the orbits of non-disc stars.  Furthermore, this predominance of disc stars in the inner region will provide the accretion flow with a preferred angular momentum direction and thus encourage a coherent formation of a disc, as we will show in \S \ref{sec:results}. Secondly, note that a majority of the innermost non-disc stars are located below the plane of the stellar disc. This introduces an inherent asymmetry about the midplane of the disc.  This asymmetry shows up in the accretion flow structure outlined in \S \ref{sec:results}.
 
\subsection{Parameters and Initialization}
\label{sec:parameters}
We perform our simulation on a base grid with physical size of $1$ pc$^3$ that is covered with $128^3$ cells. Additionally, to further resolve the innermost region, we use 9 levels of nested SMR, resulting in an inner boundary of $r_{in} \approx 6.0\times 10^{-5}$pc  $\approx  1.6 \times 10^{-3} $ $''$ $\approx 300 r_{G}$, where $r_{G} \equiv GM_{BH}/c^2 $ is the gravitational radius of the black hole.  To study the dependence of our results on $r_{in}$, we also carry out simulations for $r_{in} \approx  3.0\times 10^{-5}$pc, $1.2\times 10^{-4}$pc, and $\approx 2.4\times 10^{-4}$pc (see \S \ref{sec:boundary}).

Since we are interested in the accretion flow at the present day, we start the simulation at an initial time of $t_0 = -1.1$ kyr, that is, 1100 years in the past, starting with zero velocity and the floor values of pressure and density.  1100 years corresponds to an orbital period at $r \approx 0.3$ pc $\approx 7''$.  Since we are interested in the flow properties at radii much smaller than this, such a run time is sufficient for our purposes (as we have also checked by running larger $r_{in}$ simulations for longer times). 

Finally, we adopt the recent estimates of $M_{BH} = 4.3 \times 10^6 M_{\odot}$ and $d_{BH} =8.3 $ kpc \citep{Gillessen2017}, where $d_{BH}$ is the distance to Sgr A*.  For these parameters, 1$''\approx $ 0.04 pc $\approx $ 2 $\times$ 10$^5$ $r_g$, where $r_g \equiv $G$M_{BH}$/c$^2$, $G$ is the gravitational constant, and $c$ is the speed of light.



\subsection{Results}
\label{sec:results}
\begin{figure*}
\includegraphics[width=0.95\textwidth]{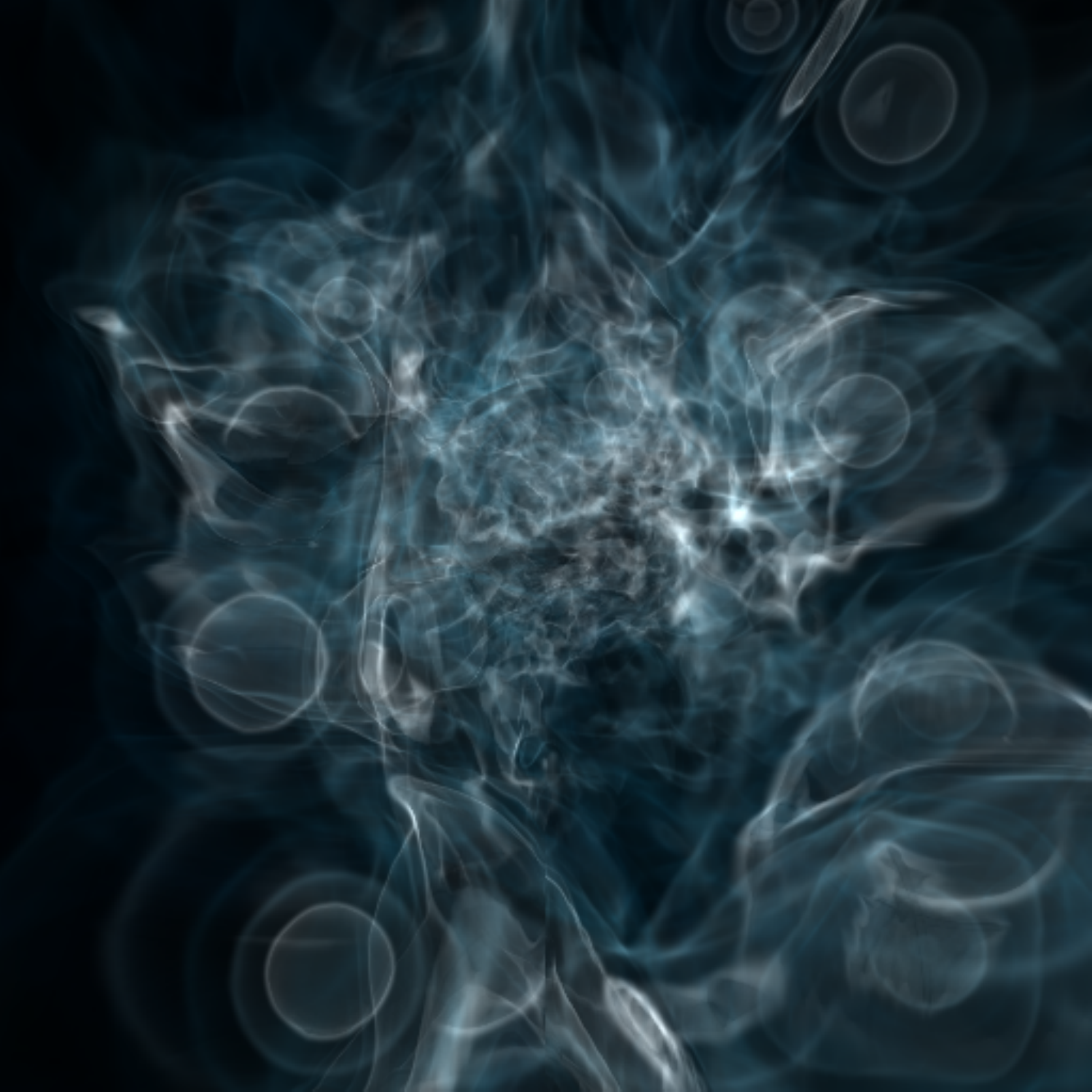}
\caption{Three-dimensional rendering of our simulation on a $0.5$ pc $
\times$ $0.5$ pc scale. This rendering was created with the {\tt YT} code \citep{yt_code} using 8 `layers' evenly spaced logarithmically in mass density between $10^{-2}$ and $10^{0.5}$ $M_\odot$/pc$^3$.   As the stellar wind sources (which appear as circular, outlined rings) plow through the material, the winds themselves form bow shocks in the direction of motion.  The interaction between these shocks causes a variety of fine scale structure to form in the flow. \emph{An animation of this figure is available online.} }
\label{fig:3D_stars}
\end{figure*}

\begin{figure*}
\includegraphics[width=0.49\textwidth]{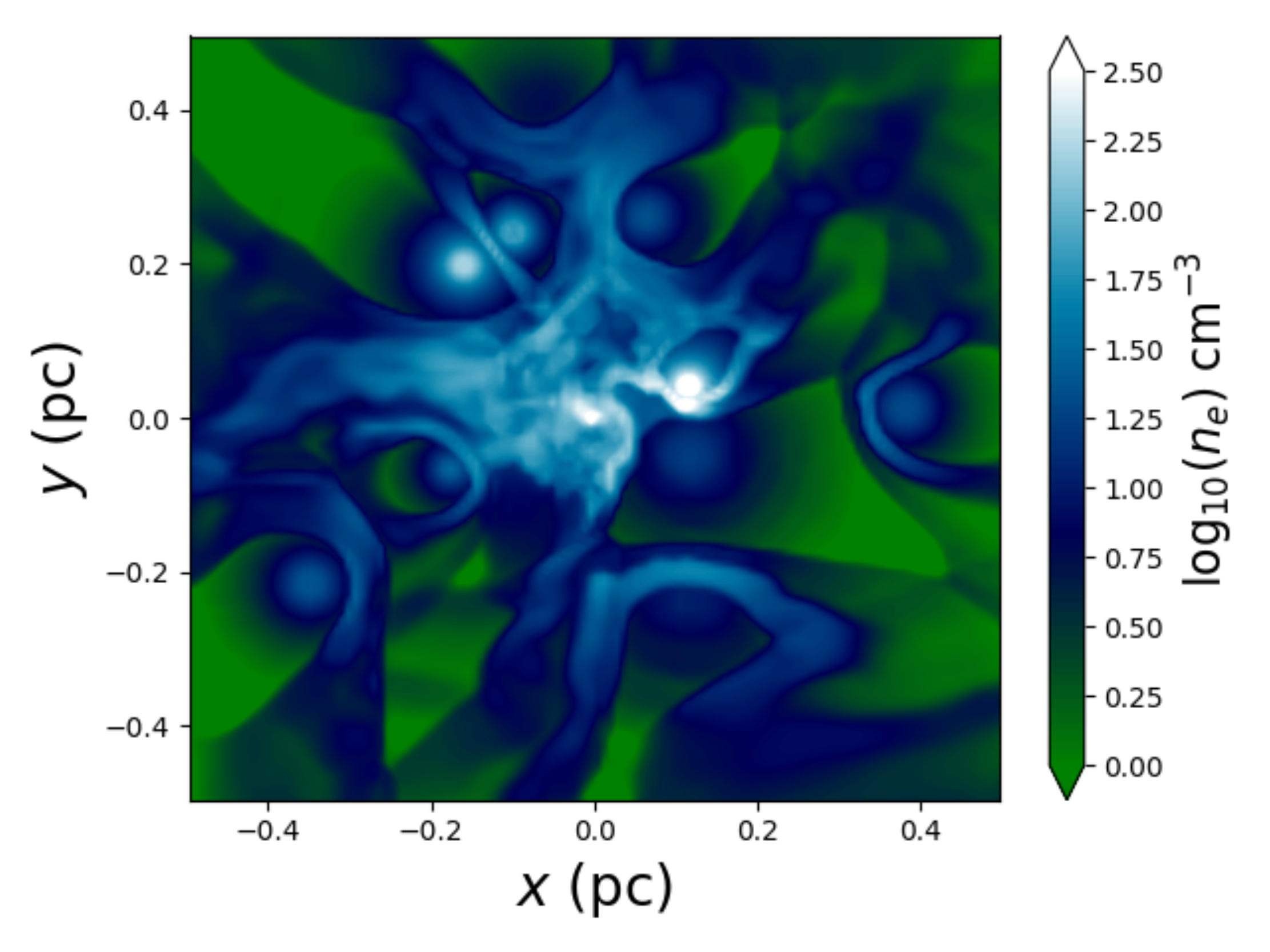}
\includegraphics[width=0.49\textwidth]{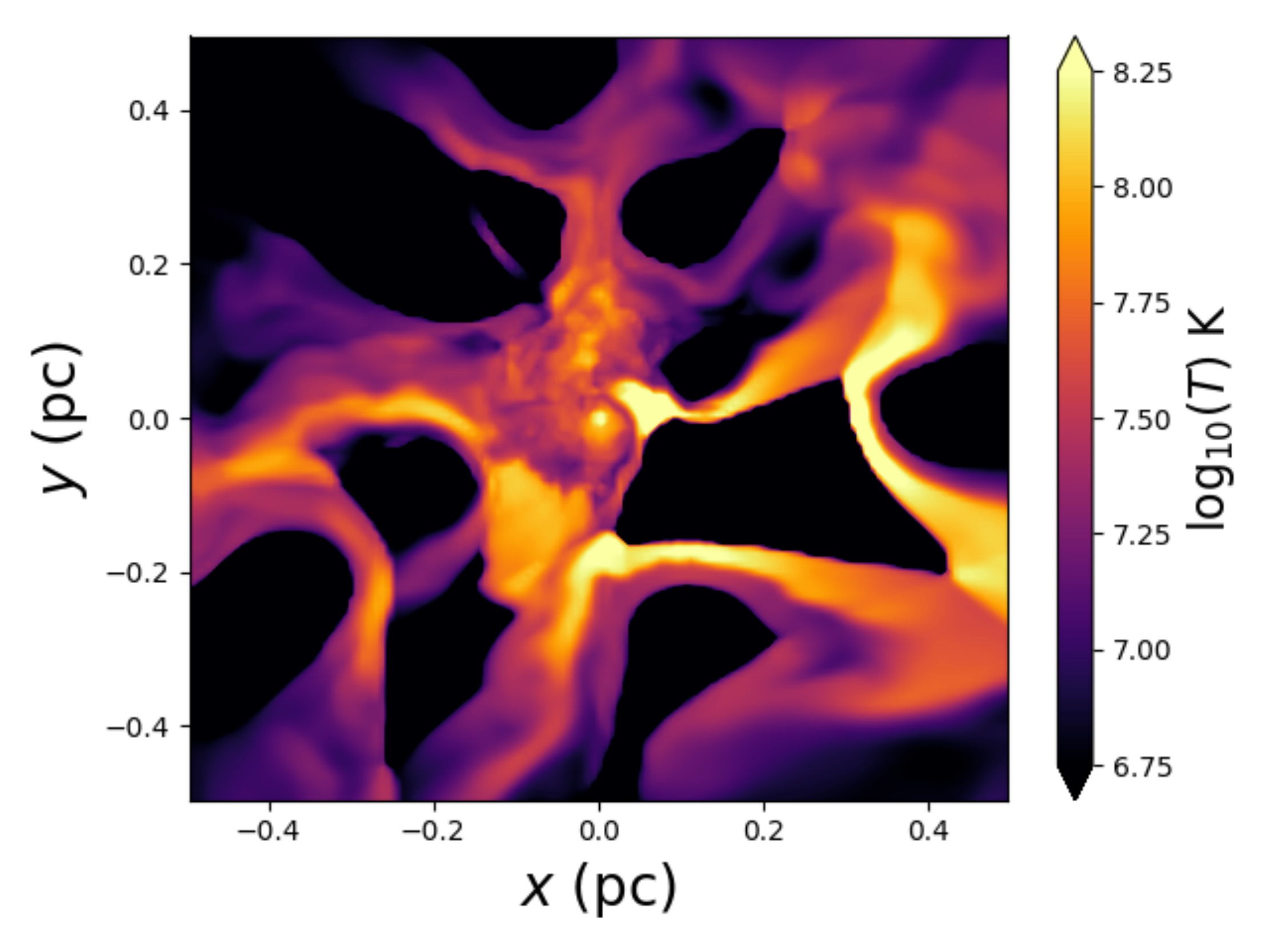}
\caption{Plane-parallel slices in the plane of the sky ($z=0$) of the electron number density, $n_e$, and Temperature, $T$, of our 3D hydrodynamic simulation, shown at the present day after running for 1.1 kyr from an initial vacuum state. The ``stars'' in our model are effectively point sources in mass, momentum, and energy (see \S \ref{sec:Model}) that travel on fixed Keplerian orbits constrained by observations. Here they appear as dense, cool, spherical regions.  The winds emitted from the stars form bow shocks as they collide with the ambient material and heat to high temperatures ($\sim 2 \times 10^8$ K). On the scale of the image, most of the stellar wind material is unbound and outflowing due to high temperature and angular momentum (see Figures \ref{fig:mdot} and \ref{fig:bernoulli}).  \emph{An animation of this figure is available online.}  }
\label{fig:stars}
\end{figure*}

\begin{figure*}
\includegraphics[width=0.49\textwidth]{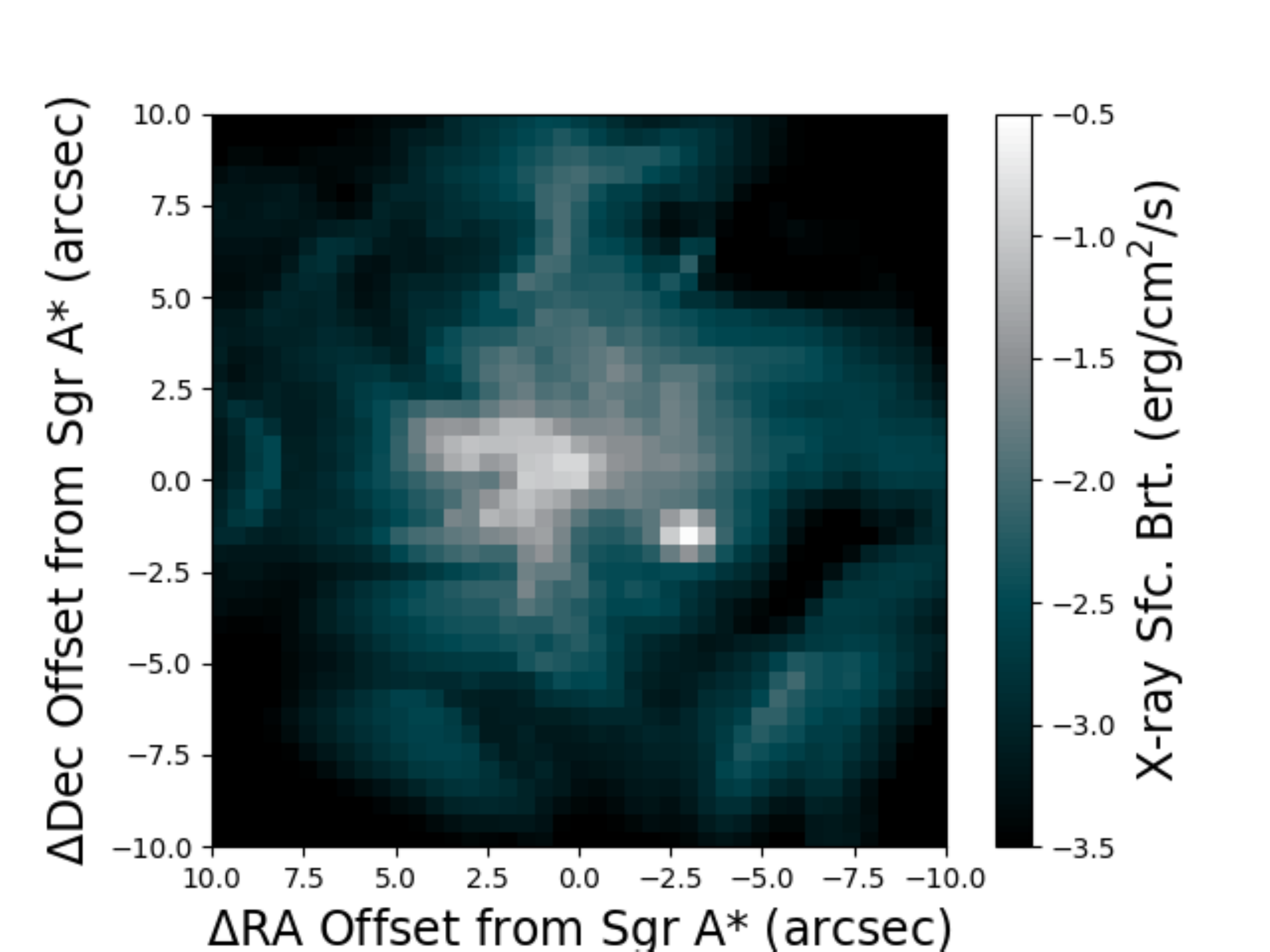}
\includegraphics[width=0.49\textwidth]{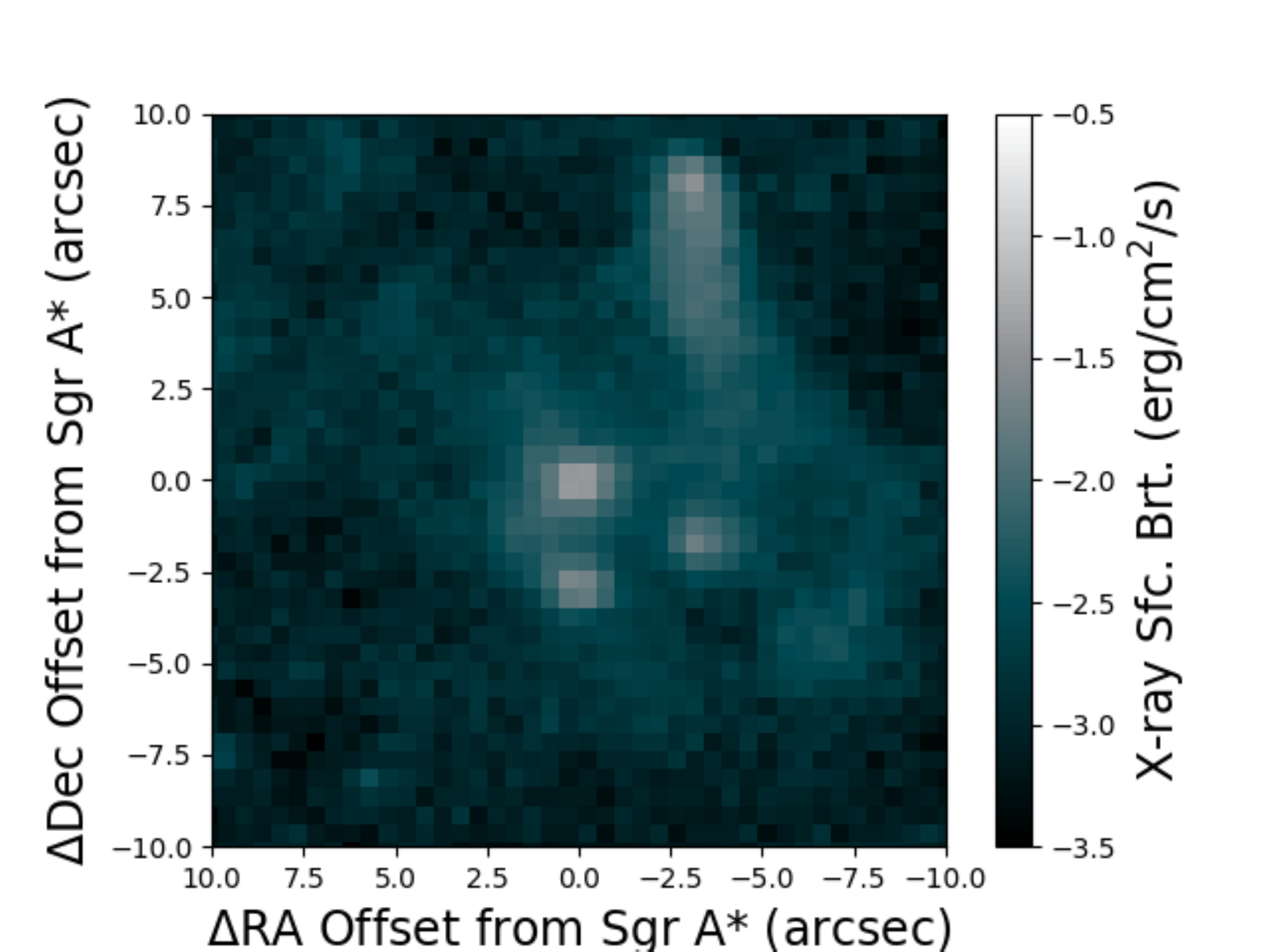}
\caption{2-8 keV surface brightness of the central $20'' \times 20''$ at the present day as calculated from our simulation (left) and as observed by $\emph{Chandra}$ (right) \citep{Li2013}. The surface brightness from the simulation has been coarsened to match the $\emph{Chandra}$ spatial resolution of $\approx$ 0.492$''$ per pixel, while the surface brightness from the observations is calculated assuming a mean photon energy of 5 keV and has not been corrected for absorption.
Both images show several point sources corresponding to the stellar wind sources in addition to an increase in surface brightness at the position of the black hole.  Note that, in our simulations, we do not the include the point source at ($\Delta$RA, $\Delta$Dec) $\approx$ (-4.5$''$,8$''$) associated with the pulsar wind nebulae seen in the \emph{Chandra} image. Integrated over the inner $1.5''-10''$, the X-ray surface brightness from our simulation agrees well with the point-source-extracted luminosity calculated from observations (see Figure \ref{fig:Lx_t}, \citealt{Baganoff2003}).  
}
\label{fig:Chandra_image}
\end{figure*}

Figure \ref{fig:3D_stars} shows a volume rendering of our simulation for the outer $\sim$ 0.5 pc ($\sim 13 ''$) of the galactic center at the present day, while Figure \ref{fig:stars} shows two-dimensional plane-of-the-sky slices of electron number density and temperature at the same time and spatial scale. The stellar winds of the ``stars'' in our simulation (which appear in the figure as dense, cool circular point-like objects) strongly interact and produce a myriad of bow shocks, heating the gas to temperatures as high as $\sim 2 \times 10^8$ K.  These point-like sources and shocks also light up in X-rays, as shown in the X-ray surface brightness profile of Figure \ref{fig:Chandra_image} (see caption and \S \ref{sec:Xray} for details of this calculation), which is overall in good agreement with the observed brightness profile seen by \emph{Chandra} (also shown in Figure \ref{fig:Chandra_image}). The largest differences (not including the pulsar wind nebula at  $\Delta$RA, $\Delta$Dec $\approx $ $-4'',8.5''$ that appears in the observations) occur a few arcseconds to the left of Sgr A* in the plane of the sky ($\Delta$RA offset from Sgr A* of $\sim$ 3$''$), where our simulations show a point-like source of emission not seen in the observational data, and around the star IRS 13E at ($\Delta$RA, $\Delta$Dec) $\approx ($ $-4'',-2''$), which is significantly brighter than the observations.   Since both of these differences were also seen by \citet{Russell2017}, who simulated the same problem\footnote{\citet{Russell2017} also included various subgrid models of feedback from Sgr A*, but the two features discussed here are seen in all of their mock X-ray images, including the ``control run'' with no feedback. } with a similar orbital configuration, mass-loss rates, and stellar wind speeds (see \S \ref{sec:comp}), they can likely be interpreted as additional constraints on the properties of the stellar winds.  For a more detailed discussion, we refer the reader to \citet{Russell2017}.

\begin{figure*}
\includegraphics[width=0.49\textwidth]{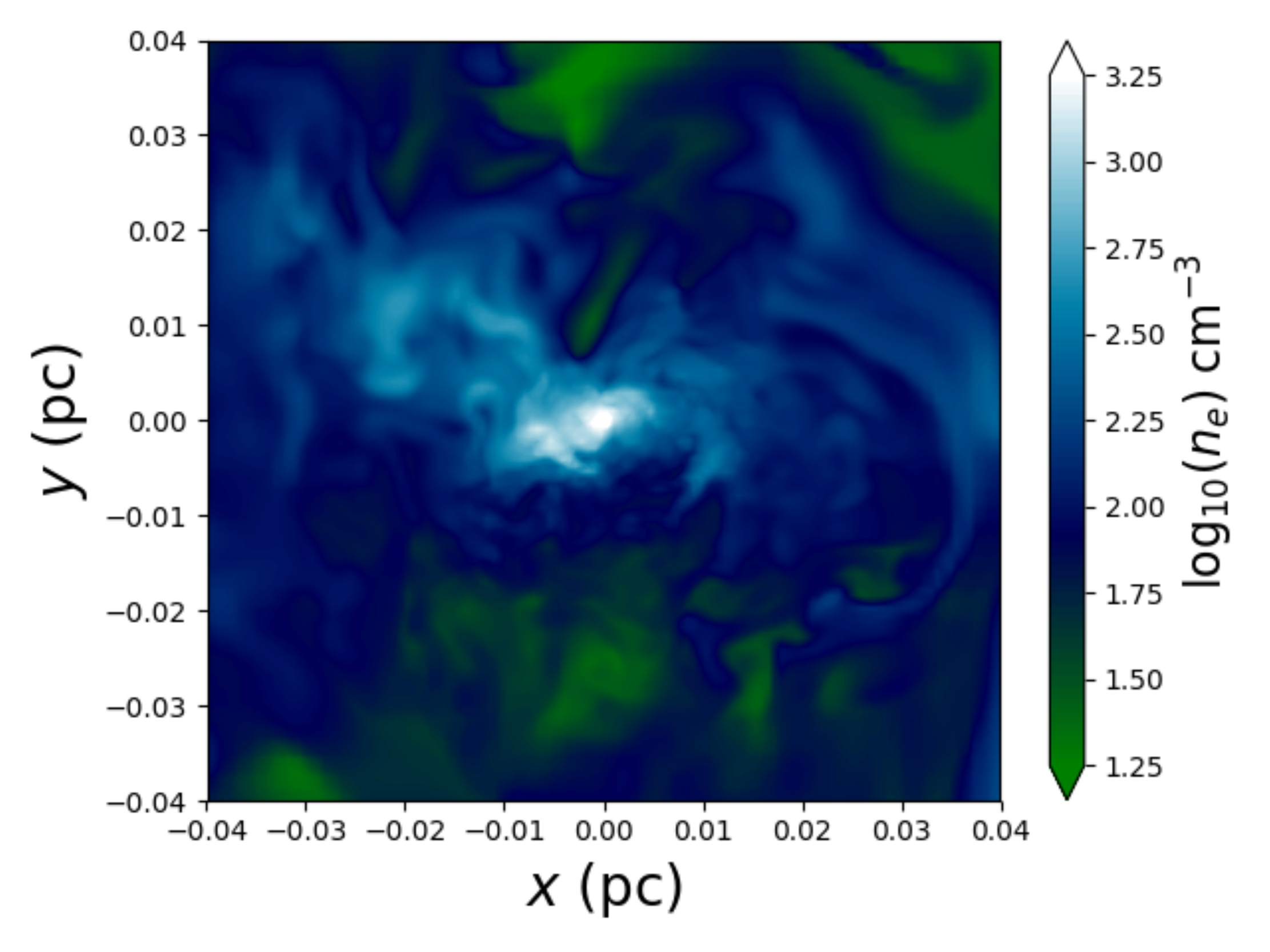}
\includegraphics[width=0.49\textwidth]{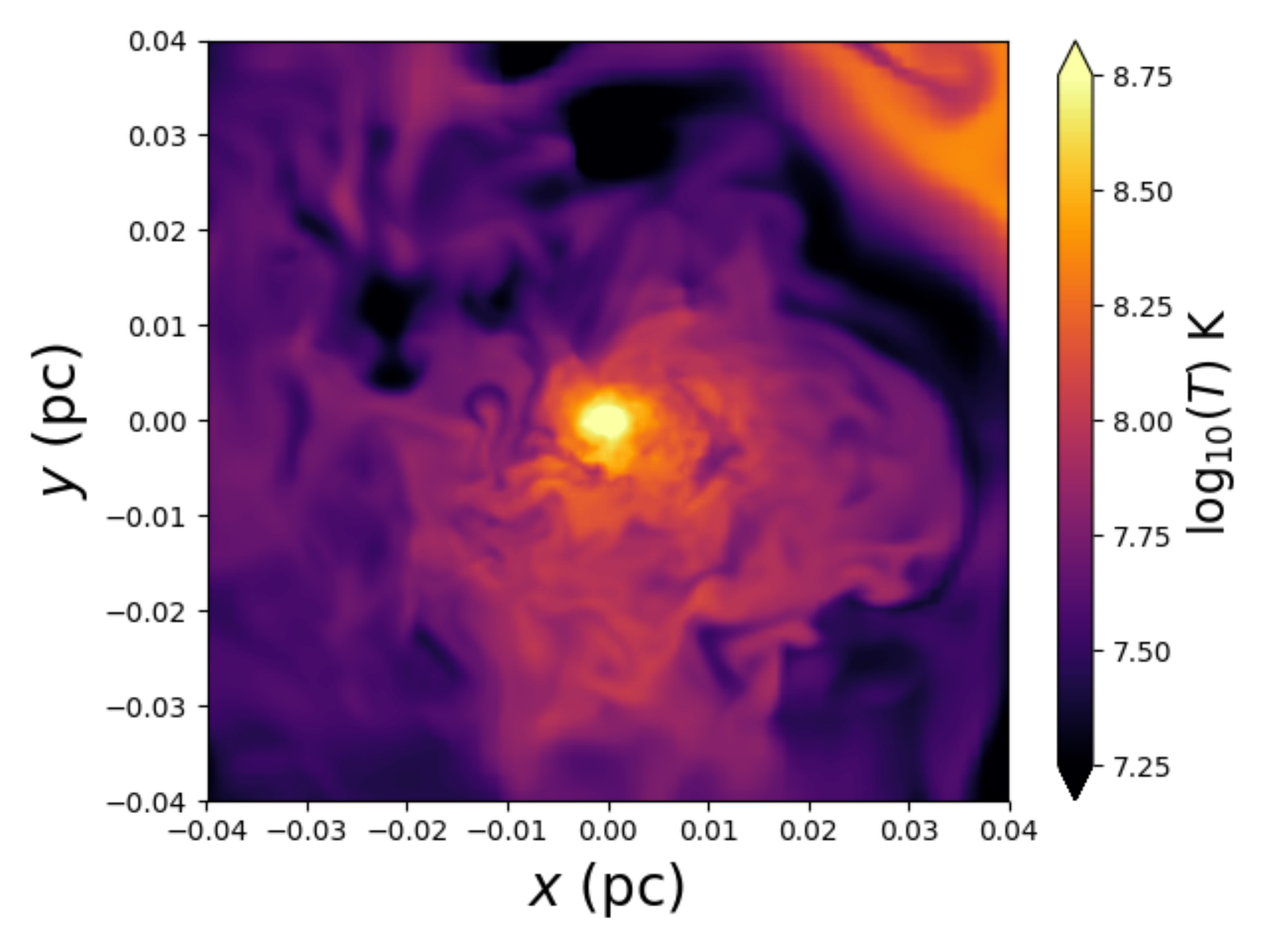}
\caption{Same as Figure \ref{fig:stars}, except zoomed in to the inner 1''. At this scale, much of the kinetic energy provided by the stellar winds has been converted into thermal energy via the shocks seen in Figure \ref{fig:stars}, resulting in a relatively smooth, hot accretion flow.  Note, however, the presence of a few relatively cold, high density clumps.  }
\label{fig:stars_inner}
\end{figure*}

As we will show, much of the material at the scale pictured in Figures \ref{fig:3D_stars}-\ref{fig:Chandra_image} is unbound outflow; only a small fraction actually reaches the inner domain. Furthermore, by the present day, the total amount of mass in the simulation contained within $r \lesssim$ 0.5 pc has saturated at a constant value of $\sim 0.2 M_\odot$.  This implies that the mass-feeding rate of the stellar winds is approximately balanced by the rate at which mass leaves the computational domain.  Figure \ref{fig:stars_inner} focuses closer in to the black hole, again showing plane-of-the-sky slices of mass-density and temperature but only in the inner 0.04 pc ($\sim 1''$).  By the time the gas has reached this scale, the shocks formed by the colliding winds have mostly dissipated, resulting in a hot, smooth flow combined with a few cooler ($T \lesssim 2 \times 10^7$ K), dense clumps.  The gas at this scale consists of roughly equal proportions of inflow and outflow (see below).  

Figure \ref{fig:history} shows the mass accretion rate through the inner boundary and the angular momentum direction vector of the inner 0.03 pc  $(\sim 0.8'')$ of the simulation as a function of time. The accretion rate varies between $\sim 2.0 \times 10^{-7} \dot M_\odot$/yr to $\sim 1.75 \times 10^{-6} \dot M_\odot$/yr on time scales as short as 10s of years.  The angular momentum vector of the flow, on average, oscillates around the normal vector of the stellar disc in which most of the innermost stars lie.  The largest deviation occurred during a period of $\sim 300$ yrs that began $\sim 500$ yrs ago when there was a rapid change from clockwise to counter-clockwise rotation with respect to the line of sight.  This event was associated with the largest spike in accretion rate that we see at $\sim 200$ yrs ago.  This was likely caused by one or two of the non-disc stars briefly providing a large source of accretion as they approached pericenter (see Figure \ref{fig:orbits}) which then temporarily disrupted whatever coherent disc may have formed. By the present day, however, the gas has settled back down to once again be aligned with the stellar disc and the flow enters a brief ``quiescent'' phase with a relatively low accretion rate that lasts for the next $\sim$200 years. One should not read too much into the latter result beyond the fact that the accretion rate could have been higher by factors of $\lesssim$7 within the recent hundreds of years.  This is because we find that our simulations are highly stochastic, and thus the exact behavior of the accretion rate as function of time can vary even with the smallest perturbation.  

With that said, to study the flow properties in more detail it is useful to study averaged fluid quantities to account for this  stochastic time variability.  We define the time and angle average of a fluid quantity $A$ as
\begin{equation}
\langle A\rangle \equiv \frac{1}{4\pi(t_{max}-t_{min})}\int\limits_{t_{min}}^{t_{max}} \int \limits _0^{2\pi} \int \limits_0^\pi A  \sin(\theta) d\theta d\varphi dt ,
\end{equation}
and the $w$-weighted time and angle average as
\begin{equation}
\langle A \rangle _w = \frac{\langle A w \rangle }{\langle w \rangle},
\end{equation}
where we use $t_{min} = -100$ yr and $t_{max} = 0$ yr.  Note that 100 yr is the free-fall time at a radius of $\approx 0.07$ pc $\approx 1.8 ''$.  We have chosen this particular time interval rather than one centered on $t=0$ because it represents a period in which the angular momentum vector of the inner regions is relatively steady (see Figure \ref{fig:history}).  Such an interval more clearly elucidates many of the general properties of the simulation while minimizing the complications inherent in describing a flow that is not in a true steady-state.  

Figure \ref{fig:primitives} shows the resulting radial profiles of the average electron number density, temperature, and radial velocity, while Figure \ref{fig:mdot} shows a radial profile of the average accretion rate, broken down into both inflow and outflow.  We define the latter two quantities as
\begin{equation}
\begin{aligned}
\dot M_{in} &\equiv - \langle 4 \pi \rho \min(v_r,0) r^2  \rangle \\
\dot M_{out} &\equiv \langle 4 \pi \rho \max(v_r,0) r^2\rangle.
\end{aligned}
\label{eq:inflow_outflow}
\end{equation} 
Figures \ref{fig:primitives} and \ref{fig:mdot} show that the flow contains four distinct regions: 
\begin{enumerate}
\item The outflow dominated region, $r\gtrsim 0.4 $ pc, which falls outside the locations of the majority of the stars and where the flow is approaching the standard Parker wind solution with $\rho \propto r^{-2}$ and $\dot M \approx $ const. $>0$.
\item The ``feeding region'' where the orbits of the stars mostly lie, $0.07$ pc $\lesssim r\lesssim 0.4$ pc, where $\dot M$ is both positive and increasing with radius due to the source term provided by the stellar winds.
\item The ``stagnation region'', $0.01$ pc $ \lesssim r\lesssim 0.07$ pc, where the mass inflow and outflow rates approximately cancel and $\dot M \approx 0$.
\item The inflow dominated region, $r\lesssim 0.01$ pc, where $\dot M\approx$ const. $<0$.  
\end{enumerate}
The transition from region 3 to region 4 is marked by an increase of inflow relative to outflow, caused by the loss of pressure support at the inner boundary leading to an accelerated radial velocity that approaches Mach 1.  The net effect of this is that, of the  $\sim 7\times 10^{-4} M_{\odot}/yr$ of material provided by the 30 stellar winds, only a small fraction of this, $\sim 7 \times 10^{-7} M_{\odot}/yr$, is accreting into the inner boundary; the rest fuels the outflow.  However, the radius at which the flow transitions from regions 3 to 4, and hence, the constant accretion rate in the innermost radii, depends on the location of the inner boundary. Larger (smaller) $r_{in}$ causes the transition to happen at larger (smaller) radii and thus results in larger (smaller) accretion across the inner boundary.   This clear dependence of our simulation results on the location of the inner boundary is not necessarily a concern; in fact, we can use it to extrapolate down to the Schwarzchild radius of the black hole where a pressure-less boundary would be appropriate.  We do this later in \S \ref{sec:boundary}. 

In Figure \ref{fig:mdot}, the inflow rate at $\sim$ 0.1 pc is $\sim$ 2-3 $10^{-5}$ $M_\odot$/yr, which is of order the canonical Bondi estimate for the rate at which gas should be gravitationally captured by the central black hole.   However, only a small fraction of this mass actually accretes to smaller radii $\ll$ 0.1 pc (and the accretion rate at small radii decreases as we decrease the innermost radius of our simulation; see Figure \ref{fig:mdot_bound}).   Thus the Bondi accretion rate estimate is not a good estimate of the accretion rate at small radii in our simulations.   This is because, as we will show in more detail below, only the low angular momentum tail of the stellar wind material can accrete to small radii in our simulations.

For a flow in which radiative cooling is inefficient, the $T \propto r^{-1}$ scaling shown in Figure \ref{fig:primitives} is expected from conservation of energy, where $T \propto GM_{BH}/r$. If the flow were adiabatic this would imply a density power law of $r^{-3/2}$ for $\gamma=5/3$, but instead we find $\rho \propto r^{-1}$. This is because the shocks generated by the accreting stellar wind material lead to an effective energy dissipation term that results in $p/\rho^\gamma \propto r^{-1/3}$, that is, an entropy profile that increases with decreasing radius. In Appendix \ref{app:single_star} we explain the precise shape of the density profile in terms of a model in which the stellar winds from only a small number of stars dominate the flow. An $r^{-1}$ density profile in the inner region of the flow implies that the total amount of mass enclosed in a spherical shell of radius $r$, $M_{enc}$, scales as $r^{2}$ in this region.   More precisely, we find that the enclosed mass at $t=0$ is well approximated by 
\begin{equation}
M_{enc} \approx 4 \times 10^{-5} M_\odot \left(\frac{r}{0.008 \textrm{pc}}\right)^2,
\label{eq:M_enc}
\end{equation}
which agrees with our simulations up to a factor of $\sim $ few for $r> 2\times 10^{-4}$ pc.

Figure \ref{fig:bernoulli} shows the mass-weighted average Bernoulli parameter and the relative contributions to the Bernoulli parameter from pressure and velocity.  We find that the flow is, on average, unbound at all radii.   For $r\gtrsim 0.04$ pc, the material is strongly unbound, that is, $\langle E \rangle \gg GM_{BH}/r$, and the Bernoulli parameter approaches a constant.  This is expected from the fact that the majority of the stars are located between $0.05-0.4$ pc and fuel a Parker wind-type solution for $r>0.4$ pc.  By contrast, the gas in the inner $r<0.05$ pc is only very slightly unbound, with the Bernoulli parameter closely mirroring the gravitational potential.  


Radiative cooling can be important in localized regions for cooling of the shocked stellar winds at large radii ($r \gtrsim 0.07$ pc), but has a negligible effect on the inner regions of the flow ($r \lesssim 0.07$ pc). To quantify this, we note that there is only $\approx 10^{-3} M_\odot$ of gas with $T <10^5$ K for $r\lesssim0.07$  pc and no gas with $T\lesssim 10^6$ K by $r\lesssim0.03$ pc. This can be understood using a simple time-scale analysis.  At $r = 0.07$ pc, the ratio between the cooling time, $\langle t_{cool} \rangle \equiv \langle P/(\gamma-1) \rangle / \langle Q_-\rangle $, and the local sound crossing time, $\langle t_{cs} \rangle \equiv r / \sqrt{\gamma \langle P \rangle/\langle \rho \rangle}$ is $\sim 300$, and increases rapidly with decreasing radius.  For $r\gtrsim 0.07$ pc, however, this ratio is typically of order $\sim 50$ and can be as small as $\sim 10$. Note that this is an angle and time averaged quantity; localized regions at $r\gtrsim 0.07$ pc can have the ratio between $t_{cool}$ and $t_{cs}$ reach $\sim 1$.\footnote{Since we include optically thin radiative cooling in the calculation, $t_{cool}/t_{cs}$ is always $\gtrsim 1$; otherwise it would quickly evolve to $t_{cool}/t_{cs} \sim 1$.} 

Figure \ref{fig:l_angle} shows the average specific angular momentum of the accretion flow, weighted both by mass and mass flux, as well as the average direction vector of the mass weighted specific angular momentum.  The bulk of the material falls into a sub-Keplerian rotation profile with $l \approx 0.5 l_{kep} = 0.5 \sqrt{GM_{BH} r}$, while the angular momentum of the material that is accreting all the way through the inner boundary is constant with radius and equal to half the Keplerian value at the inner boundary, $l \approx 0.5 l_{kep}(r_{in})$.  This indicates that only material with circularization radii $\lesssim r_{in}$ is able to truly accrete; the rest fuels outflow, as we shall show.  The reason that both the mass-weighted specific angular momentum profile and the value of the specific angular momentum at the boundary are sub-Keplerian is that the flow is predominately pressure supported, as shown in Figure \ref{fig:bernoulli}, where the rotational term comprises only $\sim 20 \%$ of the Bernoulli parameter. 

Furthermore, Figure \ref{fig:l_angle} also shows that the direction of the angular momentum vector is $\approx$ const. for the inner $r\lesssim 0.4 ''$ and is essentially aligned with the normal vector of the clockwise disc of stars.  We have shown previously in Figure \ref{fig:orbits} that five of the innermost six stars at the present day are classified as disc-stars, so it is not surprising that the resulting flow is also aligned with the disc if we consider that most of the material is provided by these nearby stars. The fact that this direction is $\approx$ constant with radius makes it convenient to define a new coordinate system in which the $z$-direction is aligned with the angular momentum.  In this new coordinate system we can make 2D, $\varphi$-averaged contour maps to better study the disc structure. 

In these new coordinates, Figure \ref{fig:stream} shows contour maps of $\varphi$-averaged mass accretion rate overplotted with velocity streamlines, Bernoulli parameter, and density, in addition to $\varphi$-averaged $\theta$ profiles of density, angular velocity, temperature, and accretion rate at $0.04$ pc $\approx 66 r_{in}$.   Though we do find a disc-like structure with the density peaked in the midplane, the scale height of this disc is large, with only a factor of $\sim$ 2-3 contrast between the midplane density and the polar density.   This is because the disc is hot and mostly pressure supported (see Figure \ref{fig:bernoulli}), which causes the disc to puff up and reach a scale height, $H$, of $H \approx r$.    Additionally, we find that accretion primarily occurs by bound material along the southern polar region, while the midplane and northern pole are moderately unbound and generally outflowing.  The asymmetry in $\theta$ is a direct result of the asymmetry in the distribution of non-disc stars at the present day (Figure \ref{fig:orbits}), where a majority of the inner-most non-disc stars are located below the midplane of the stellar disc. The somewhat counter-intuitive result that the midplane is predominantly outflowing and not inflowing is caused by the stellar wind material having a wide range of angular momentum.   The significant population of low angular momentum material would naturally accrete spherically, but the material with larger angular momentum can only reach a radius $\sim l^2/(GM_{BH})$, at which point it scatters off of the effective potential, preferentially towards the midplane.  The presence of both of these components results in the accretion structure shown in Figure \ref{fig:stream}, where a combination of both high and low angular momentum material inflow along the southern pole until the circularization radii of the high angular momentum material is reached.  At this point the unbound, high angular momentum material ``turns aside'' to the midplane and feeds outflow while the bound,low angular momentum material continues on until it either reaches the inner boundary or feeds the outflow along the northern pole. 

These two very different components to the accretion flow are additionally seen in the fact that the midplane and the southern polar regions have very different dynamics.  This is highlighted in Figure \ref{fig:pole_v_mid}, where we show $\varphi$-averaged radial profiles of accretion rate, radial velocity, and angular  velocity for $\theta=90^\circ$ and $\theta = 170^\circ$.  At $\theta = 170^\circ$ (southern pole), the material is essentially in free-fall with an accretion rate that nicely matches the $\propto \sqrt{r}$ predicted from feeding by a few isolated stars (Appendix \ref{app:single_star}).  At $\theta = 90^\circ$ (midplane), on the other hand, $v_r \ll v_{ff}$ and the material is nearly Keplerian with velocity predominantly in the $\varphi$-direction.\footnote{Note, however, that by comparing Figure \ref{fig:bernoulli} to Figure \ref{fig:pole_v_mid}, $\langle v_r \rangle^2 \ll  \langle v_r^2 \rangle$, meaning that there can exists large instantaneous radial flows that cancel out when averaged over time.} This means that the flow can be roughly described as a superposition of a low angular momentum, spherical-Bondi type solution with a high angular momentum, Keplerian thick disc type solution.  In our simulations, the former dominates the accretion rate while the latter dominates the mass.

\begin{figure}
\includegraphics[width=0.45\textwidth]{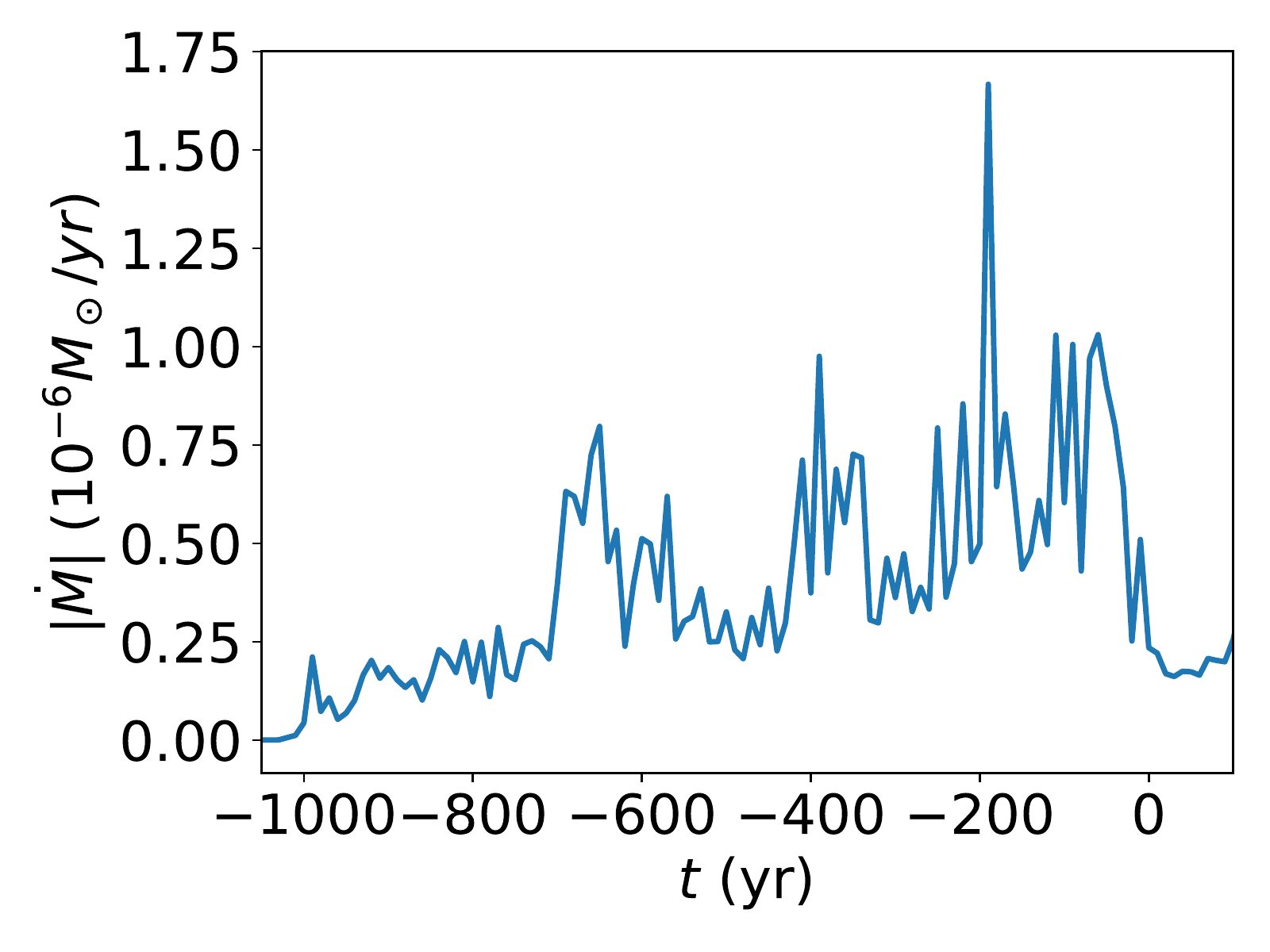}
\includegraphics[width=0.45\textwidth]{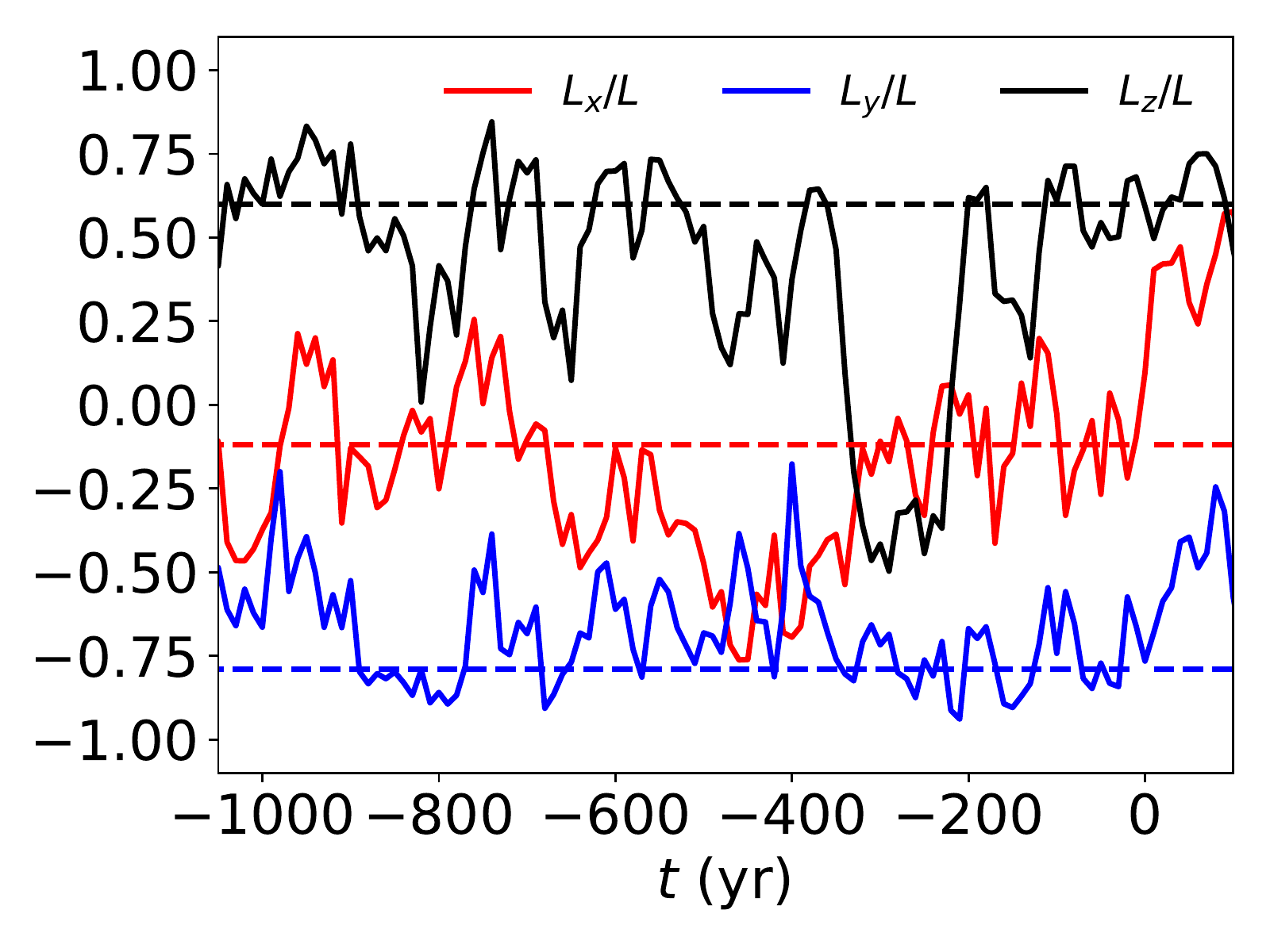}
\caption{Top: Accretion rate as a function of time in our simulation, measured at $2.5 r_{in} \approx 1.5 \times 10^{-4}$ pc $\approx 740  r_g$. Bottom: Angular momentum direction vector averaged over the inner $10r_{in} \approx 6 \times 10^{-4}$ pc $<r<0.03$ pc.
  Dashed lines represent the angular momentum vector of the stellar disc in which a majority of the innermost stars orbit. The largest spike in the accretion rate is associated with a rapid change in the angular momentum vector of the flow.   }
\label{fig:history}
\end{figure}

\begin{figure}
\includegraphics[width=0.45\textwidth]{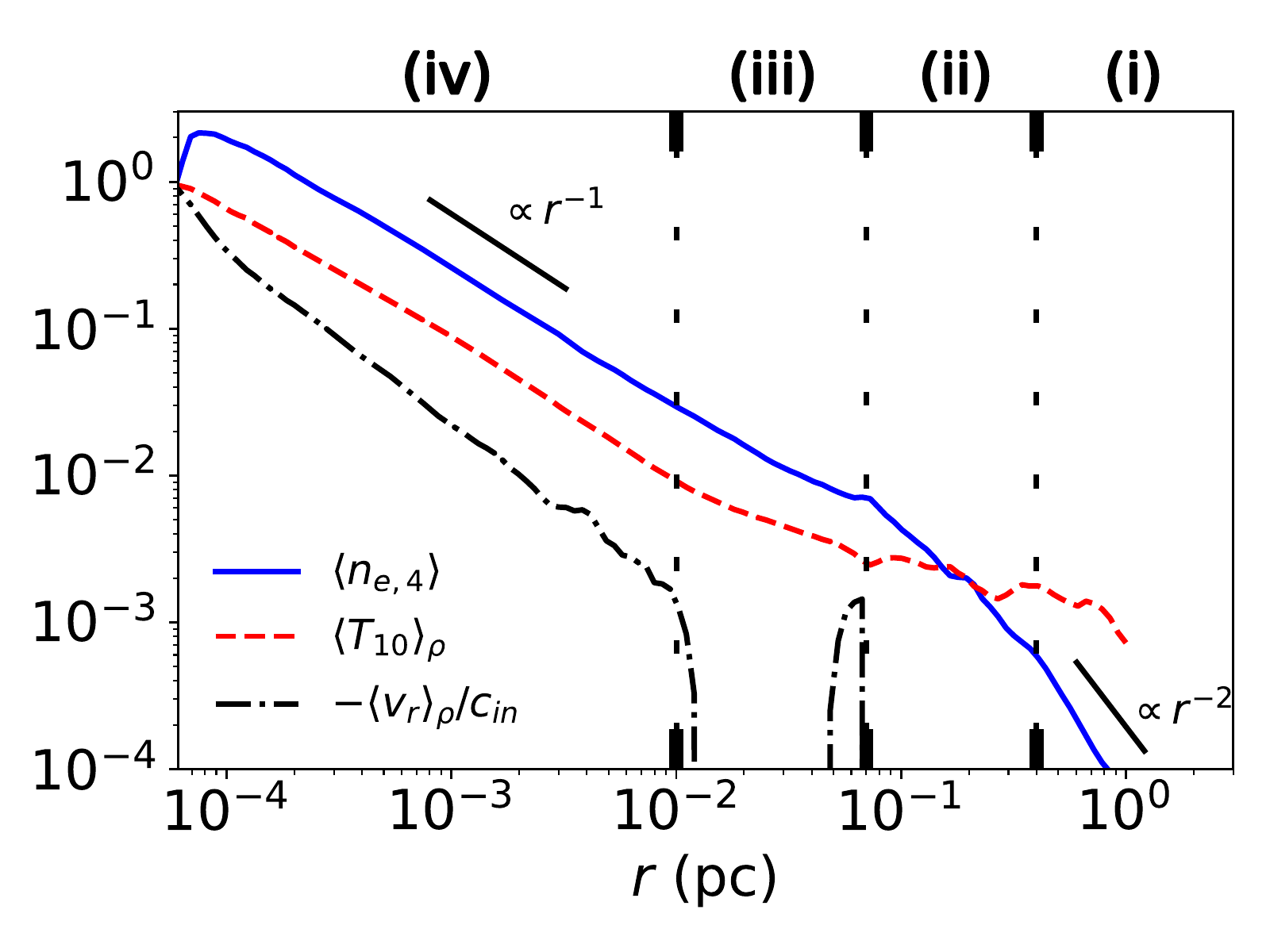}
\caption{Time and angle averaged electron number density in units of $10^4$ cm$^{-3}$, $\langle n_{e,4}\rangle$, temperature in units of $10^{10}$ K, $\langle T_{10} \rangle_\rho$, and radial velocity normalized to the average sound speed at the inner boundary, $\langle v_{r}\rangle_\rho/ c_{in}$, where $c_{in} \equiv \langle c_s\rangle_\rho(r_{in})$. Vertical lines demarcate regions (i)-(iv) as defined in \S \ref{sec:results}.  In the inner accretion flow, $r \lesssim 0.07$ pc (regions iii and iv), all three variables follow power-laws in radius of $\propto r^{-1}$, while for $r\gtrsim 0.4$ pc (region i) the density follows an $r^{-2}$ profile, as expected for a Parker wind-type solution.      }
\label{fig:primitives}
\end{figure}

\begin{figure}
\includegraphics[width=0.45\textwidth]{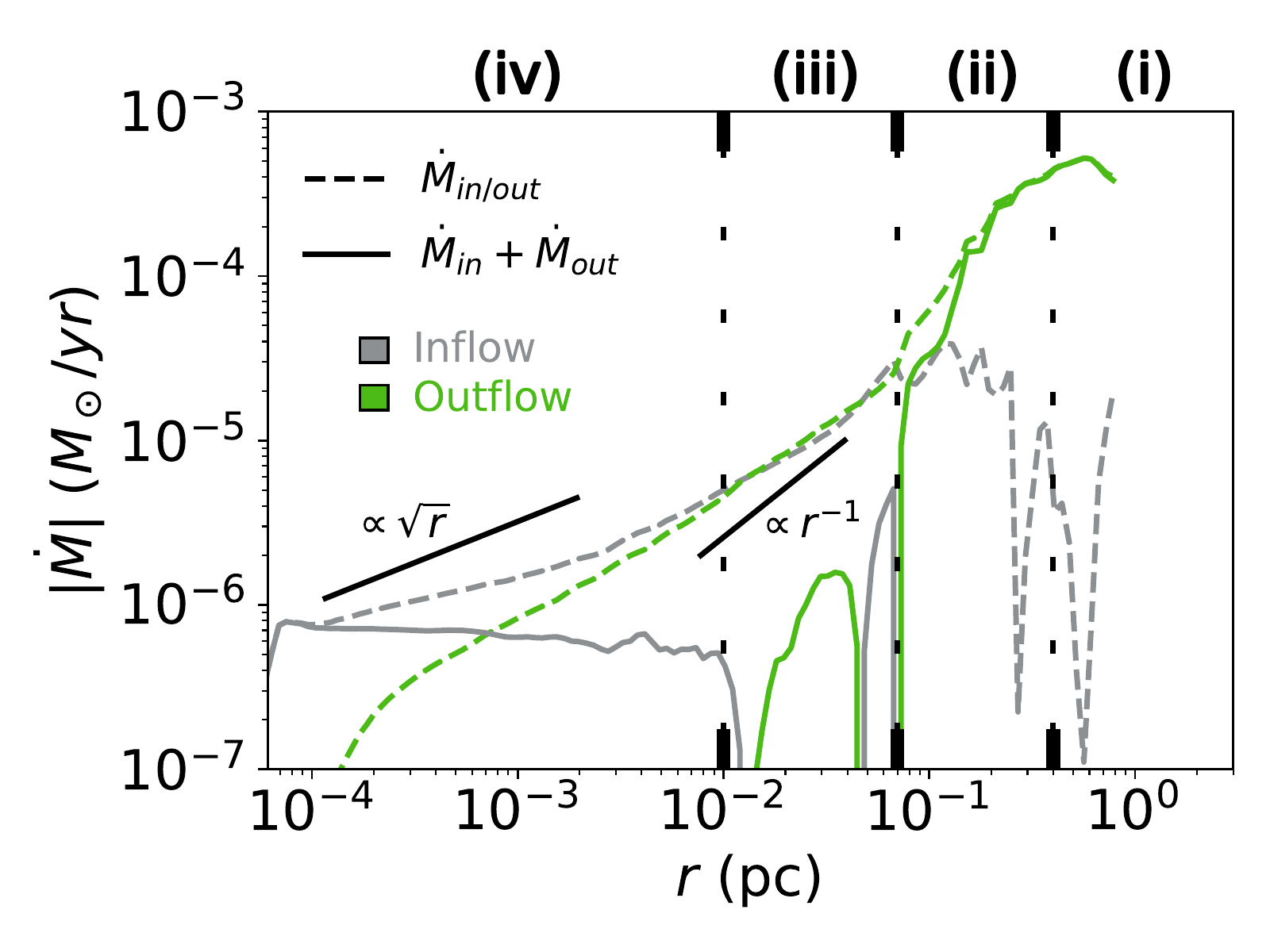}
\caption{Time and angle averaged accretion rates as a function of radius, including the net accretion rate (solid) and inflow and outflow rates computed separately (dashed, Equation \ref{eq:inflow_outflow}).  Here Kelly green lines denote outflow while silver lines denote inflow. Vertical lines demarcate regions (i)-(iv) as defined in \S \ref{sec:results}.  Outflow dominates for $r\gtrsim 0.07$ pc (region i), where most of the stellar winds are located, while inflow dominates for $r\lesssim 0.01$ pc (region iv).
In between, the rates are nearly equal in magnitude. Of the total $\approx$ 7 $\times 10^{-4}$ $M_\odot$/yr added to the simulation from the stellar winds, only a small fraction, $\approx $ 6 $\times 10^{-7}$ $M_\odot$/yr, flows into the inner boundary. In addition, the accretion rate at the inner boundary, $r_{in}$, decreases with smaller $r_{in}$ (\S \ref{sec:boundary}, Figure \ref{fig:mdot_bound}).   }
\label{fig:mdot}
\end{figure}

\begin{figure*}
\includegraphics[width=0.45\textwidth]{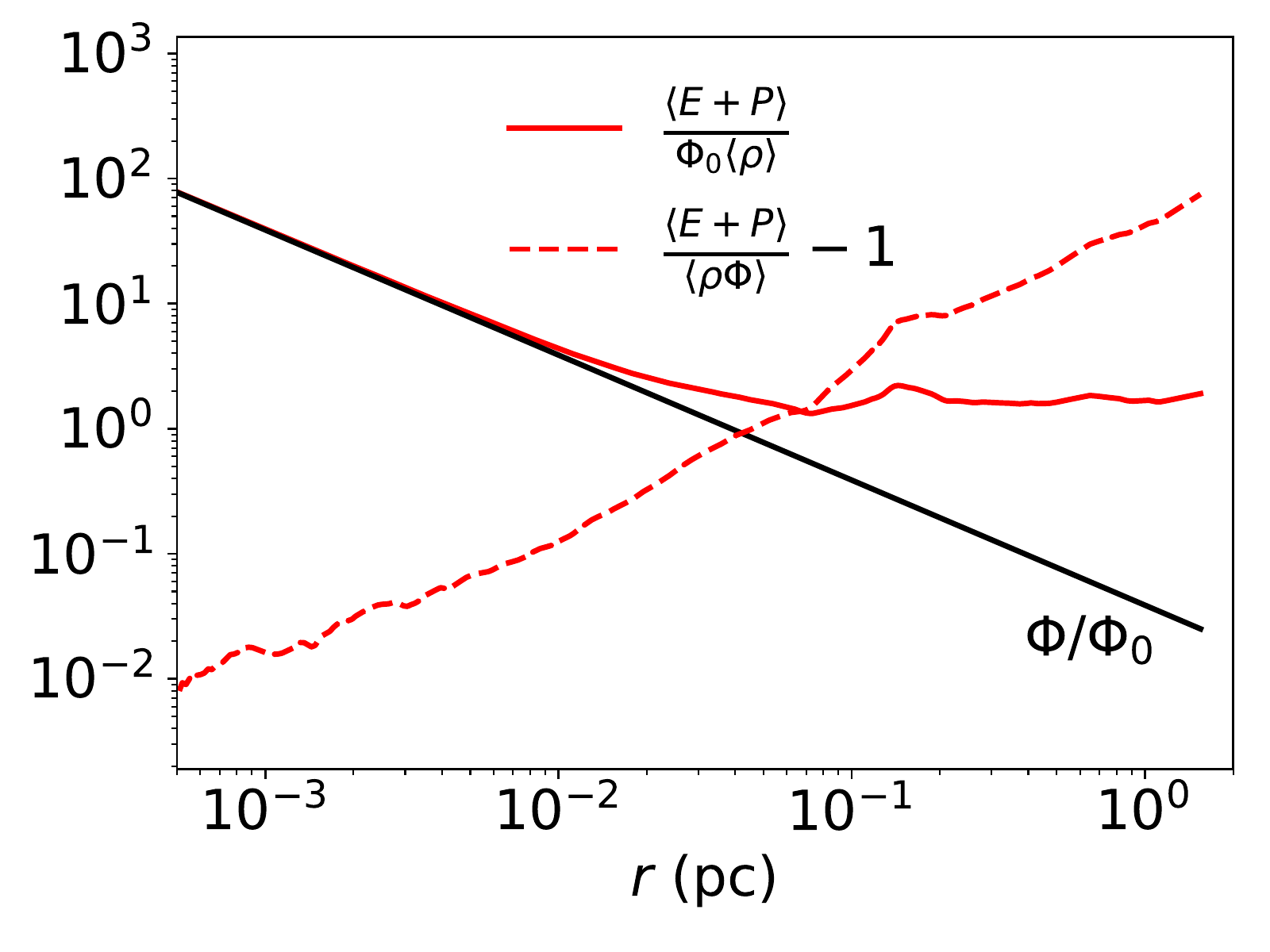}
\includegraphics[width=0.45\textwidth]{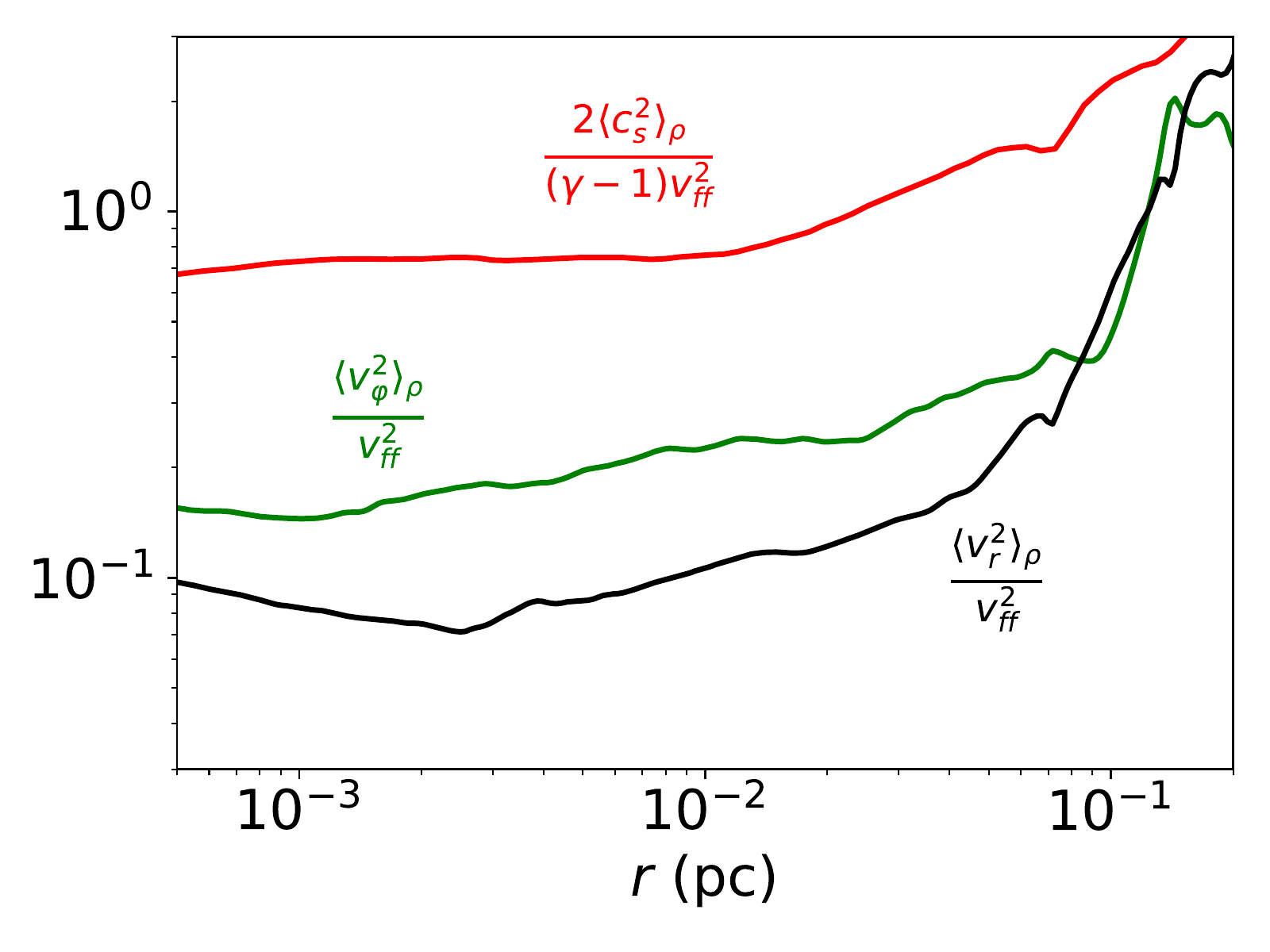}
\caption{Left: Time and angle-averaged total specific energy (solid red), $(E+P)/\rho$, where $E \equiv 1/2 \rho v^2 + P/(\gamma-1) $, gravitational potential (solid black), $\Phi \equiv GM_{BH}/r$, and the Bernoulli parameter (dashed red), $(E+P)/\rho - \Phi$. The latter quantity is normalized to the gravitational potential, while the former two are normalized to the gravitational potential at $r_0 = 0.04$ pc [i.e. $\Phi_0 = \Phi(r_0)$].  Right: Time and angle-averaged components of the Bernoulli parameter, including the pressure term (top line), the orbital kinetic energy term (middle line), and radial kinetic energy term (bottom line), all plotted as fractions of the gravitational potential, $GM_{BH}/r$. Here the azimuthal, $\varphi$ direction is defined with respect to a coordinate system which has $\hat z$ aligned with the average density weighted angular momentum axis.  The inner accretion flow is slightly unbound, with Bernoulli parameter $\gtrsim 0$, and predominantly pressure supported. Note that, comparing to Figure \ref{fig:primitives}, $\langle v_r^2\rangle_\rho\propto r^{-1}$, while $\langle v_r \rangle_\rho \propto r^{-1}$, which is due to the cancellation of both inflow and outflow reducing the average of $v_r$.   }
\label{fig:bernoulli}
\end{figure*}

\begin{figure*}
\includegraphics[width=0.45\textwidth]{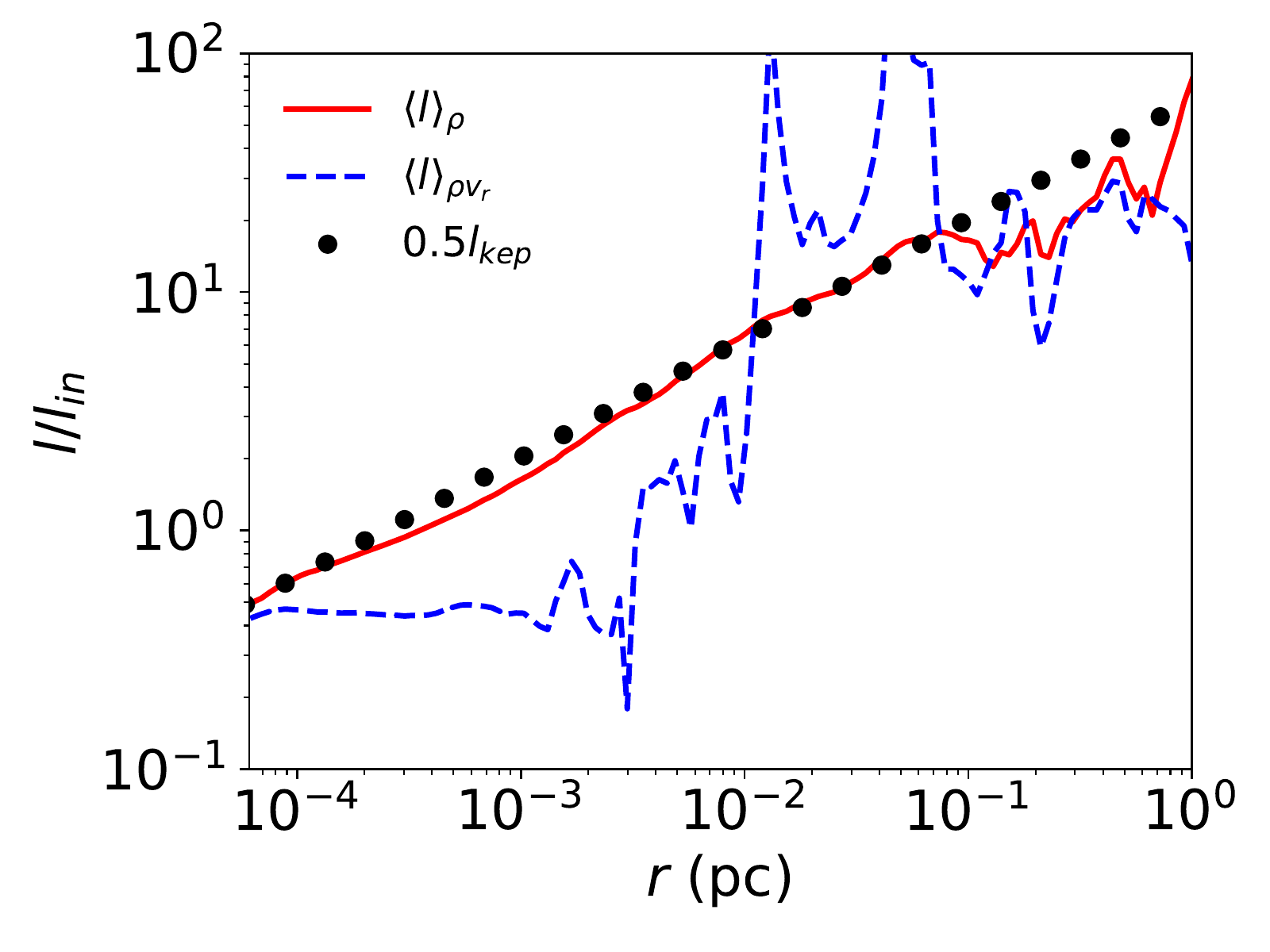}
\includegraphics[width=0.45\textwidth]{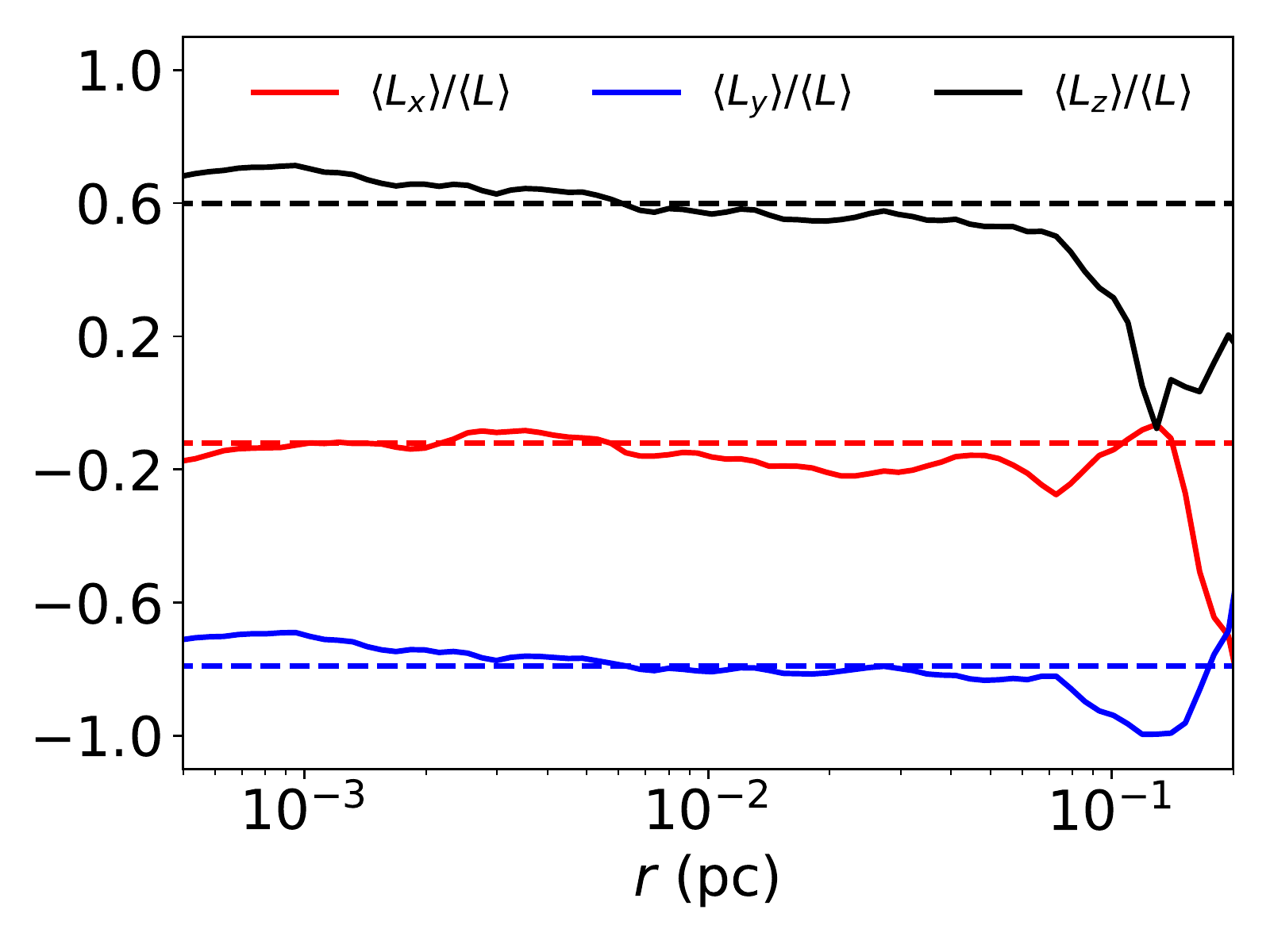}
\caption{Left: Comparison between the density-weighted and the mass flux-weighted averages of the angular momentum, both normalized to the Keplerian value at the inner boundary, $l_{in}$. Right: Time and angle averaged angular momentum direction vector of the inner region of our simulation.  Dashed lines represent the normal vector of the clockwise stellar disc taken from \citet{Paumard2006}.  Here we define $\langle L\rangle^2 \equiv \langle L_x\rangle^2 + \langle L_y\rangle^2 + \langle L_z\rangle^2$, where $L_i = \rho l_i$ is the angular momentum per unit volume in the $i$th direction.  Most of the mass lies in a slightly sub Keplerian distribution with $l \approx 0.5 l_{kep}$ with a well-defined direction that is constant in the inner $r\lesssim 0.1$ pc and aligned with the clockwise disc of stars.   The material flowing into the inner boundary, on the other hand, has a nearly constant angular momentum of $l \approx 0.5 l_{in}$ (left panel; dashed blue line), which shows that only material that has circularization radii $\lesssim r_{in}$ is able to accrete. }
\label{fig:l_angle}
\end{figure*}

\begin{figure*}
\includegraphics[width=0.49\textwidth]{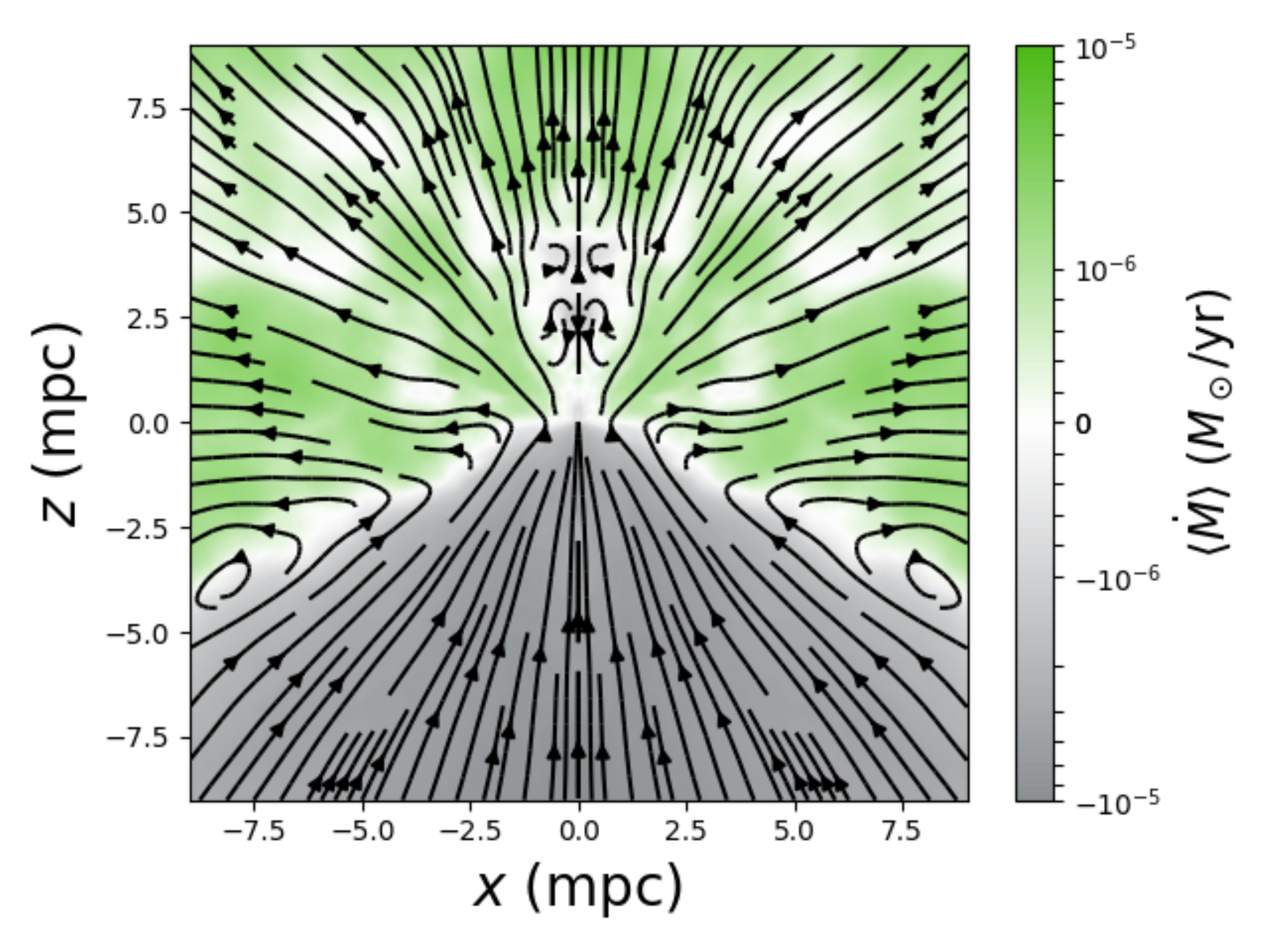}
\includegraphics[width=0.49\textwidth]{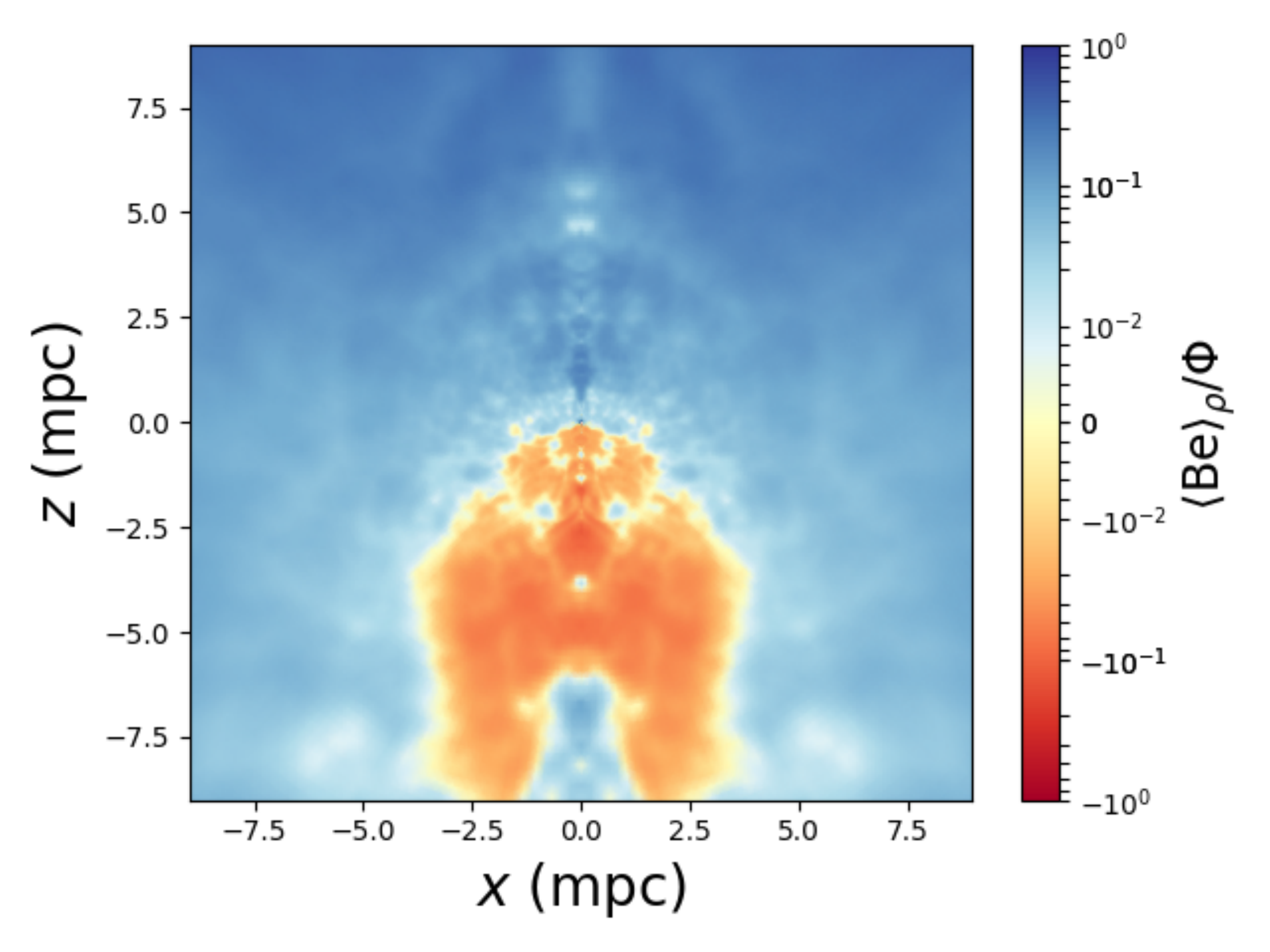}
\includegraphics[width=0.49\textwidth]{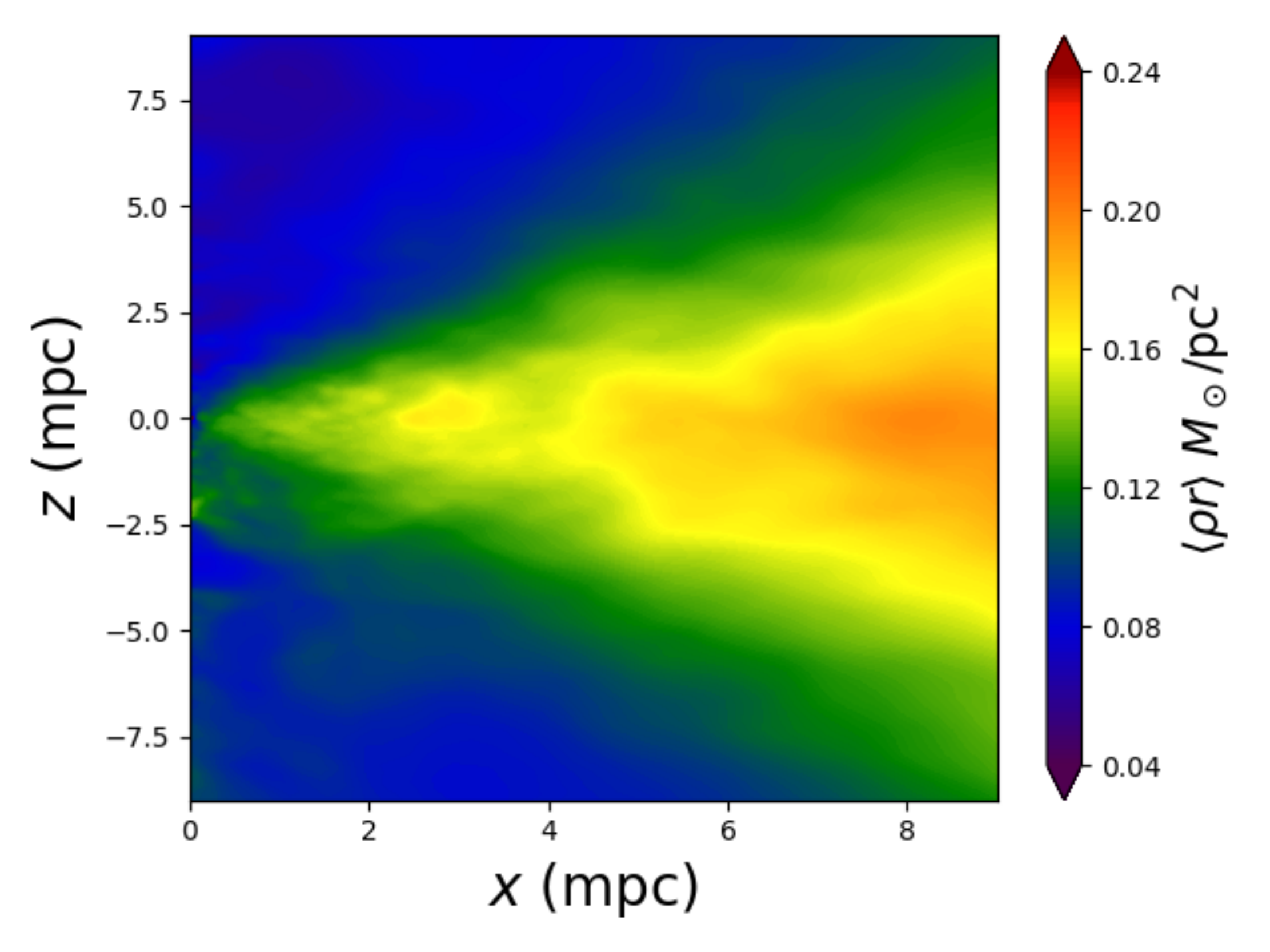}\\
\includegraphics[width=0.45\textwidth]{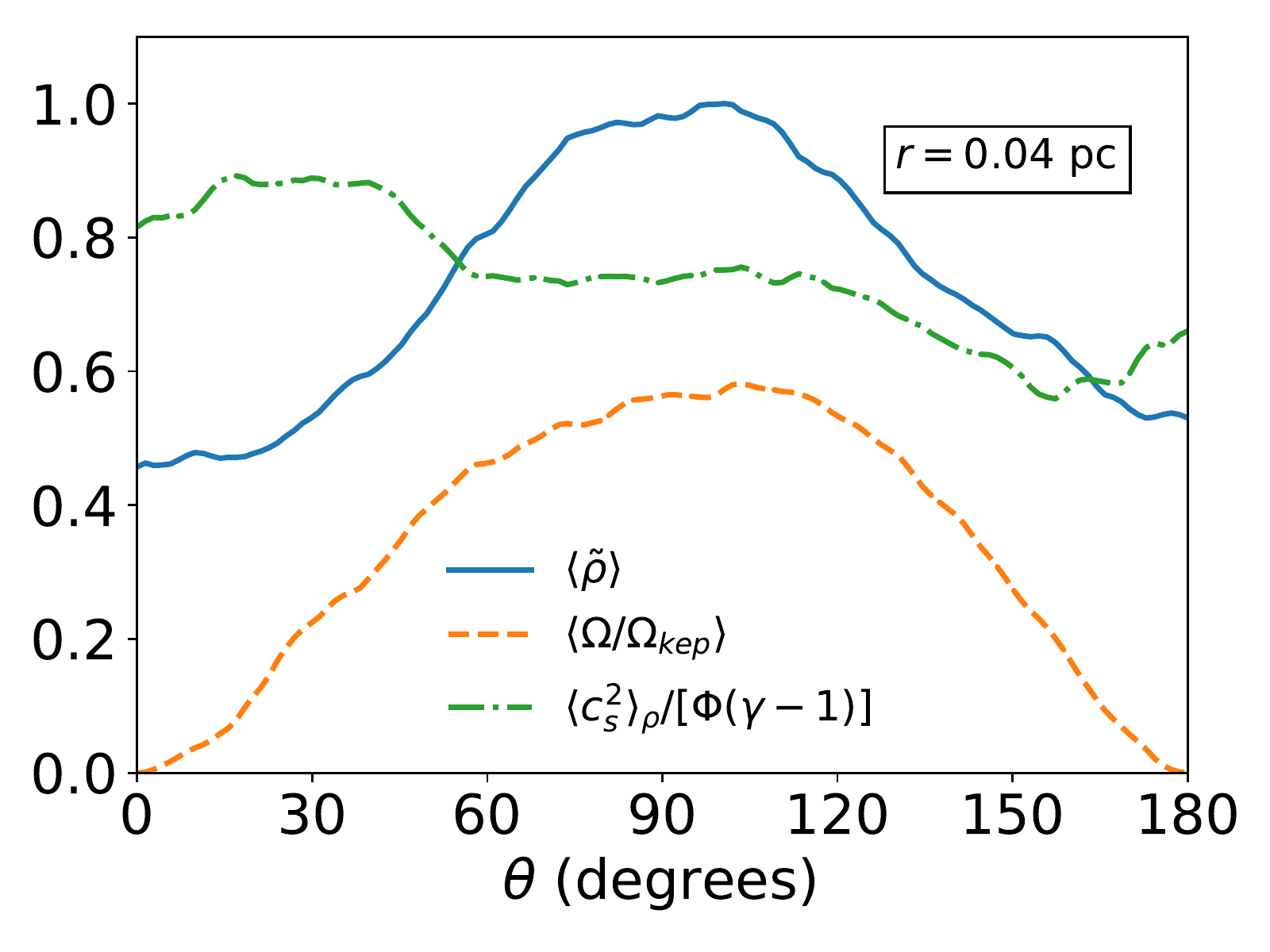}
\includegraphics[width=0.45\textwidth]{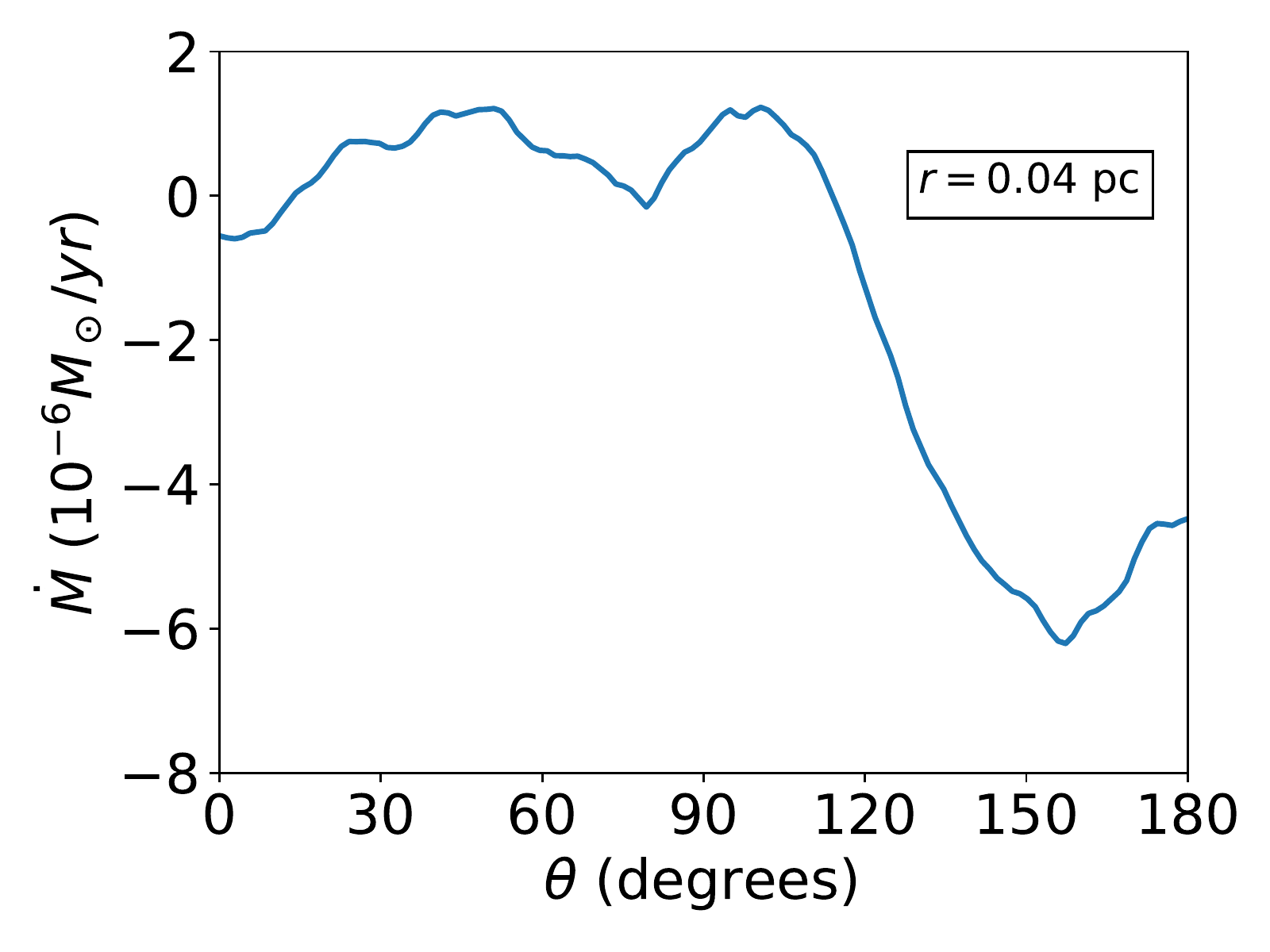}
\caption{Top Left: Time and $\varphi$ averaged mass accretion rate, $\dot M = 4 \pi \rho v_r r^2$, where green denotes outflow and silver denotes inflow, overplotted with streamlines of velocity. Top Right: Time and $\varphi$ averaged Bernoulli parameter relative to the gravitational potential, $\textrm{Be}/\Phi$. Middle: Time and $\varphi$ averaged mass density, multiplied by the spherical radius $r$ to account for the $\rho \propto r^{-1}$ scaling we show in Figure \ref{fig:primitives}.  Bottom Left: Angular profiles at $0.04$ pc $\approx 0.1''$ of the time and $\varphi$ averaged mass density (normalized so that the peak density is 1), $\tilde \rho$, angular velocity in units of the Keplerian rate, $\Omega/\Omega_{kep}$, and the ratio between the thermal component of the Bernoulli parameter and the gravitational potential, $c_s^2/[\Phi(\gamma-1)]$. Here $\Omega \equiv v_\varphi/[r \sin(\theta)]$ and $\Omega_{kep} \equiv \sqrt{GM_{BH}/[r\sin(\theta)]^3}$  Bottom Right: Time and $\varphi$-averaged accretion rate as a function of polar angle  at 0.04 pc.    Here $\varphi$ is defined as the azimuthal angle with respect to the angular momentum axis shown in the right panel of Figure \ref{fig:l_angle}. The disc that forms is very thick and pressure supported, with only a small contrast between the density in the midplane compared to the density at the poles (note the linear scale on the density contour and angular profile). Furthermore, accretion occurs primarily by bound material (Be$<$0) in the southern polar regions, while the midplane and northern pole are predominately composed of unbound(Be$>$0) outflow.  This is caused by the asymmetry of the location of the non-disc stars (see Figure \ref{fig:orbits}) and the fact that material can only inflow for $r>r_{circ}$, where $r_{circ}$ is the circularizaiton radius, at which point it is preferentially ``scattered'' towards the midplane (defined with respect to the angular momentum axis).   }
\label{fig:stream}
\end{figure*}

\begin{figure}
\includegraphics[width=0.45\textwidth]{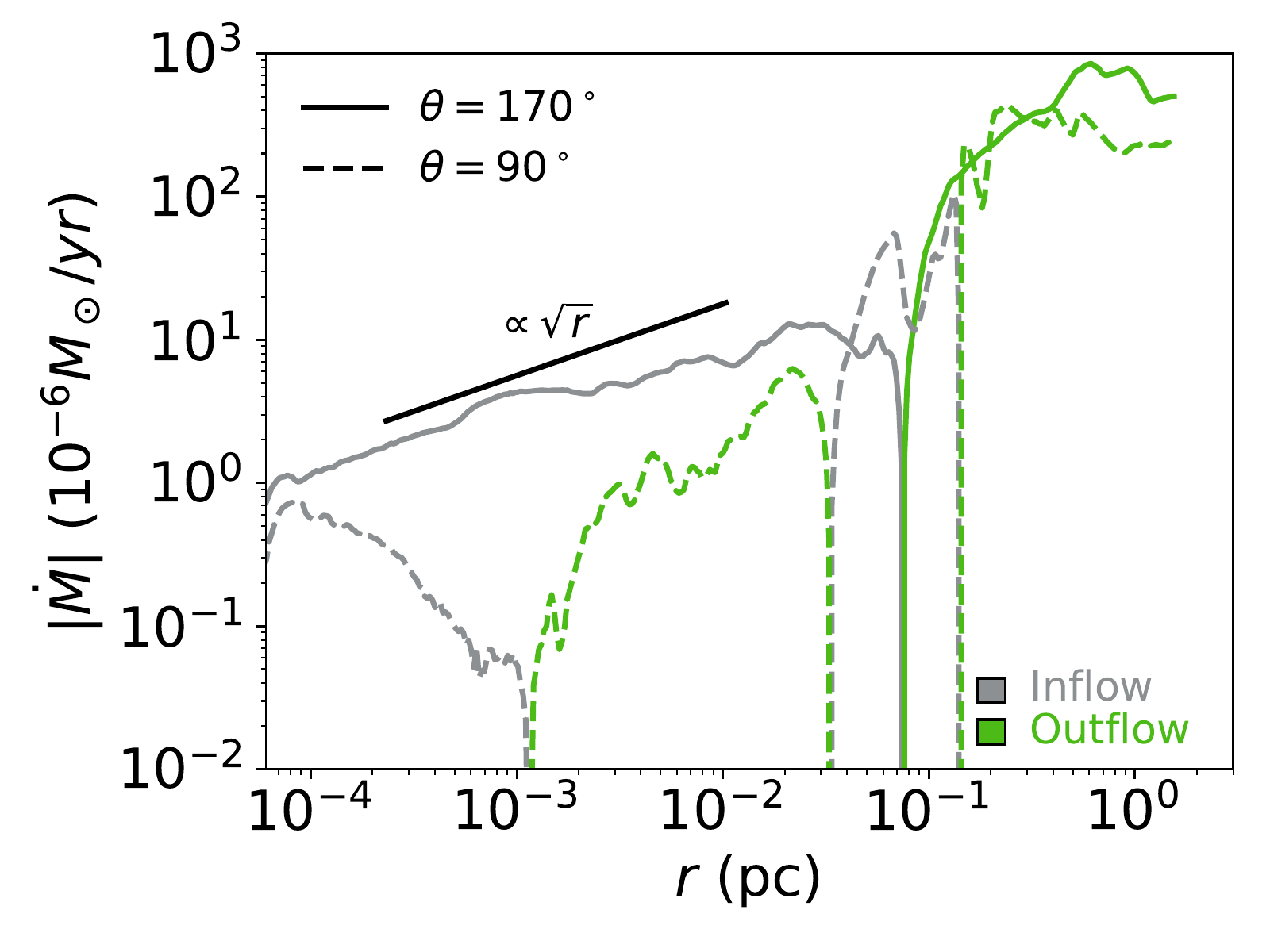}
\includegraphics[width=0.45\textwidth]{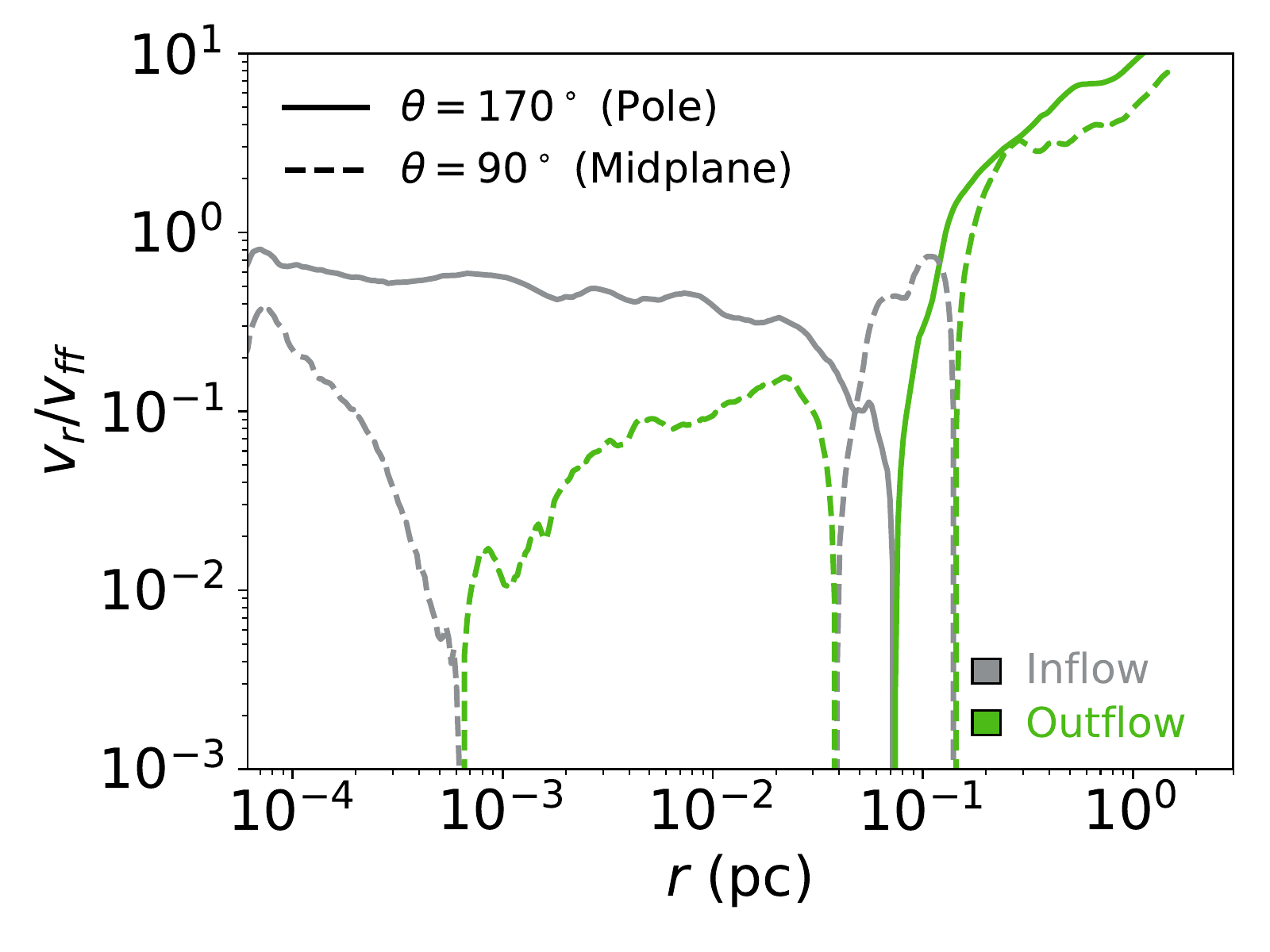}
\includegraphics[width=0.45\textwidth]{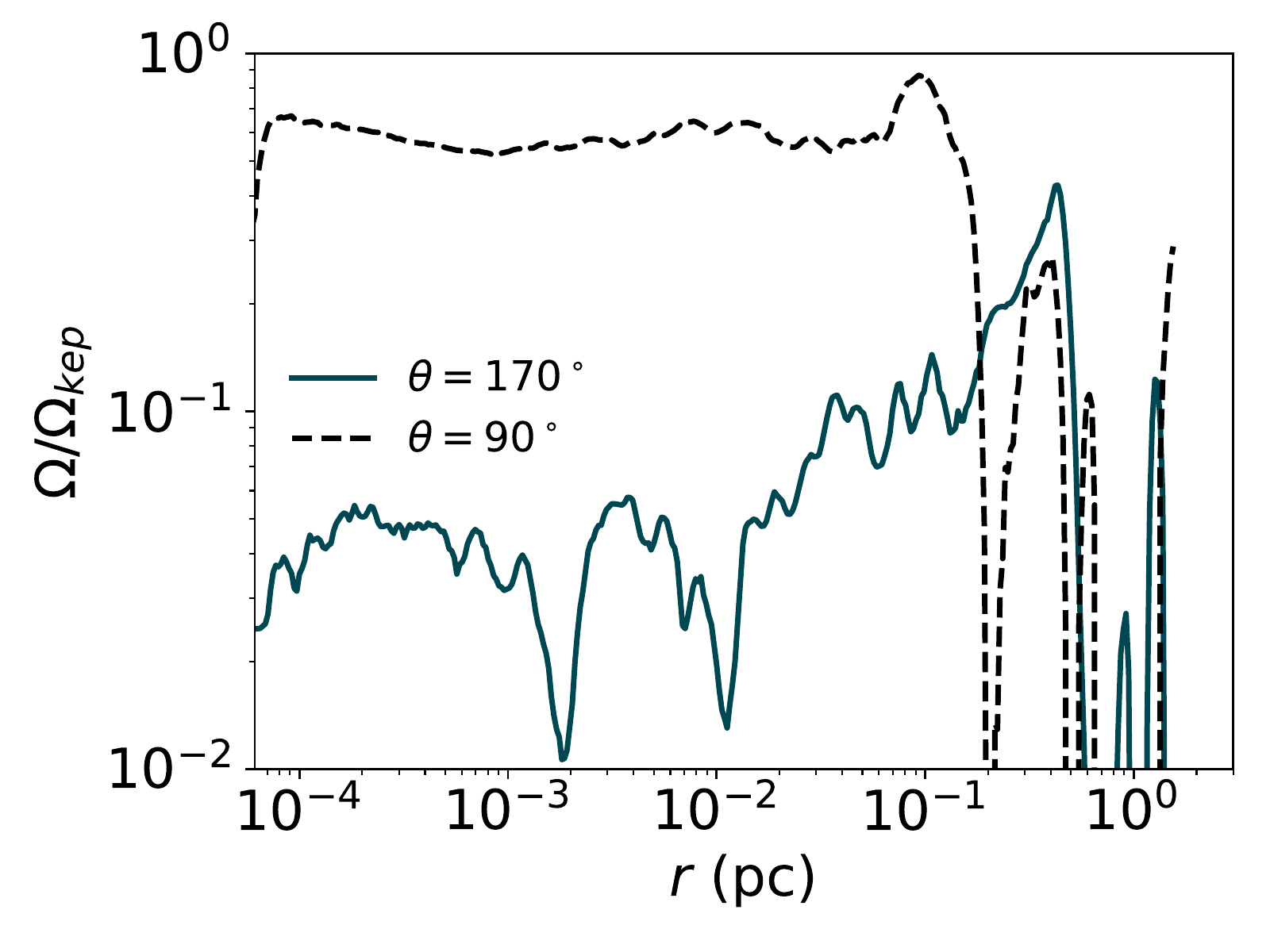}
\caption{Time and $\varphi$-averaged radial profiles of accretion rate (top), radial velocity in units of the free fall speed (middle), and angular velocity in units of the Keplerian rate (bottom) along $\theta$ slices, where $\Omega \equiv v_\varphi/[r \sin(\theta)]$, $\Omega_{kep} \equiv \sqrt{GM_{BH}/[r\sin(\theta)]^3}$, and $\theta$ is defined with respect to the angular momentum axis shown in the right panel of Figure \ref{fig:l_angle}. Solid lines are profiles along the southern pole, dashed lines are profiles along the midplane (see Figure \ref{fig:stream}), while green denotes outflow and silver denotes inflow.  The bound material in the southern pole is essentially in free-fall, with an accretion rate that nicely follows the $\sqrt{r}$ power-law predicted from   isolated star accretion (Appendix \ref{app:single_star}).  The material in the mid-plane, on the other hand, is dominated by pressure and rotational support with a much smaller radial velocity and inflow/outflow rates.    }
\label{fig:pole_v_mid}
\end{figure}

\subsubsection{Extrapolating Down To The Event Horizon}
\label{sec:boundary}

As discussed in the previous section, the amount of matter that accretes through the inner boundary depends on the value of $r_{in}$.  This is for two reasons. First, the ``absorbing'' boundary condition that we use removes radial pressure support, leading to an increased inflow rate in the innermost region.  Second, in order for material to accrete, it must have $l \lesssim 0.5 l_{in}$, where $l_{in} \equiv \sqrt{GM_{BH}r_{in}}$ is the Keplerian angular momentum at the inner boundary.  Both of these effects would be physically reasonable if $r_{in}$ represented the event horizon of the black hole, but unfortunately such a small $r_{in}$ is too expensive for our current computational resources, which use $r_{in} \approx 300 r_{G}$.  On the other hand, we have found that our simulation quantities roughly obey power laws over much of the inner domain, so we can reasonably extrapolate down to smaller radii.

\begin{figure}
\includegraphics[width=0.45\textwidth]{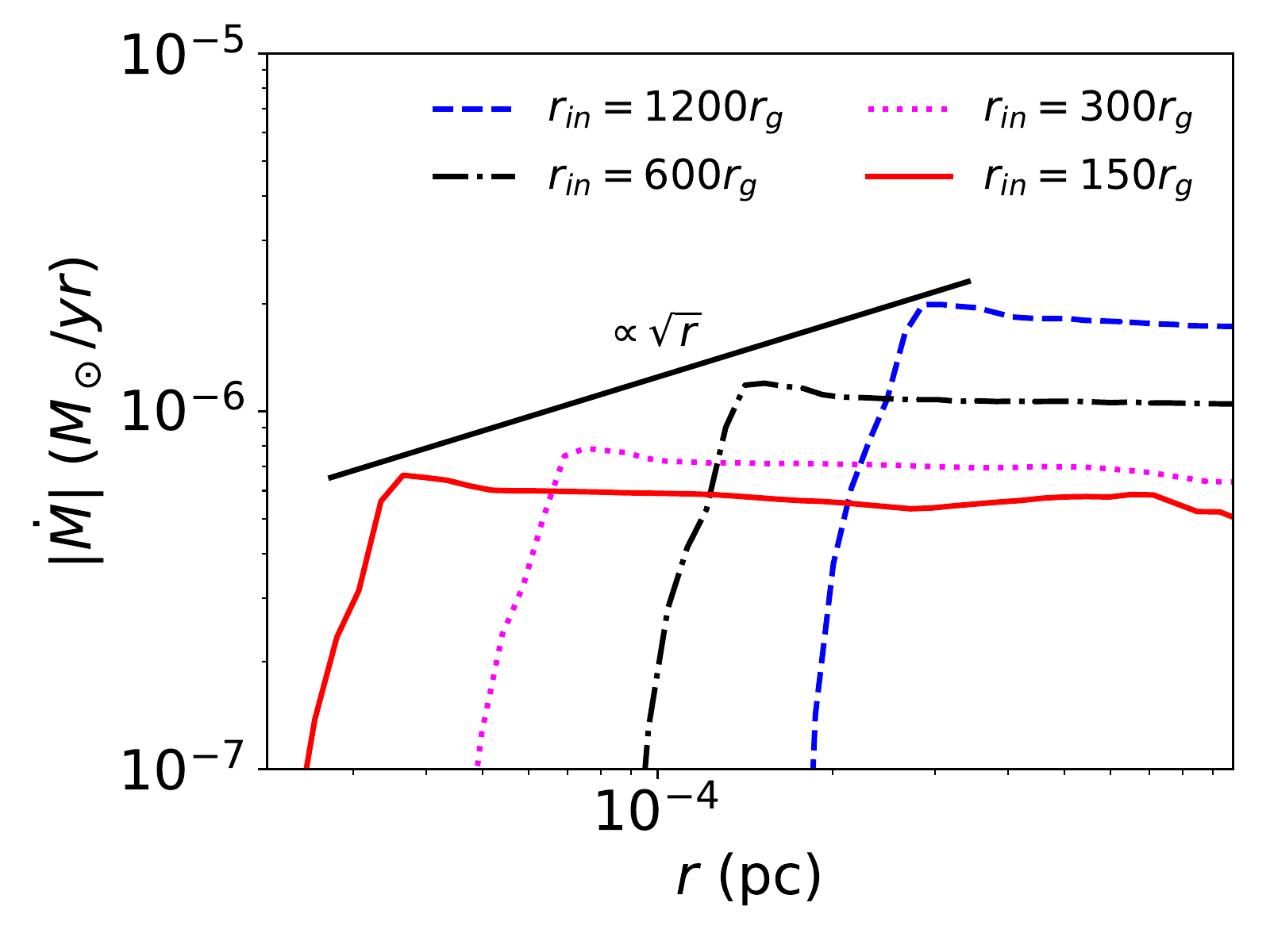}
\includegraphics[width=0.45\textwidth]{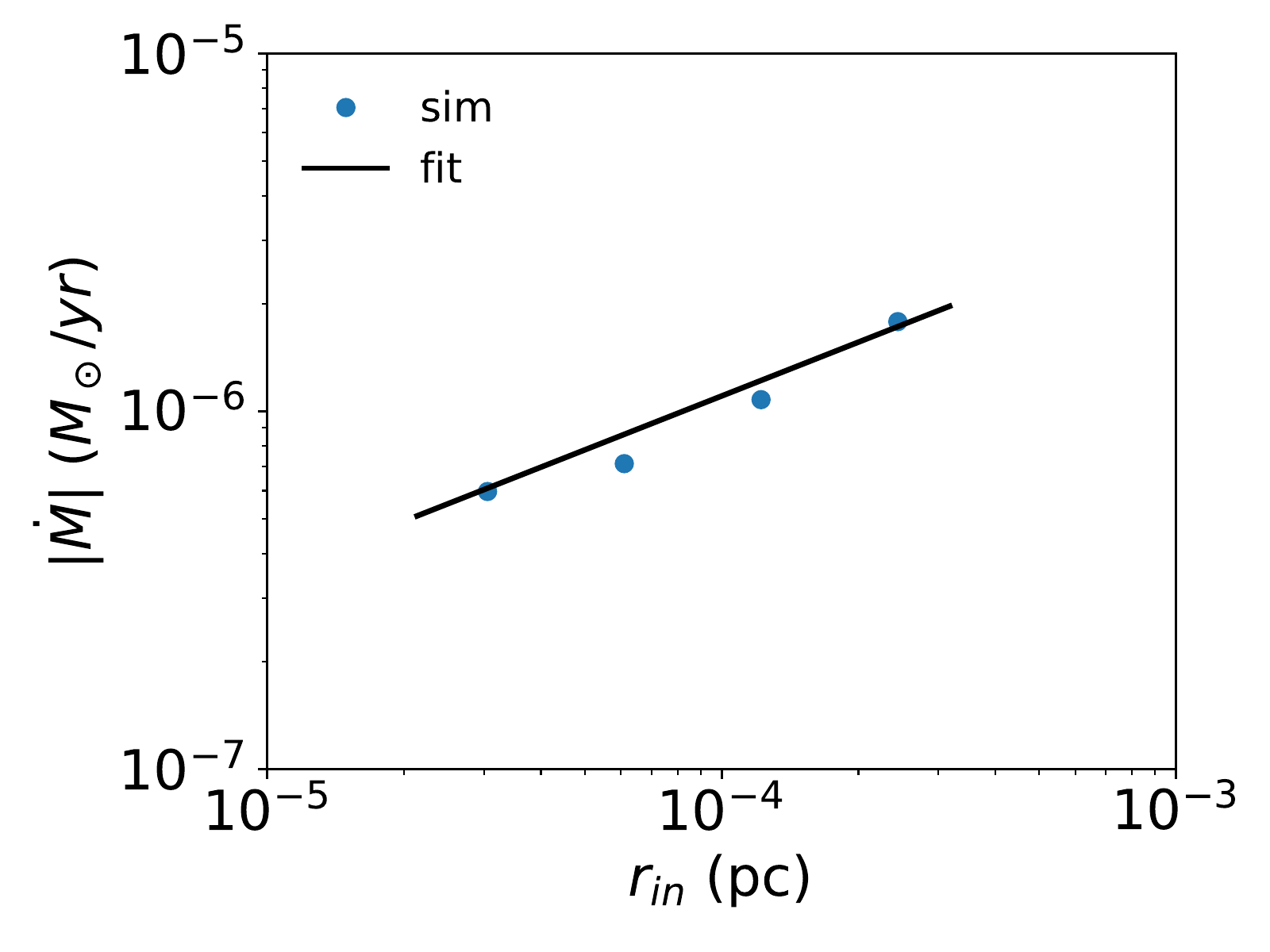}
\caption{Dependence of the accretion rate on the inner boundary.   Top: Time-averaged accretion rate as a function of radius for four different values of the inner boundary.  Bottom: The fit presented in Equation \ref{eq:mdot_fit}  plotted vs. the time averaged accretion rate at $r =2.5 r_{in} $ for the same four values of the inner boundary radius.  Using the $\sqrt{r}$ dependence of the accretion rate that holds well over this range of $r_{in}$, we estimate $\dot M \approx 3 \times 10^{-8}$ $M_\odot$/yr at the horizon of the black hole.}
\label{fig:mdot_bound}
\end{figure}

The effect of the inner boundary is to force $v_r(r_{in})$ to be $\approx - c_{in} \sim - v_{ff}(r_{in}) \propto r_{in}^{-1/2} $, while we have shown that $\rho \propto r^{-1}$ (Figure \ref{fig:primitives}).  Thus, we expect $\dot M \propto \sqrt{r}$, which is the natural result of only handful of stars that have wind speeds comparable to their orbital speeds dominating the accretion supply (see Appendix \ref{app:single_star}). We have already shown that the accretion rate measured along the southern pole that dominates the inflow nicely matches this scaling relation (Figure \ref{fig:pole_v_mid}).  For further confirmation of this result, in the top panel of Figure \ref{fig:mdot_bound}, we plot the accretion rate as function of radius for four different values of $r_{in}$ compared to an $r^{-1/2}$ power law. The agreement is fairly good.  By setting the constant of proportionality using the accretion rate in the $r_{in} \approx 300 r_{G}$ simulation, we find that
\begin{equation}
\dot M \approx 2.4 \times 10^{-8} M_\odot /\textrm{yr}\sqrt{\frac{r_{in}}{r_{G}}},
\label{eq:mdot_fit}
\end{equation}
which is shown in the bottom panel of Figure \ref{fig:mdot_bound} to be an excellent representation of our simulations. 
For a non-rotating black hole, the horizon is located at 2$r_{G}$ and thus Equation \ref{eq:mdot_fit} predicts an accretion rate of $\approx  3.4 \times 10^{-8} M_\odot/$yr.  The Bondi rate that would be inferred from the density and temperature at $2''$ in our simulation is $2.4 \times 10^{-5} M_\odot/$y. Our estimated $\dot M$ at the horizon is a factor of $\sim 700$ lower due to the presence of rotationally-driven outflow.  Remarkably, the prediction of Equation \eqref{eq:mdot_fit} is entirely consistent with the observational limits inferred from polarization measurements \citep{Marrone2007} as well as previous estimates of the accretion rate based on models of the horizon-scale accretion flow (e.g., \citealt{Shcherbakov2010,Ressler2017}). It is unclear if this result will hold in MHD simulations, however, since angular momentum transport in rotationally supported material may modify $\dot M$ from the value set by the low angular momentum tail in our hydrodynamic simulations. 

\subsubsection{The Effect of Including S2}
\label{sec:S2}
The star S2, which has an orbit that reaches $\sim 3000$ $r_g$ (or $\approx$ $0.01''\approx 4$ mpc, \citealt{Gillessen2017}), is of particular interest for many studying the galactic center. Its exceptionally well-constrained orbit has been used to constrain the mass and distance to Sgr A*, and high-precision measurements of its next pericenter passage will be used to test the theory of General Relativity \citep{Grould2017,Hees2017,Chu2018}. 
Though S2 is much fainter and thus expected to have a much weaker stellar wind than the typical WR star surrounding Sgr A*, its proximity to the black hole could increase its potential effect on feeding and/or disrupting the accretion flow in the innermost radii \citep{Loeb2004,Nayakshin2005,Giannios2013,Schartmann2018}. This would be especially true at pericenter, which is expected to occur in the year 2018.  To test this hypothesis, in this section we briefly consider the effect that the wind from this star could have on accretion onto Sgr A*. Note that S2 is among the most massive of the S stars \citep{Habibi2017} and thus the most likely to have a strong wind.

For the observed properties of S2 (e.g. \citealt{Habibi2017}), the theoretical model of \citet{Vink2001} predicts a mass-loss rate of  $\approx 2 \times 10^{-8}$ $M_\odot$/yr for a fiducial wind speed of $2000$ km/s. Note that this mass-loss rate is $\sim$ 3 orders of magnitude less than the typical WR star in our simulation. Using said mass-loss rate and wind speed in addition to the precisely known orbit as given by \citet{Gillessen2017},  we performed a second simulation that is identical to the the first except that it included S2 as an additional wind source term. 

\begin{figure}
\includegraphics[width=0.45\textwidth]{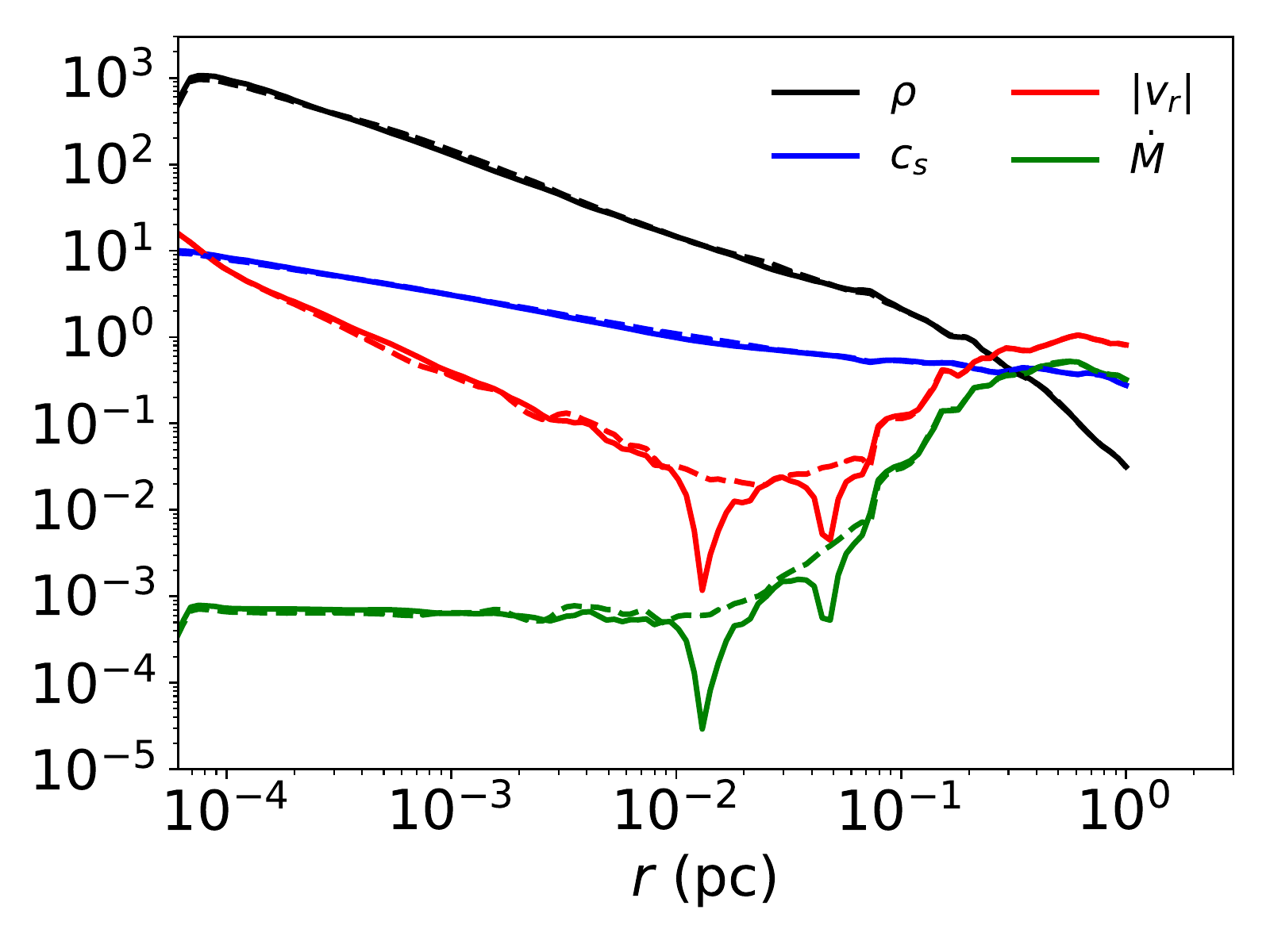}
\caption{Comparison between the time and angle averaged radial profiles of mass density, sound speed, radial velocity, and accretion rate for simulations including the stellar wind provided by the star S2 (solid) and without S2 (dashed).  The units of $\rho$, $c_s$ ($v_r$), and $\dot M$ are $M_\odot$/pc$^3$, pc/kyr, and $M_\odot$/kyr, respectively.
Due to its low mass loss rate compared to the WR stars,  including S2 has a negligible effect on the average radial profiles of the flow around Sgr A*. }  
\label{fig:S2}
\end{figure}

Figure \ref{fig:S2} shows that including S2 has essentially no effect on the time and angle averaged flow properties. This is because, even at $\sim$0.01'', the inflow and outflow rates shown in Figure \ref{fig:mdot} are still almost 2 orders of magnitude larger than the mass loss rate of S2.   This is consistent with the results of \citet{Amuse2016}, who found that a simulation that included the winds of the S-stars alone could only provide significant accretion if their mass-loss rates were $\sim$ 10-100 times larger than those inferred from observations (e.g., the values quoted above for S2).

\section{Constraining Stellar Wind Mass-loss Rates and Wind Speeds with X-ray Observations}
\label{sec:Xray}
The simulations presented in this paper used the mass-loss rates presented in \citet{Martins2007} that were obtained by fitting stellar wind models to infrared spectra. Their models included the effects of clumping, which reduce the inferred mass-loss rates for some of the stars (but not all) by a factor of $\sim3$.  Estimates of the mass loss rates of the same stars derived from radio observations, however, are, on average, smaller by a factor of $\sim 2$ \citep{YZ2015}. This is even without including the effects of clumping, which would reduce the radio-inferred  mass-loss rates of some stars by another factor of $\sim3$.  The infrared and radio data probe different spatial scales of the winds and use different modeling techniques so it is not clear which is a better representation of the true mass-loss rates.   

To obtain an additional constraint, we turn to \citet{Baganoff2003}, who presented spatially resolved X-ray observations of Sgr A* that measured the total 2-10 keV luminosity at two different scales, namely, between $1.5''$-$10''$ and also $<1.5''$. As in previous work \citep{Baganoff2003,Quataert2004,Rockefeller2004},  
we propose that the hot gas responsible for both of these emission components is provided by the stellar winds of the WR stars.  In that case, these measurements of the X-ray luminosity help determine the stellar wind mass-loss rates, since the luminosity scales as $\propto n^2 \propto \dot M_{wind,tot}^2$ where $\dot M_{wind,tot}$ is the total mass-loss rate of all the stars.  The constraint is even stronger when we consider that the two measurements probe regions in the flow with very different dynamics.  $10 ''$ (0.4 pc) lies outside most of the stellar winds where the solution approaches a large scale Parker wind whose properties are primarily determined by the total mass-loss rate of the WR stars and the stellar wind velocities.   $1.5''$ (0.06 pc), on the other hand, is inside most of the stellar winds and falls within the ``stagnation region'' described in \S \ref{sec:results}.  Here the hydrodynamic solution depends more strongly on the distribution of the stellar wind mass-loss rates with radius.  For a fixed total mass-loss rate, a uniform distribution of mass-loss with radius results in a higher density at $1.5''$ than if most of the mass-loss is provided by stars at larger radii \citep{Quataert2004}.  

In order to compare our results to the \emph{Chandra} observations, we again use SPEX \citep{SPEX} exactly as described in \S \ref{sec:cool}, except we consider only the contributions to $\Lambda$ from photon frequencies corresponding to the 2-10 keV range, denoting this as $\Lambda_X$.  The total X-ray luminosity    of our simulation within a cylindrical radius $s$, $L_{X}(s)$, is then computed by integrating: 
\begin{equation}
L_X(s)= \int\limits_0^{2\pi} \int\limits_{-z_{max}}^{z_{max}} \int \limits_{r_{in}}^s  \frac{\rho^2}{\mu_e \mu_{H,\odot}} \Lambda_X  s ds dz d\varphi,
\label{eq:Lx}
\end{equation}
where $z_{max}$ is half of the box length of our simulation and $r_{in}$ is the radius of the inner boundary.   Doing this, we find that at the present day, $L_X(10'') - L_X(1.5'') \approx 2.5 \times 10^{34}$erg/s and $ L_X(1.5'') \approx 7.3 \times 10^{33}$ erg/s.  These are to be compared with the \emph{Chandra} measurements of $2.4$ (1.8-3.2) $\times 10^{34}$ erg/s and $2.4$ (1.8-5.4) $ \times 10^{33}$ erg/s, respectively.  The agreement between our models and the \emph{Chandra} data is overall quite good, particularly accounting for uncertainties in massive star mass-loss rates (e.g. \citealt{Smith2014}).  In more detail, the X-ray luminosity between $1.5''$ and $10''$ is in excellent agreement with the \emph{Chandra} data, but the X-ray luminosity of the inner $1.5''$ of our simulation is overproduced by a factor of $\sim$ few.  As discussed above, this suggests that the overall mass-loss rate is roughly the right value 
  but that the distribution of the mass-loss rates with radius (i.e. the location of the stars) is perhaps too spread out in radius such that there is an over-density at $1.5''$ (see \citealt{Quataert2004}).  This is consistent with the fact the orbits of several of the 30 stars in our simulation are uncertain due to the lack of information about the line-of-sight position.  Since the orbits directly determine the mass-loss distribution, we hypothesize that a better knowledge of the line of sight positions of the WR stars would bring our simulations into better agreement with observations.  

\begin{figure}
\includegraphics[width=0.45\textwidth]{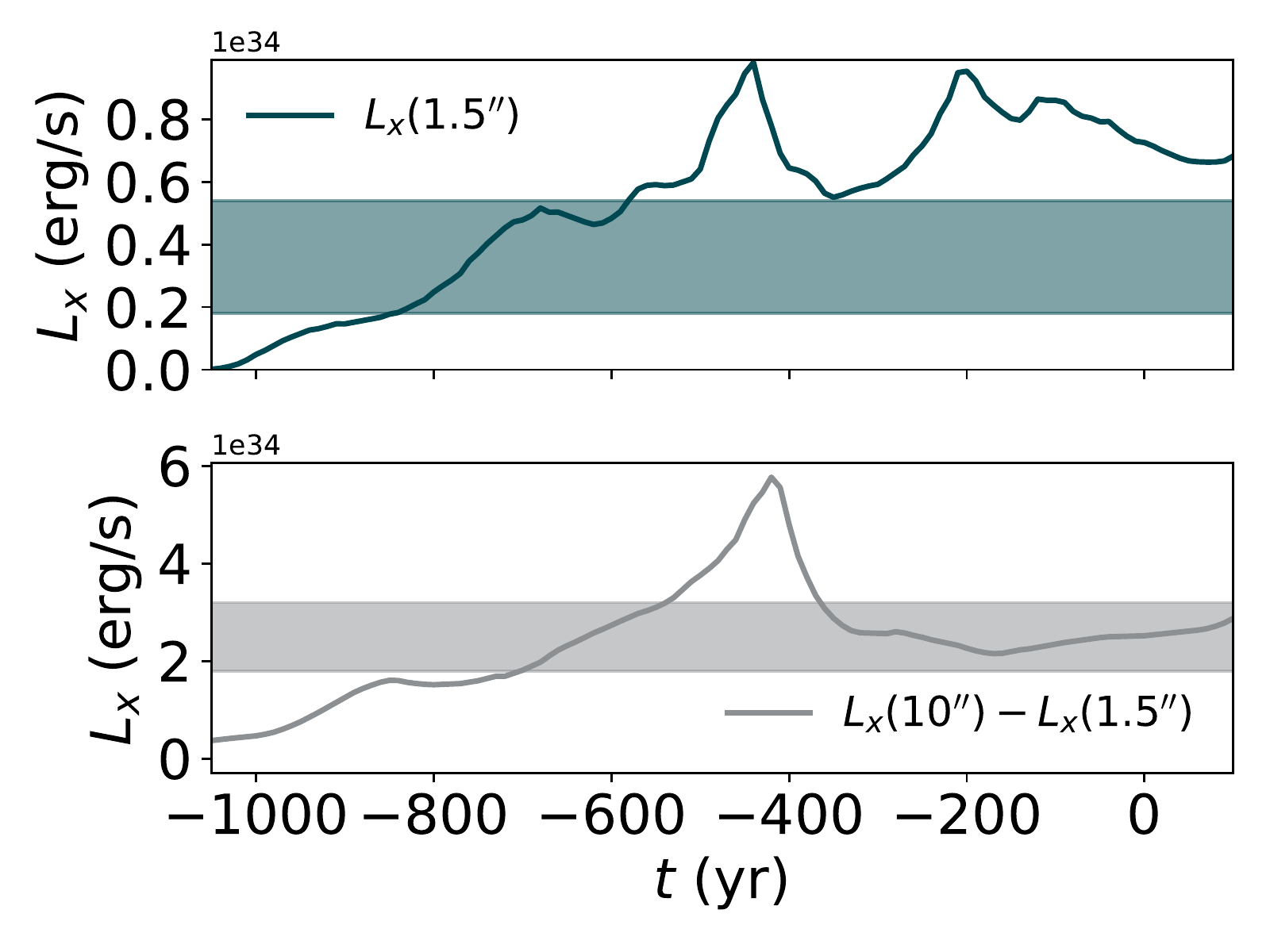}
\caption{Total X-ray (2-10 keV) luminosity produced by our simulations as a function of time within a cylindrical radius of $1.5''$, $L_X(1.5'')$, and within cylindrical radii $1.5''$ and $10''$, $L_X(10'')-L_X(1.5'')$. The shaded regions represent the 90\% confidence level intervals of \emph{Chandra} observations of Sgr A*.  At the present day, our simulation accurately reproduces the emission between $1.5''$ and $10''$ but has a luminosity within $1.5''$ that is a factor of 2-3 too large.  This  discrepancy is likely caused by the uncertainty in the stellar orbits or wind properties of the WR stars in our simulation leading to a moderate $(\sim 50 \%)$ over-density of gas at $\sim 1.5''$.}
\label{fig:Lx_t}
\end{figure}

The X-ray light curves shown in Figure \ref{fig:Lx_t} support this argument, which show that the X-ray luminosity between $1.5''$ and $10''$ has been relatively steady over the past $\sim 400$ years despite the fact that the stellar wind distribution has changed significantly (Figure \ref{fig:orbits}). $L_X(1.5'')$ on the other hand, does display slightly more pronounced variation with time over the same interval,  suggesting that is is more dependent on the instantaneous orbital configuration of the stars. 

Even with the configurations of stellar winds adopted in our simulation, however, the discrepancy with the X-ray measurements is small enough that these results argue in favor of the \citet{Martins2007} mass-loss estimates as opposed to those of \citet{YZ2015}, which would decrease $L_X$ by a factor of $\gtrsim 4$. 

Additionally, \citet{Baganoff2003} also provide a best fit temperature for the gas at $\sim 10''$, namely, 1.3 keV, which agrees very well with the value we find in our simulation ($\approx 1.5 keV$, Figure \ref{fig:primitives}).  The temperature at this scale is predominately set by the stellar wind speeds (also taken from \citealt{Martins2007}), scaling roughly as $v_{wind}^2$.  This agreement then implies a confirmation of the wind speeds to the $\sim 10 \%$ level.    

These two results together give us confidence that our simulation is capturing all of the hot gas observed by \emph{Chandra} and that the resulting flow is a reasonable representation of observations.


\section{Implications for Horizon-Scale Accretion Modeling}
\label{sec:imp}
One of the goals of this work is to use the observationally constrained simulation of the accretion provided by WR stars to assess the ``right'' initial conditions for GRMHD simulations that model Sgr A*.  In this section we summarize the properties of the accretion flow presented in \S \ref{sec:results} by comparing and contrasting them to the standard initial torus structure used in past work.

Like the results of our simulation presented in \S \ref{sec:results}, the typical initial torus used by many GRMHD simulations is thick and pressure supported. Unlike our results, however, this torus is usually well contained within some polar opening angle, beyond which there is a sharp cut-off in density.  Here we have shown that the `disc' is much less sharply defined, with a contrast of only a factor of $\sim$ a few between the midplane and the polar regions (Figure \ref{fig:stream}).  Even more striking is the presence of a significant amount of low angular momentum inflow along the pole, which we estimate being as high as $3.4 \times 10^{-8} M_\odot/$yr when extrapolated to small radii (Figure \ref{fig:mdot_bound}), comparable to the accretion rate estimated at the horizon in Sgr A*. The presence of such an inflow could potentially inhibit the formation of jets, suppress outflow, and increase the net accretion rate onto the black hole in horizon-scale simulations.

This polar inflow also has the effect of driving a pressure supported outflow (Figure \ref{fig:stream}), so that the matter in the midplane of the disc is continually being recycled.  In GRMHD simulations this could suppress the MRI if the growth rate is smaller than the inflow/outflow rate.  That is, though the configuration is indeed unstable to the MRI with angular velocity decreasing with radius, it is possible that before the instability can grow significantly the fluid will be swept away and either accreted or propelled to large radii. Since our simulations produce a disc with root-mean-squared radial flow timescale
that is comparable to (i.e., $\sim 1-2$ times longer than) the rotational period, a simple timescale analysis is inconclusive; a full treatment of MHD is required to determine the importance of the MRI on angular momentum transport in the disc.

Finally, we find that the angular momentum distribution follows a (sub) Keplerian profile of $\approx 0.5 l_{kep}$ as opposed to the constant angular momentum tori used by GRMHD simulations. This difference, however, is likely less important because the horizon-scale simulations quickly evolve to a similarly sub-Keplerian distribution after the onset of accretion.  

Future work will seek to directly take the results of our simulations and implement them as initial and boundary conditions for a GRMHD simulation in order to determine the significance of these differences on the properties of the resulting flow.

\section{Comparison to Previous Work}
\label{sec:comp}
Several groups have studied the fueling of Sgr A* with 3D simulation using several different models for the stellar-wind emitting source terms \citep{Rockefeller2004,Cuadra2005,Cuadra2006,Cuadra2008,Amuse2016}.
Here we focus on the work of \citet{Cuadra2008} (C08), which is the most similar to ours in that they included the most up-to-date stellar mass-loss rates, stellar wind speeds, and current day star locations while also evolving the position the stellar wind sources with time. In particular, C08 focused on three different orientations of the accreting stars.  The orientation of our stars at the present day is equivalent to their ``1-disc'' model with the exception of the star S97, whose orbit has been more precisely determined by \citet{Gillessen2017}.  The major differences between the two simulation are
\begin{itemize}
\item Computational methods:  We use a conservative grid based hydrodynamic code while C08 used a smoothed-particle hydrodynamic (SPH) code.  Conservative, finite-volume schemes excel in capturing shocks, an area in which SPH schemes can have trouble, particularly when the gas is diffuse and low temperature.  
\item The inner boundary radius.: The inner boundary of the C08 simulation was set at 0.05$''\approx 2$ mpc, while our inner boundary is $\approx 32$ times smaller, at $\approx  1.6 \times 10^{-3}$ $'' \approx 0.06$ mpc.
\item The cooling function: C08 used a three-part piece wise cooling curve that approximates \cite{Sutherland1993} for a solar hydrogen abundance with $Z=3 Z_{\odot}$.   We use a larger number of power law segments to approximate the cooling curve appropriate for stellar wind material of WR stars that are largely bereft of hydrogen (also with $Z = 3 Z_\odot$). Furthermore, the {\tt SPEX} code that we use to calculate the cooling function includes more lines than the \citet{Sutherland1993} calculation (see \citealt{Schure2009}), which enhances the peak of the curve at $\sim 10^{5}$ K.  These differences, however, have a relatively small effect on the cooling curves, which are plotted in Figure \ref{fig:cc}. Except at the very highest ($T \gtrsim 10^9$ K) and lowest ($T \lesssim 3 \times 10^4$ K) temperatures, the cooling curves are within a factor of 2 of each other.
\end{itemize}
C08 focused on the effect of the different stellar orbital distributions on the accretion history and X-ray luminosity, while this work is primarily focused on modeling the structure of the innermost accretion flow at the present day.  Thus much of the information presented here is not in C08 for comparison and vice-versa.  We can, however, compare the mass accretion rate history and the radial profile of the angular momentum to C08, while noting that the radial profiles of density, temperature, and radial velocity of the C08 simulation are presented in a later work by the same group (\citealt{Cuadra2015}, C15) where it is labeled as the ``control run.'' 

The level of variability seen in the accretion rate history is comparable in both simulations, with the average accretion rate in C08 being a factor of $\sim$3-4 times higher as expected from the larger inner boundary radius used in their simulation.  Also similar is the level of variability seen in the angular momentum vector as a function of time, which is primarily determined by the time-evolving configuration of the stellar winds.  At the present day, however, our angular momentum vectors are in two different directions, forming an angle of $\sim 50^\circ$ with each other. This is not necessarily surprising due to the high level of temporal variability in this vector (Figure \ref{fig:history}) and the stochastic nature of the inner accretion flow.
C08 also found that the stellar winds of only 3 stars contributed significantly to the accretion near the inner boundary which is consistent with the model we propose in Appendix \ref{app:single_star} to explain the $\dot M \propto \sqrt{r_{in}}$ dependence we find in our simulation.

While the accretion and angular momentum histories are broadly similar in the two simulations, there are striking differences seen in the radial profiles of fluid quantities, particularly in the inner region of the flow.  These include:
\begin{itemize}
\item Temperature: we find $T \propto \approx r^{-1}$ with $T\approx 2 \times 10^8$ K at $0.1''$ ($ 4$ mpc) while C08 found $T \propto \approx r^{-0.4}$ with $T\approx 4 \times 10^7$ K at $0.1''$ ($ 4$ mpc), almost an order of magnitude lower.  Note that part of the difference in magnitude is caused by our assumption that $X=0$, corresponding to a larger mean molecular weight than that used by C08..  
\item Radial velocity: the radial Mach number of our simulation is $\approx 7 \times 10^{-2}$ ($v_r \approx 2 \times 10^8$ cm/s) at $0.1''$ ($ 4$ mpc), while in C08 at the same radius the radial flow is supersonic at ($v_r \gtrsim 10^8$ cm/s).
\item Density: our density profile is much steeper, $\rho \propto \approx r^{-1},$ compared to C08, $\rho \propto \approx r^{-0.5}$. 
\item Angular momentum: we find $l\approx 0.5 l_{kep} \propto \sqrt{r}$, while C08 found $l\approx l_{kep}(0.05'' \approx 2$ mpc$)=$ const.
\end{itemize} 

What causes such large differences between the two simulations? We can only speculate.   By varying the cooling function, we have found that our results are not strongly dependent on the particular choice of $\Lambda$, so it is not likely that this is the source of disagreement. On the other hand, we have found that the inner boundary condition can cause artificial effects out to $\sim $ a few $ r_{in}$, namely, reducing pressure support and increasing radial velocity, so at $0.1''$ (4 mpc) C08's results may still be affected by their boundary at $0.05''$ (2 mpc).  Also, as evidenced by their Figure 8, around $0.1''$ (4 mpc) they have only a handful of particles in their simulation meaning that the inner region of their simulation may be under-resolved.  

We note that \citet{Cuadra2015} followed up on the work of C08 by including subgrid models to account for feedback from the black hole.  Their models are motivated by the fact that, in reality,most of the material accreting at the radius corresponding to the inner boundary of their simulations may ultimately be ejected in an outflow.  Indeed, our simulations that probe smaller radii generally support this expectation, though we find that the outflow proceeds in a direction \emph{perpendicular} to the angular momentum axis (that is, in the orbital plane) of the gas, as opposed to their ``instantaneous'' feedback models that eject material either isotropically, in some fixed opening angle, or parallel to the angular momentum axis.   With feedback, none of their models significantly improve the agreement between the flow properties in our simulations.  Though the radial velocity and accretion rate are lower with feedback, their density and temperature profiles still have different scalings with radius than those seen in our simulations. In the same work, \citet{Cuadra2015} also present an ``outburst'' model where a large amount of mass is injected through the inner boundary over a 300 yr period some time in the past.  Again, this model does not bring our simulations into any closer agreement, and, in fact, the density profile post-outburst is even flatter than their ``control'' run with no feedback.

\section{Conclusions}
\label{sec:conc}

We have presented the results of 3D hydrodynamic simulations that track the accretion of stellar winds in the galactic center from the stars at distances of $\sim 0.1$ pc all the way down to $\sim$300 gravitational radii of Sgr A*, roughly 32 times further in than in previous work.  These are also the first grid-based finite volume simulations of the fueling of Sgr A*. Our simulations include radiative cooling in collisional ionization equilibrium, and adopted the observationally constrained stellar orbits, mass loss rates, and wind speeds of the 30 WR stars that dominate the accretion budget.  We find reasonable agreement between our predicted diffuse X-ray luminosity and \emph{Chandra} X-ray observations (Figure \ref{fig:Lx_t}). This demonstrates that the mass-loss rates and wind speeds from \citet{Martins2007} must be of order the true values (probably within a factor of 2).  Our goal in this work is to detail the flow properties at the innermost radii in order to better motivate initial and boundary conditions for GRMHD simulations of Sgr A*.  These will be used to interpret not only EHT and GRAVITY observations, but also the wealth of observational data that exists across the electromagnetic spectrum.   

We find that the gas at small radii (well inside the orbits of the mass-losing stars, i.e., $r\lesssim 0.01$ pc) develops a 2-component structure.  Most of the mass is moderately unbound in an equatorial rotation supported disc while most of the accretion proceeds by bound material along the poles via the low angular momentum tail of the stellar winds. Only a small fraction ($\lesssim 0.1 \%$) of the stellar wind material is captured by the black hole, leading to a hot, pressure-supported, sub-Keplerian `disc' of gas at small radii that is at most times (though not always) aligned with the clockwise stellar disc (Figure \ref{fig:history}). The accretion rate at small radii is much less than the Bondi rate due to the finite angular momentum of the stellar wind material (Figure \ref{fig:mdot}). While radiative cooling can be significant in the vicinity of the stellar winds, it has a negligible effect at smaller radii and thus cannot remove pressure support.  Due to the pressure support and broad angular momentum distribution, there is only a mild contrast in density between the polar regions and the midplane (a factor of $\sim$ a few), much more akin to spherical accretion than even very geometrically thick RIAF models (see Figure \ref{fig:stream} and \ref{fig:mdot_bound}). 

Accretion in our simulations is dominated by bound, low angular momentum material that flows in from the southern pole, feeding both accretion and outflow that is primarily directed along the midplane and northern pole (Figure \ref{fig:stream}).  This structure is due to both the asymmetry of the distribution of non-disc stars about the midplane of the stellar disc at $t=0$ (Figure \ref{fig:orbits}) and also to the gas possessing a wide range of angular momentum, as expected when the stellar winds of only a few stars contribute to accretion and their wind speeds are comparable to their orbital velocities.  Further evidence for this picture is that the accretion rate through the inner boundary scales as $\propto \sqrt{r_{in}}$, which is identical to the distribution of mass-loss rate vs. circularization radius produced by the wind of a single star (Appendix \ref{app:single_star}).  Using this scaling relation to extrapolate down to the horizon of Sgr A* we find an accretion rate of $\approx 3.4 \times 10^{-8} M_\odot$/yr for a non-spinning black hole (Figure \ref{fig:mdot_bound}), consistent with observational limits on the horizon-scale accretion rate \citep{Marrone2007,Shcherbakov2010,Ressler2017}.    

We find that our results are not altered by including the star S2 as an additional wind source, despite its proximity to Sgr A* (Figure \ref{fig:S2}).  This is because its mass-loss rate is $\sim 3$ orders of magnitude lower than most of the other stars in our simulation, so its effect on the time-averaged flow is negligible.  
Since S2 is the brightest of the `S-stars' \citep{Gillessen2017} and thus likely has the strongest wind of the S-stars, this result confirms that the $\lesssim 100$ other S-stars can safely be neglected in calculations of accretion in the galactic center.

The flow structure at the innermost radii that we have outlined here could have a significant impact on GRMHD simulations of accretion onto Sgr A* and their predicted observational properties.  Polar inflow might directly oppose jet formation, while the outflow/inflow structure might be less susceptible to the build up of MRI turbulence if the inflow/outflow times are short compared to the MRI growth time. On the other hand, it is possible that the opposite will occur if an outflow from small radii disrupts the polar inflow we find in our simulations. Directly incorporating the flow properties found here as initial and boundary conditions in future horizon scale simulations will be a primary focus of future work.

A key possible limitation of our simulation is the neglect of magnetic fields, which can be a significant source of angular momentum transport in accretion discs and might alter the picture presented here.  Magnetic fields might be important both by generating magnetic braking of the inflow and/or via the MRI.  Furthermore, anisotropic conduction and viscosity along field lines may significantly alter the dynamics of the flow and suppress the accretion rate (e.g., \citealt{Johnson2007,Shcherbakov2010}).  MHD simulations of the problem studied here with and without conduction/viscosity will be carried out in the near future. 

This is a particularly exciting time to be studying the galactic center, in which both observations and theory are rapidly pushing the boundary of what is feasible.  In the not too distant future, as computational resources continue to improve, we may be able to simulate the entire dynamical range of accretion from the parsec scale of the WR stars all the way down to the event horizon of Sgr A*, even while including some of the non-ideal, collisionless physics important in this hot, low density plasma.  Moreover, as the EHT and GRAVITY continue to take data, we will be able to compare directly to spatially resolved observations of the event horizon while self-consistently making predictions about the X-ray, infrared, and radio data at larger radii.    Such a wealth of information combined with the computational and theoretical horsepower already being put in place will bring us that much closer to solving many of the outstanding questions related to the emission from the galactic center and inform our knowledge of low-luminosity AGN more generally.

\section*{Acknowledgments}
We thank the anonymous referee for several helpful comments.  We thank C. Gammie, J. Cuadra, J. R. Lu, R. Genzel, S. Gillessen, and K. El-Badry for useful discussions, as well as all the members of the horizon collaboration, \href{http://horizon.astro.illinois.edu}{http://horizon.astro.illinois.edu}, for their advice and encouragement.  We thank F. Baganoff for making the \emph{Chandra} X-ray data available to us for use in comparison to our simulations.  We thank D. Fielding for advice and help on plotting formats and give a double portion of thanks to C. J. White for his frequent aid in using the {\tt Athena++} code. Finally, we thank N. M. Lemaster and P. F. Hopkins for freely providing the preliminary coding framework that we built upon to calculate the stellar wind source terms.  This work was supported in part by NSF grants AST 13-33612, AST 1715054, AST-1715277, \emph{Chandra} theory grant TM7-18006X from the Smithsonian Institution, a Simons Investigator award from the Simons Foundation, and by the NSF
through an XSEDE computational time allocation TG-AST090038
on SDSC Comet.   SMR is supported in part by the NASA Earth and Space Science Fellowship.  This work was made possible by computing time granted by UCB on the Savio cluster.

\bibliographystyle{mn2efix}
\bibliography{stellar_wind}

\appendix

\section{Angular Momentum Distribution of a Single Accreting Star}
\label{app:single_star}
In this Appendix we briefly describe a toy model of a single accreting star that can be used to qualitatively explain the scalings of $\dot M_{in}$ and $\rho$ observed in our simulation.

Consider the case of a single wind-emitting star in a circular orbit around the black hole at a distance of $r_{orbit}$.  Let the coordinate system be aligned such that the $z$-direction is aligned with the angular momentum of the orbit and consider the time at which the star is located at $(x,y) = (r_{orbit},0)$. By construction, the $y$-component of the specific angular momentum of the emitted gas is thus 0.  We define a spherical polar coordinate system $(r,\theta,\phi)$ centered on the star so that $\theta = 0$ corresponds to the $y$-direction, or equivalently the direction of the orbital velocity vector. For a given $\theta$, assuming that pressure effects are negligible, stellar wind material that is emitted at some $\phi_0$, with $x$-component of the angular momentum $l_x = \sin(\phi_0)\sin(\theta)v_{orbit}r_{orbit}$, will travel around the black hole and eventually collide with the material emitted at $-\phi_0$, with $l_x = -\sin(\phi_0)\sin(\theta)v_{orbit}r_{orbit}$.  This collision results in a shock that converts the angular momentum in the $x$ direction to internal energy. On the other hand, the $z$-component of the specific angular momentum of the stellar wind material can be written as 
\begin{equation}
l_z = \left(v_{orbit} + \cos(\theta)v_{wind}\right)r_{orbit}, 
\end{equation} 
which is bounded by $l_{min} = (v_{orbit}-v_{wind})r_{orbit}$ and $l_{max} = (v_{orbit}+v_{wind})r_{orbit}$.

If we define the ``distribution function,'' $f_x(x)$, of the mass ejected from the star per unit time with respect to some variable $x$ as 
\begin{equation}
\dot M_{wind} = \int \limits_{x_{min}}^{x_{max}} f_x(x) dx,
\end{equation}
then, using the relation $dl_z = - v_{wind}r_{orbit} \sin(\theta) d \theta$, we have:
\begin{equation}
f_{l_z}(l_z) = \frac{1}{2} \frac{\dot M_{wind}}{v_{wind}r_{orbit}} = \textrm{const.},
\end{equation}
with the limits $l_{min} < l_z <l_{max}$. This implies that the rate at which material with specific angular momentum $l_z$ is emitted from the star is proportional to $ l_z$.   For the case of $v_{wind}>v_{orbit}$, some material will have $l_z<0$ and will ultimately collide with an equal amount of material containing $-l_z$.  Therefore, we define a ``net'' distribution function as
\begin{equation}
f_{l_{z,net}}(l_{z,net}) = 
\left\{
\begin{array}{ll}
      \frac{1}{2} \frac{\dot M_{wind}}{v_{wind}r_{orbit}} & |l_{min}|\leq l_{z,net} \leq l_{max} \\
      \delta(l_{z,net})\frac{1}{2} \frac{\dot M_{wind}}{v_{wind}r_{orbit}} \left(|l_{min}| - l_{min}\right) & \textrm{else,} \\
\end{array} 
\right.
\label{eq:dmdlnet}
\end{equation}
where $\delta(l_{z,net})$ is the Dirac-delta function.  Equation \eqref{eq:dmdlnet} has three interesting extremes.  First, when $v_{wind} \ll v_{orbit}$, the stellar wind material has angular momentum predominately equal to the orbital angular momentum of the star with very little scatter.  This would result in the formation of a ring of material co-rotating with the star and almost no accretion.  Second, when $v_{wind} \gg v_{orbit}$, the net $z$-component of the angular momentum is essentially 0, and Bondi-Hoyle-Lyttleton \citep{Hoyle1939,BHoyle1944} type accretion is expected, with an accretion rate onto the black hole $\propto (v_{orbit}/v_{wind})^2 \dot M_{wind}$.  Finally, the case of interest for this work is when $v_{wind} \sim v_{orbit}$, which results in an extended distribution of angular momentum with $ 0< l_z< 2 l_{orbit}$ (where we have dropped the ``net'' subscript because $l_z$ is everywhere $>0$). This wide range of angular momenta directly corresponds to a wide range in circularization radii, $r_{circ} = l_z^2/(GM_{BH})$, with distribution function
\begin{equation}
f_{r}(r_{circ}) = \frac{1}{4} \frac{v_{orbit}}{v_{wind}} \frac{\dot M_{wind}}{r_{orbit}} \sqrt{\frac{r_{orbit}}{r_{circ}}},
\end{equation}
over the range $0<  r <4 r_{orbit}$.  If we assume that material can only accrete until it reaches $r = r_{circ}$, at which point it either settles into a disc or is converted into outflow, then we expect $\dot M_{in} \propto \sqrt{r}$.  Finally, if the radial velocity of the in-falling matter is essentially free-fall, $v_r \propto r^{-1/2}$, we obtain a power law scaling for the mass density: $\rho \propto r^{-1}$. 

Even when the stellar winds of multiple stars are contributing to the accretion flow, this picture should give a reasonable qualitative understanding as long as the stars are sufficiently isolated from one another.  That is, for a given pair of stars, as long as the time for the two winds to collide is longer than the shortest free fall time, then the winds will not have a chance to shock and alter the angular momentum profile before plunging to smaller radii.  

The true problem is, of course, more complicated, as, for example, the majority of the orbits are eccentric and pressure effects are non-negligible.  However, this simple picture provides an intuitive understanding of our simulation results with a physical justification for the mass density and accretion rate scalings, directly relating them to the angular momentum distribution of a single star and the small number of accreting stars.

\end{document}